\documentclass[fleqn,usenatbib]{mnras}

\usepackage[T1]{fontenc}
\usepackage{booktabs}
\usepackage{graphicx}

\DeclareRobustCommand{\VAN}[3]{#2}
\let\VANthebibliography\thebibliography
\def\thebibliography{\DeclareRobustCommand{\VAN}[3]{##3}\VANthebibliography}

\usepackage{graphicx}	
\usepackage{amsmath}	
\usepackage{amssymb}	
\usepackage{color}
\usepackage{xcolor}
\usepackage{caption}
\usepackage{subcaption}
\usepackage{newtxtext,newtxmath}
\usepackage{deluxetable}
\usepackage{threeparttable}

\newcommand{\oiii}{[O~{\footnotesize III}]}
\newcommand{\siii}{[S~{\footnotesize III}]}

\title[K2 Binary PN Central Stars]{Binary Central Stars of Planetary Nebulae Identified With {\it Kepler/K2}}

\author[Jacoby et al.]{
George H. Jacoby,$^{1,2}$\thanks{E-mail: george.jacoby@noirlab.edu}
Todd C. Hillwig,$^{3}$
David Jones,$^{4,5}$
Kayla Martin,$^{6,7}$
Orsola De Marco,$^{6,7}$
\newauthor
Matthias Kronberger,$^{8}$
Jonathan L. Hurowitz,$^{9}$ Alison F. Crocker,$^{10}$
and Josh Dey$^{10}$
\\
$^{1}$Lowell Observatory, 1400 W Mars Hill Road, Flagstaff, AZ 86001, USA\\
$^{2}$NSF's NOIRLab, 950 N. Cherry Ave., Tucson, AZ 85719, USA\\
$^{3}$Department of Physics and Astronomy, Valparaiso University, Valparaiso, IN 46383, USA\\
$^{4}$Insituto de Astrof\'isica de Canarias, E-38205 La Laguna, Tenerife, Spain\\
$^{5}$Departamento de Astrof\'isica, Universidad de La Laguna, E-38206 La Laguna, Tenerife, Spain\\
$^{6}$Department of Physics and Astronomy, Macquarie University, Sydney, NSW, 2109, Australia\\ 
$^{7}$Astronomy, Astrophysics and Astrophotonics Research Centre,
Macquarie University, Sydney, NSW, 2109, Australia\\
$^{8}$Deep Sky Hunters Collaboration\\
$^{9}$Department of Earth, Atmospheric and Planetary Sciences, Massachusetts Institute of Technology, Cambridge, MA 02139, USA\\
$^{10}$Department of Physics, Reed College, Portland, OR 97202, USA\\
}

\date{Accepted XXX. Received YYY; in original form ZZZ}

\pubyear{2021}

\begin{document}
\label{firstpage}
\pagerange{\pageref{firstpage}--\pageref{lastpage}}
\maketitle

\begin{abstract}
We present the identification of 34 likely binary central stars (CSs) of planetary nebulae (PNe) from {\it Kepler/K2} data, seven of which show eclipses. Of these, 29 are new discoveries. Two additional CSs with more complicated variability are also presented. We examined the light curves of all `possible', `likely' and `true' PNe in every {\it Kepler/K2} campaign (0 through 19) to identify CS variability that may indicate a binary CS. For Campaigns 0, 2, 7, 15, and 16 we find 6 likely or confirmed variables among 21 PNe. Our primary effort, though, was focused on Campaign 11 which targeted a Galactic bulge field containing approximately 183 PNe, in which we identified 30 candidate variable CSs. The periods of these variables range from 2.3~h to 30~d, and based on our analysis, most are likely to be close binary star systems. We present periods and preliminary classifications (eclipsing, double degenerate, or irradiated systems) for the likely binaries based on light curve shape. From our total sample of 204 target PNe, with a correction for incompleteness due to magnitude limits, we calculate a binary fraction of PN central stars to be 20.7 percent for all the observed PNe, or 23.5 percent if we limit our sample only to `true' PNe. However these fractions are almost certainly lower limits due to the large angular size of the \emph{Kepler} pixels, which leads to reduced sensitivity in detecting variability, primarily as a result of dilution and noise from the nebula and neighbouring stars. We discuss the binary population of CSs based on these results as part of the total known sample of close binary CSs.
\end{abstract}




\section{Introduction}

A planetary nebula (PN) represents a stage in stellar evolution in which nearly half the star (or more) is ejected into the ISM along with various nucleosynthesis products, especially He, C, and N. PNe also produce significant quantities of dust  \citep{otsuka17} that play an important role in subsequent star formation and Galactic evolution. Due to their intrinsic brightness, PNe are key probes of extragalactic distances \citep[via the PN luminosity function, or PNLF;][]{jacoby89}, kinematics \citep{coccato09} and chemical abundances \citep{corradi15b}.

When PNe form from a binary stellar system, spectacularly complex morphological shapes can develop \citep{boffin19, decin20}. And the central binary has been shown to have a clear connection to the nebular shaping \citep[e.g.,][]{hillwig16}. In fact, the binary pathway may be the most common formation mechanism for the PN phenomenon \citep{demarco09}. If correct, then several mysteries can be explained: the complex and non-spherical shapes for bright PNe represent 80--85 percent of PNe \citep{parker06} while spherical shell PNe, which represent a small fraction of PNe, are always faint \citep{soker97, jacoby2010}; the low number of PNe in the Galaxy relative to that expected if all eligible stars become PNe \citep{moe2006}; the existence of globular cluster PNe \citep{jacoby17, minniti19}; the cause of H-deficient PNe and their associated high abundance discrepancy factors \cite[ADFs;][]{wesson18, jacoby2020}; and, the unexpectedly high brightness of the PNLF cutoff in E/S0 galaxies \citep{davis18} having no young stellar population. 

Beyond understanding the properties of the general population, close binary PNe are invaluable probes of the common-envelope phase of binary evolution, considered a critical point in the formation of all close binaries involving at least one evolved component \citep[white dwarf, neutron star, or black hole;][]{iaconi19,jones20}.

However, the statistics to support the dominance of binaries in PNe are lacking. The only binary fraction that is reasonably well quantified is that of short period binaries, at $\sim$15 percent \citep{miszalski09, bond00}. This fraction is a strict lower limit because there are a number of ways in which even a short period companion can elude detection \citep[e.g., by the system's inclination, or companion's size;][]{demarco08,jones17}. An attempt at determining the fraction of binaries, including longer periods, was carried out with a near-infrared excess method with tantalising, but uncertain, results due to the faintness of typical companions when compared with the brightness of central stars \citep{DeMarco13,Douchin15,Barker18}. 

Space-based photometric surveys can be used to find short period binaries in a more complete way because they are
unaffected by atmospheric effects that limit the precision of the measurements. In this way it should be possible to obtain a more accurate estimate of the short period binary fraction, one that may even allow one to calculate a completion factor. A first attempt to use {\it Kepler} to find some of those short period binaries that would be missed from ground-based surveys \citep{demarco15} resulted in yet another promising result, with most detections below the limit for ground-based variability detection, but a valid binary fraction was hampered by the very low number of PN in the original {\it Kepler} field (six objects). 

Initial survey work in this direction has also been carried out with the {\it TESS} space telescope, again with an indication of a higher incidence of photometric variability consistent with central star duplicity \citep{aller20}. Gaia data has also been used to identify variable central stars of planetary nebulae (CSPN) \citep{chornay21}. While there are not enough data points to identify variability type for most targets, they do recover a large fraction of the known variable, binary CSPNe in their sample. They also identify four new periodic and likely binary CSPNe and give a final list of 58 new candidate binaries. 

With the {\it Kepler/K2} missions, the {\it Kepler} satellite was able to monitor 19 additional fields beyond the original field. Of these, six fields included known PNe, thereby providing an opportunity for significantly enhancing the identification statistics for PNe having binary central stars (CS). This study therefore picks up where the pioneering work of \citet{demarco15} leaves off. Now, with 30 times as many PN targets, the photometric statistics for binary central stars can be improved.

We describe the observational data in Section~\ref{sec:obs} and our reduction tools and procedures in Section~\ref{sec:analysis}. The results of the search for variables are presented in Section~\ref{sec:nonvar-PN} (non-variables) and Section~\ref{sec:var-PN} (variables). We discuss the implications of the variable detections for the frequency of PNe with close binary central stars in Section~\ref{sec:discuss} along with the observational and astrophysical uncertainties in estimating that fraction, including a comparison to previous binary fraction estimates.


\section{Observations}
\label{sec:obs}

Our principal source of observational material derives from the {\it Kepler/K2} mission.
{\it Kepler/K2}  observations offer a rare space-based opportunity for observing variable stars of all types with excellent sensitivity. We obtained data for PNe in six fields, including Campaign~11, a PN-rich field in the direction of the Galactic Bulge, positioned at RA(J2000)$=$260$\fdg$38750, Dec(J2000)$=-$23$\fdg$97583 (b$=$1$\fdg$2600, l$=$7$\fdg$2119).

\subsection{Pros and Cons of Using {\it Kepler/K2}}
\label{subsec:pros-cons}

The {\it Kepler/K2} missions offer considerable advantages over ground-based surveys, but also have a number of downsides. 

Compared to ground-based surveys for variability, the chief advantages of using the Kepler satellite for detecting CS variability are that issues due to weather, scheduling, cadence, and diurnal and annual cycles can be avoided, leading to dramatically improved amplitude sensitivity. The value of that sensitivity advantage is magnified by the lack of an atmosphere that introduces color-dependent extinction systematics. In addition, the {\it K2} target selection is completely unbiased in the sense that campaign fields are selected with no {\it a priori} awareness of the PN population in the field. 

There are, however, several disadvantages of {\it K2}, including:

\begin{itemize}
    \item {a campaign duration that is limited to about 90 days and is not reproducible;}
    \vspace {0.5ex}
    
    \item {poor spatial resolution of the observations (pixels span 3.98\arcsec\ and  the measuring aperture is $\sim$10\arcsec\ ) leading to significant dilution of the light curve amplitude and an uncertain identification of the true variable in crowded fields;}
    \vspace {0.5ex}
    
    \item {a single choice of observing bandpass (410-850 nm) that may not optimize the sensitivity to variability;}
    \vspace {0.5ex}
    
    \item {PNe in the \emph{Kepler/K2} fields were all observed at 30 min cadence. The shortest period binary CSPNe have periods of only four to five times the exposure time leading to potential amplitude dilution and loss of resolution in the shape of the light curve, but see individual object descriptions below for more discussion;}
    
    \item {the presence of significant data systematics due to noise and artifacts arising from spacecraft motion, primarily the 5.885h periodic thruster bursts and their harmonics.}
\end{itemize}

The last issue has been largely, though not completely, addressed with extensive and meticulous reduction software and careful analysis by the Kepler team. Nevertheless, the  sensitivity to variability suffers somewhat relative to the original {\it Kepler} mission.

Table~\ref{tab:campaigndata} summarizes the number of PNe observed during each of the {\it K2} campaigns. PNe were explicitly requested as targets only during campaigns 0, 2, 7, and 11. Me~2-1 was part of the very large Campaign~15 GO program 15009 (Charbonneau) regarding planets around FGKM stars. Abell~30, which was observed in Campaign~16, was requested as part of GO programs 16071 (Boyd) and 16901 (Lewis) for studies of x-ray sources and white dwarfs. The table also lists the number of variable  CSPNe that we identified, although two variable objects were removed from consideration as binaries.

\begin{table*}
\centering
\caption{Number of PNe in {\it K2} Campaign Fields }
\label{tab:campaigndata}
\begin{tabular}{llcrr}
\hline
Campaign & Observation & Useful   & Number & Number      \\
         & Period      & Duration (days) & of PNe & of Variables \\
\hline
0   & 2014 Mar 08 to 2014 May 27 & 36 &   3 &  0 \\
2   & 2014 Aug 23 to 2014 Nov 13 & 79 &   4 &  1 \\
7   & 2015 Oct 04 to 2015 Dec 26 & 83 &  12 &  3 \\
11a & 2016 Sep 24 to 2016 Oct 18 & 23 & 183 & 30 \\
11b & 2016 Oct 21 to 2016 Dec 07 & 48 & -   & -  \\
15  & 2017 Aug 23 to 2017 Nov 20 & 90 &   1 &  1 \\
16  & 2017 Dec 07 to 2018 Feb 25 & 80 &   1 &  1 \\
\hline
\end{tabular}
\end{table*}


\subsection{Ground-based Observations}
\label{subsec:ground}

We examined every variable candidate using data from The Asteroid Terrestrial-impact Last Alert System (ATLAS), when available, and a few selected candidates using the Las Campanas 1-m Swope telescope, the Kitt Peak 4-m Mayall telescope, and the SARA 0.6~m telescope at CTIO in order to independently measure periods and amplitudes. We also accessed the PanSTARRS database to estimate CS magnitudes and, in some cases, the geometry of the nebula. We describe below the characteristics of those observations.

\emph{ATLAS:} the ATLAS sky survey telescope (Tonry et al. 2018) observes the available sky every few nights in two filters: cyan (420--650 nm) and orange (560--820 nm). Typically, there have been $\sim$1000 visits for each object over the course of 3--4 years, but the frequency of visits is low compared to the frequencies of the variability of our targets. Consequently, finding a clear period from ATLAS data alone is compromised by aliasing. Nevertheless, we attempted to validate the known {\it K2} periods with ATLAS. In principle, the period's precision can be improved over the {\it K2} value with the leverage of the extended ATLAS time base.

Unfortunately, we were able to phase only a few targets with the ATLAS data, because, like K2, the targets tend to be intrinsically faint and in crowded fields.
\vspace{1ex}

\emph{Las Campanas Swope 1-m:} we used the Las Campanas Observatory (LCO) 1-m Swope telescope to image many of the Campaign~11 targets on five nights beginning with UT 2018 August 04. The goals were to obtain images of the nebulae for morphological classification, to verify the central star identifications, and to obtain some multi-epoch photometry to confirm that the central star was the true variable. The camera on the Swope uses an E2V 4K x 4K CCD~231-84 having a field of view of 29.7×29.8 arcmin with 0.435\arcsec\  pixels. Image quality was typically 1.0\arcsec\ during the run.

Due to weather and technical issues, we were able to obtain multiple epoch (3-4 visits) observations for only five of the {\it K2} variable PNe (M~3-42, PHR~1713-2957, PHR~1734-2000, PTB~26, and Th~3-12), primarily on the first night.
\vspace{1ex}

\emph{Kitt Peak Mayall 4-m:} we observed the Campaign~11 targets Terz~N19 and V972~Oph on the nights of 2017 August 17 and 2017 August 18, respectively, at the Kitt Peak Mayall 4-m telescope using the KOSMOS spectrograph \citep{martini2014} in imaging mode. Exposure times were 180~s in the {\it r}-band yielding sampling intervals of 240~s--300~s.
\vspace{1ex}

\emph{SARA-CT:} the Southeastern Association for Research in Astronomy (SARA) consortium operates three telescopes at different sites around the world \citep{keel17}. One of those is the 0.6-m telescope at CTIO designated as SARA-CT. Observations of the Campaign 11 target PTB~26 were conducted with the SARA-CT telescope using the FingerLakes Instruments CCD in the \emph{V} and \emph{I} bands. Details for this telescope/CCD combination are available in \citet{keel17}. We calculated differential magnitudes for the CS using three comparison stars, which were all found to be photometrically stable to within 0.02 magnitudes.


\section{K2 Data Analysis} 
\label{sec:analysis}

\subsection{Special Considerations in Analyzing {\it Kepler/K2 Data}}
\label{subsec:special}

As noted in Section~\ref{subsec:pros-cons}, {\it Kepler/K2} has 3.98\arcsec\ pixels, leading to several analysis issues. These are important in all fields, but are most severe in the Campaign~11 field near the Galactic Center.
\vspace{1ex}

\emph{Crowding and dilution:} with large measuring apertures ($\sim$10\arcsec), crowding from nearby stars is always a concern. The situation is extreme in the dense stellar field of Campaign 11 where photometry of every object is contaminated by the light from many neighboring stars in addition to the inevitable contribution from the nebula.  Neighboring stars also add photon noise to the measurements, further limiting the detection for variability.

The effects of dilution are exacerbated by the large distances for many PNe, especially for Campaign~11 where most PNe are near the Galactic Center. Also for Campaign~11, there is dust along the line of sight, and therefore these PNe tend to be fainter than PNe in other campaigns. In a few test cases where ground-based observations are available, dilution factors of 12 or greater were measured, and so, modeling the binary systems is not possible with {\it K2} data alone.

Other studies have addressed the crowding problem for {\it K2} with particular science cases in mind. \citet{poleski2019} and references therein describe several targeted schemes (e.g., for transits or microlensing). Without prior knowledge of the light curves, and the unavoidable effects of nebula dilution, we did not attempt to apply those techniques.
\vspace{1ex}

\emph{Target confusion:} the large pixels also lead to a confusion problem where the true source of the variability may be a nearby star unrelated to the PN. As noted in \ref{subsubsec:checkscript}, we encountered a clear case for the target MGE~003.0836+01.643 which appeared to be an RR Lyr star and therefore was easily flagged as a false variable CS identification early in the analysis.

In principle, we should be able to mitigate some of the effects of dilution from neighboring stars using point-spread-function fitting (PSF) photometry. We found, however, that the {\it Kepler/TESS} tool, \textsc{K2SC}, \citep{aigrain16} did not perform well in this application. Rather, we obtained the best results using the `Pre-search Data Conditioned Simple Aperture Photometry' (PDCSAP) light curve files provided through the {\it Kepler/K2} pipeline and obtained via MAST. Unfortunately, the PDCSAP algorithm is not available in the open-source software, but the \textsc{LightKurve} `Self Flat-Fielding' (SFF) algorithm \citep{vanderburg2014} yields good results and is what we use in our scripts described below in Section~\ref{subsec:tools}.

\subsection{Tools}
\label{subsec:tools}

A variety of publicly available tools have been applied to the reduction and analysis of this {\it K2} survey. Specifically, we employed the following tools extensively:

\begin{itemize}
    \item {The Mikulski Archive for Space Telescopes (MAST) was used to access the {\it K2} images and light curves using the {\it K2} Data Search \& Retrieval tool, as well as the Python interface provided by \textsc{LightKurve} \citep{lightkurve}. MAST was also used to access data from ATLAS \citep{tonry18} via the Casjobs query tool \citep{heinze18} for selected targets.}
    \vspace{1ex}
    
    \item {The Hong Kong/AAO/Strasbourg H\,$\alpha$ planetary nebula database (HASH) served as our reference for classifying PNe as `true', `possible', etc, and also as a source of coordinates and images to assess crowding. See \citet{parker16}}.
    \vspace{1ex}

    \item {The SIMBAD database \citep{wenger2000} was used to check for previously known variable stars in the vicinity of our targets.}
    \vspace{1ex}
    
    \item {We derived the periodicities of the targets using the Periodogram Service provided by the NASA Exoplanet Archive, the algorithms within \textsc{LightKurve}, and our own software tools.}
    
\end{itemize}


\subsubsection{The Script to Find Variables} 
\label{subsubsec:findscript}

The extensive Python package, \textsc{LightKurve}, provided the basis for two scripted tools that we developed to expedite the analysis of the survey sample. 

The first script performs the following services for each of the {\it K2} targets in the following order:

\begin{enumerate}
    \item {Collects the pixel file for the target.}
    \vspace{1ex}
    
    \item {Displays images from the pixel files with and also without the default photometry aperture mask. These images allow us to assess the crowding from neighboring stars and nebulae, the accuracy of the mask placement, and the appropriateness for the pixel selection used in the photometry.}
    \vspace{1ex}
    
    \item {Collects the light curves for the target from MAST, including both light curves for Campaign~11a and Campaign~11b\footnote{The release notes for Campaign 11 state that it was broken into two segments, Campaign~11a and Campaign C11b, as a result of an error in the initial roll-angle used to minimize solar torque on the spacecraft. This error would lead to targets rolling out of their pixel apertures by the end of the campaign. To avoid that, a correction was applied twenty-three days into the campaign by applying a $-0.32^{\circ}$ roll offset. The correction would then require new target aperture definitions for the second part of the campaign. The observing durations for the two campaigns were 23 days and 48 days, respectively, for Campaign 11a and 11b.}.  We explored the use of several different pipelines for our light curve analysis, including EPIC Variability Extraction and Removal for Exoplanet Science Targets or EVEREST \citep{luger2016} and Self Flat-Fielding, or K2SFF \citep{vanderburg2014}, but we ultimately adopted the Kepler ``ktwo'' Pre-search Data Conditioning Simple Aperture Photometry, or PDCSAP, light curves as being the most effective for this application.}
    \vspace{1ex}
    
    \item {Displays the light curve, as downloaded from MAST.}
    \vspace{1ex}
    
    \item {Runs a Lomb-Scargle \citep{vanderplas18} search for periodicities in each light curve.}
    \vspace{1ex}
    
    \item {Plots a phased light curve for each significant period found in the Lomb-Scargle search, as well as a binned version of the phased light curve, which can improve the visibility of variability in some cases and generally yields a better estimate of the peak-to-peak amplitude of the variability excursion.}
    \vspace{1ex}
    
    \item {Saves all the figures and images, as well as the downloaded light curve, in local files.}
\end{enumerate}

Despite the 30 min cadence observations, the large number of data points (of order 2000) allowed us to search down to periods of only a few times 30 min. We found that our shortest recovered period (0.09503~d for Terz~N19) is well reproduced relative to short cadence ground observations due to the extreme oversampling of the phase-folded light curves. Broad, smoothly changing features experience some dilution in amplitude due to being undersampled in individual exposures, and sharp features are both diluted and broadened in orbital phase. However, even sharp features that are shorter than the exposure time can be recovered, though diluted in both amplitude and phase.

Because Campaign~11 was broken into two short segments, referred to as C11a and C11b, the script operates on both `sub-campaigns' in parallel, while treating them as independent data sets. In this way, we could assess the reliability of a period very easily by comparing the two periodograms.

\subsubsection{The Script to Check Variables} 
\label{subsubsec:checkscript}

The second script was developed in order to validate the origin of variability. That is, having detected a variable in the {\it K2} light curve for a target, we need to ensure that the intended target is truly the source of the variability rather than a nearby star whose light is spilling into the $\sim$10\arcsec\  photometry aperture. 

This check was motivated, in part, when the CS for MGE~003.0836+01.6435 in Campaign~11 appeared to have the signature of a classic RR~Lyr star. That situation seemed rather extraordinary, warranting further investigation. In fact, SIMBAD reports that OGLE~BLG-RRLYR-28966 \citep{soszynski14}, just 8.2\arcsec\ northwest of the PN, is an RR Lyr with the same period as our script yielded.

As a check, we ran the second script on all candidate variables. This script allows us to adjust the size and position of the photometry aperture mask anywhere within the pixel file. It then runs through the same processes as the first script and displays the periodograms and phased light curves derived from the default aperture mask and the adjusted aperture mask. By moving the mask off the PN and on to nearby pixels (i.e., stars), we can manually maximize the variability signal. If the signal falls off as we move the mask off the PN, then we build confidence that the PN is the source of variability. In only one other case did we find the situation where the signal increased as we moved significantly away from the PN central star, and we discuss that case (PHR~J1738-2419) in Section~\ref{sec:var-PN}. In seven other instances, we did find a slight improvement in the variability signal when moving the aperture mask by a single pixel or adjusting its shape; we attribute these improvements to minor coordinate/pointing errors and reductions in dilution.

It is important to note that although we did not identify any other known variables within the photometric apertures, this does not mean that the photometry is not contaminated or diluted by other stars inside the aperture.  To the contrary, a rudimentary check of Gaia eDR3 \citep{gaia_eDR3} indicates that the photometry of essentially all the candidates identified in this work is, to some extent, contaminated by nearby stars of similar brightnesses.  This serves to emphasise the need for ground-based confirmation of the candidate binaries identified.

\subsection{Procedure}
\label{subsec:procedure}

We processed every {\it K2} PN target through our Python variable finder script described in Section~\ref{subsubsec:findscript}. These data were also checked using the NASA Exoplanet Archive Periodogram Service tool and \textsc{Period04} \citep{lenz05}. Each light curve was explored visually to determine a most-likely classification in order to determine the orbital periods reported in Table~\ref{tab:var-PN-list}. This is because most period-search tools preferentially return periods with one cycle of variability. However, this photometric period is not always the orbital period. Ellipsoidal variability, for example, exhibits two cycles per orbit and many of these have identical, or nearly identical, maxima and minima. So, the proper orbital period for these objects is twice the photometric period that is found in period searches. Having multiple team members apply multiple tools provided independent checks on the most likely period.

We checked each variable candidate to be sure that the variability signal truly originated at the pixel location of the PN central star using the script to check variables.

\begin{figure}
\centering
\includegraphics[width=0.3\textwidth]{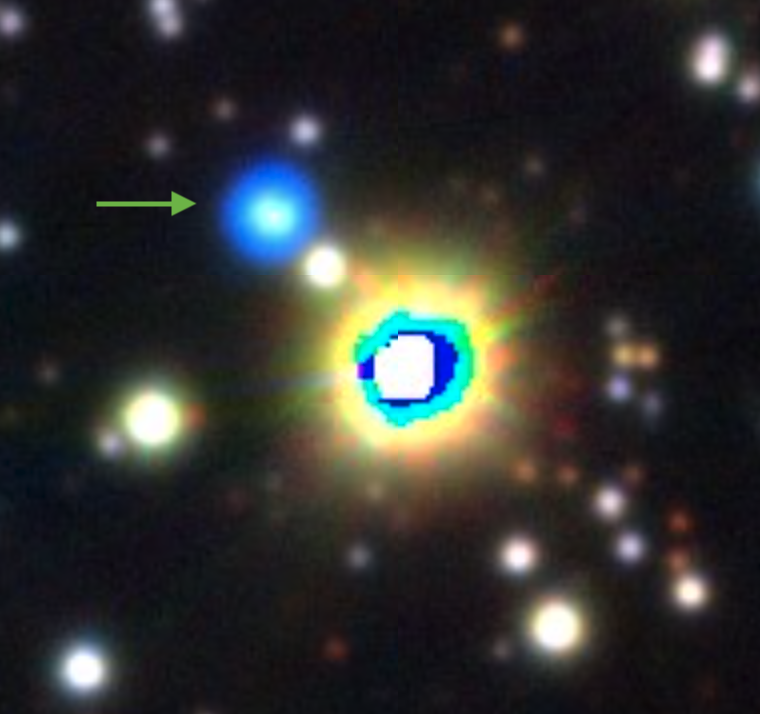}
\caption{A Sloan {\it gri} (blue, green, red) image of the PN candidate Kn~J1857.7-2349 from PanSTARRS. The faint nebula indicated by the arrow is about 4\arcsec\  in diameter and also appears in our LCO H\,$\alpha$ and [\ion{O}{iii}] images, but not in our {\it V} or {\it R} data. Coordinates (J2000) for this object are RA=284$\fdg$44333, Dec$=-$23$\fdg$82764. North is up and east is to the left; the image size shown is 45\arcsec\  (north-south) by 49\arcsec\  (east-west).}
\label{fig:KnJ1857_ps}
\end{figure}

\begin{figure}
\centering
\includegraphics[width=0.475\textwidth]{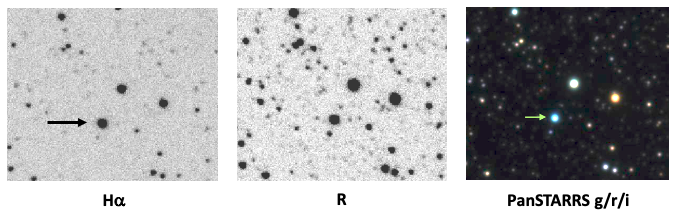}
\caption{Images of the PN candidate Kn J1838.0-2202 taken with the LCO 1-m Swope telescope in H\,$\alpha$ and {\it R}. Although subtle, the H\,$\alpha$ image is brighter than the {\it R}-band image relative to the other field stars. Coordinates (J2000) for this object are RA=279$\fdg$50667, Dec$=-$22$\fdg$04689. North is up and east is to the left; the image size shown is 83\arcsec\  (north-south) by 105\arcsec\  (east-west).}
\label{fig:KnJ1838_image}
\end{figure}


\begin{table*}
\centering
\resizebox{\textwidth}{!}{%
\begin{threeparttable}
\caption{Variable PNe in K2 Fields} 
\label{tab:var-PN-list}
\setlength\tabcolsep{3.6pt} 
\begin{tabular}{clccrccllc}
\hline
Campaign & PN Name & {\it K2} ID & RA (2000) & Dec (2000) & Period & {\it K2} Amplitude & Variable & Companion & HASH    \\
  &                   & (KIC)     &              &              & (Days).     & (Percent) & Type      & Type  & Status \\
\hline
 2 &  PC 12             & 205189451 & 250$\fdg$9741 & -18$\fdg$9533 &   0.5915(5)       & 4.3 &  ellipsoidal  &  DD     &   True  \\
\\
 7 &  CGMW 4-3783       & 215825200 & 283$\fdg$7703 & -23$\fdg$4702 &   0.6982(3)       & 2 &  ellipsoidal  &  DD     &   True  \\
 7 &  H 2-48            & 215837631 & 281$\fdg$6463 & -23$\fdg$4467 &   4.10(3)         & 3 &  wind:        &  none?  &   True  \\
 7 &  Y-C 2-32          & 216730154 & 283$\fdg$8778 & -21$\fdg$8277 &   30(3)           & 0.1 &  irradiation  &  ?      &   True  \\
\\
11 &  JaSt 52           & 222918303 & 266$\fdg$1551 & -26$\fdg$7904 &   3.12(8):        & 2.3 &  uncertain    &   ?     &   True  \\
11 &  PHR J1724-2302    & 235356385 & 261$\fdg$2217 & -23$\fdg$0430 &   4.55(15)        & 0.4 &  irradiation  &   MS    &   True  \\
11 &  Th 3-12           & 235658593 & 261$\fdg$2755 & -29$\fdg$7547 &   0.2640(5)       & 1.6 &  ellips/eclipses      &   ?   &   True \\
11 &  M 2-11            & 235899761 & 260$\fdg$1387 & -29$\fdg$0109 &   0.5746(8)       & 1.1 &  irradiation  &   MS    &   True  \\
11 &  Th 3-14           & 236512646 & 261$\fdg$4337 & -26$\fdg$9633 &   1.15:           &  15:: &  wind/IGW:    &  none?  &   True  \\
11 &  PTB 26*           & 238276515 & 262$\fdg$3052 & -16$\fdg$7954 &   0.152184(2)     & 22 &  ellips/eclipses  &   ?   &  Likely \\
11 &  Th 3-35           & 240616801 & 264$\fdg$6758 & -28$\fdg$7127 &   3.330(10)       & 1.2 &  eclipsing    &   ?     &   True  \\
11 &  H 1-17            & 240626683 & 262$\fdg$4192 & -28$\fdg$6728 &   2.07(2):        & 0.3 &  uncertain    &   ?     &   True  \\
11 &  M 4-4             & 242185530 & 262$\fdg$2095 & -30$\fdg$1292 &   12.5(9):        & 0.4 &  uncertain    &   ?     &   True  \\
11 &  H 2-13*            & 251248502 & 262$\fdg$7838 & -30$\fdg$1744 &   0.896255(10)        & 1.5 &  irradiation:  &   MS:    &   True  \\
11 &  MPA J1711-3112    & 251248513 & 257$\fdg$9042 & -31$\fdg$2059 &   0.9825(10)      & 1.7 &  ellipsoidal: &   DD    &   True  \\
11 &  PHR J1713-2957    & 251248519 & 258$\fdg$3954 & -29$\fdg$9608 &   0.1667(2)       & 0.8 &  irradiation  &   MS    &   True  \\
11 &  MPA J1714-2946    & 251248520 & 258$\fdg$7012 & -29$\fdg$7797 &   1.94(5):        & 0.2 &  uncertain    &   ?     &   True  \\
11 &  MPA J1717-2356    & 251248526 & 259$\fdg$2875 & -23$\fdg$9415 &   0.2073(5)       & 0.7 &  irradiation  &   MS    &   True  \\
11 &  PHR J1718-2441    & 251248531 & 259$\fdg$6783 & -24$\fdg$6903 &   2.07(2):        & 0.6 &  uncertain    &   ?     &   True  \\
11 &  M 3-39            & 251248537 & 260$\fdg$2979 & -27$\fdg$1936 &   4.75(15)        & 0.2 &  irradiation  &   MS    &   True  \\
11 &  K 5-31            & 251248541 & 260$\fdg$5542 & -29$\fdg$8767 &   11.9(6):        & 1.3 &  uncertain    &   ?     &  Likely \\
11 &  Terz N19          & 251248550 & 261$\fdg$3483 & -26$\fdg$1981 &   0.0950349(4)      & 6.2 &  ellips/eclipses  &   ?   &  True \\
11 &  PHR J1725-2338    & 251248552 & 261$\fdg$4242 & -23$\fdg$6422 &   0.20985(7)      & 5.4 & irradiation/eclipsing &   MS   &   True \\
11 &  M 3-42            & 251248557 & 261$\fdg$7491 & -29$\fdg$2589 &   0.3202(3)       & 0.3 &  ellipsoidal: &   DD    &   True  \\
11 &  Th 3-15*          & 251248559 & 261$\fdg$7942 & -27$\fdg$7326 &   0.1507715(3)      & 3.1 &  ellips/eclipses   &   ?   &  True \\
11 &  K 5-3             & 251248570 & 262$\fdg$6717 & -23$\fdg$7501 &   2.56(4):        & 0.2 &  uncertain    &   ?     &   True  \\
11 &  PHR J1734-2000    & 251248581 & 263$\fdg$5571 & -20$\fdg$0140 &   0.3367(5)       & 1.2 &  ellipsoidal: &   DD    &   True  \\
11 &  PPA J1734-3015    & 251248582 & 263$\fdg$6083 & -30$\fdg$2550 &   0.1702(3)       & 1.4 &  irradiated:  &   MS    &   True  \\
11 &  V972 Oph*         & 251248586 & 263$\fdg$6829 & -28$\fdg$1766 &   0.2800(13)      & 1.5 &  irradiation  &   MS    &   Possible \\
11 &  MGE 359.2412+02.3353 & 251248587 & 263$\fdg$6913 & -28$\fdg$3390 &   0.2682(4)       & 0.4 &  irradiation  &   MS    &   Possible \\
11 &  JaSt2 2           & 251248610 & 264$\fdg$6850 & -28$\fdg$1118 &   0.3716(6)       & 15 &  ellipsoidal: &   DD    &   Possible \\
11 &  PHR J1738-2419*    & 251248612 & 264$\fdg$7154 & -24$\fdg$3289 &   0.418839(6)     & 0.5 &  ellipsoidal  &   DD:    &   Possible \\
11 &  JaSt 12           & 251248613 & 264$\fdg$7471 & -28$\fdg$7783 &   4.357(13)       & 3.9 & eclipsing/ellipsoidal: &   DD:   &   Possible \\
11 &  Terz N 1567       & 251248634 & 266$\fdg$3679 & -25$\fdg$6368 &   0.1714(3)       & 0.2 &  irradiation  &   MS    &   True  \\
\\
15 &  Me 2-1            & 249356906 & 230$\fdg$5805 & -23$\fdg$6254 &   22(1)           & 0.1 &  irradiation  &  ?      &   True  \\
\\
16 &  Abell 30          & 211839836 & 131$\fdg$7228 & +17$\fdg$8795 &   1.060(3)        & 1.7 &  irradiation  &  MS     &   True  \\
\hline
\end{tabular}
\begin{tablenotes}
\item[*]Previously known variables: PTB~26 \citep{chornay21}, H~2-13 \citep{hlabathe15}, Th~3-15 \citep{hlabathe15,soszynski15}, V972~Oph \citep{Tappert2013}, PHR~J1738-2419 \citep{soszynski15}
\end{tablenotes}
\end{threeparttable}}
\end{table*}


\section{The Non-Variable PN Central Stars}
\label{sec:nonvar-PN}

Table~\ref{tab:nonvar-PN-list} lists the 168 PNe for which we found no variability. The table lists the PNe in order of the {\it K2} campaign, and sub-ordered by the {\it K2} identification numbers (KIC).

We include their PN status classifications adopted from HASH \citep{parker16}. Not all objects in the survey are confirmed as `true' PNe. Some may be classified as candidate PNe, or are described with a similar cautionary term, e.g., `likely', or `possible'. Nevertheless, we include them in the discussions below because those targets may eventually be shown to be true PNe. The tables include the designation status in the last column.

We adopt the classifications from the HASH database in which a PN is defined to be `true' (or T), `likely' (or L), or `possible' (or P). \citet{parker16} state their criteria as follows: `we assign objects as T that are confirmed PN with indicative multi-wavelength PN-type morphologies, PN spectral features and sometimes presence of an obvious CSPN. L indicates an object which is likely to be a PN but whose imagery or spectroscopy are not completely conclusive or unavailable. P indicates a possible PN where the morphology and spectroscopy are insufficiently conclusive, usually due to a combination of low S/N spectra, insufficient wavelength coverage, very low surface brightness or indistinct nebulosity.'

Table~\ref{tab:nonvar-PN-list} includes an additional status option -- `Poss-pre' for Possible (as defined earlier) -- but as applied to a `pre-PN' (or proto-PN) as noted in HASH. These are usually not-quite-PN that are very young objects on the evolutionary path to becoming a PN, sometimes displaying only Balmer emission due to the low temperature of the parent star relative to a typical PN.  The central star may appear as a very red star. There are nine objects in the `Poss-pre' class in our sample. 

There are a variety of reasons why we may not have seen periodic signals in the light curves for these objects with the most likely explanation being that they are not variable. But, other factors may come into play: insufficient sensitivity to variability due to dilution from nearby stars in these crowded fields, dilution from the nebula emission, incorrect identification of the central star, faintness of the central star because it (a) is inherently faint, (b) is distant, (c) suffers extinction along the line of sight and internal to the nebula, and (d) experiences a combination of these, or finally, the target is not a PN at all. In addition, objects can be mis-targeted because the coordinates were imprecise, but this issue does not appear to be a problem with K2's large pixels.

Three PN candidates deserve special note. The first two are not in current catalogs: Kn~J1857.7-2349 ($=$~Kn~79) is not confirmed as a PN. It was suggested as a compact PN candidate by \citet{kronberger2016}. Based on PanSTARRS and our  LCO H\,$\alpha$ and \oiii\ images taken with the 1-m Swope telescope, it has a very faint and small ($\sim$4\arcsec) nebula and a blue central star. The nebulosity likely arises from emission lines since it is not seen in our LCO {\it V} or {\it R} images. See Figure~\ref{fig:KnJ1857_ps}.

Kn~J1838.0-2202 ($=$~Kn~77) is also not confirmed as a PN. It was suggested as a PN candidate by \citet{kronberger2016} because it appears to be brighter in H\,$\alpha$ than in broadband {\it R} (see Figure~\ref{fig:KnJ1838_image}) by $0.95\pm0.05$ mag, and it has a blue star having {\it g}$\sim$14.7. Since it has no nebula visible in either H\,$\alpha$ or \oiii\  based on our LCO data, it cannot be confirmed as a PN.  The H\,$\alpha$ excess suggests that it is an emission-line object, possibly due to compact ionized gas surrounding the blue star (e.g., a young PN), or a disc, or stellar emission. 

Another object of note among the non-variables central stars, V~348~Sgr is a compact object with a small ($\sim$10\arcsec) faint extended H\,$\alpha$ nebula. It is considered to be a pre-PN, although listed as a `likely' PN in the HASH catalog. These conditions present a challenge for detecting CS variability because any CS signal is highly diluted by the bright, compact nebula. V~348~Sgr has a [WC11] central star \citep{Crowther1998} with R~Coronae~Borealis-like, semi-regular variability \citep{heck1985,DeMarco2002} exhibiting multi-mag excursions over time scales of tens of days to hundreds of days. The ATLAS data also capture variations on a scale of 4--5 mags over several hundred days, but we don't see any evidence for variability in the {\it K2} light curve. Only two [WR] central stars have ever been detected to be short period binaries \citep{hajduk10,manick15} so it would have been unusual to find this object to be a short-period variable. 

For completeness when preparing for the observations, we included all PN candidates known at the time, based on HASH, that were expected to fall on a detector in the field of view. In retrospect, we could have omitted several objects because their properties, as known today, are not consistent with measuring a central star flux. We comment on a few of these below as being illustrative of the challenges with these targets.

\emph{JaSt 51:} there is no optical counterpart to this PN candidate or its central star in either the HASH or PanSTARRS images. It was identified as a PN candidate in the \citet{jacoby2004} survey in the \siii\  line at 953.2 nm, but not recovered in H\,$\alpha$. It is likely heavily reddened, and so, detecting a variable central star in the optical is not practical.

\emph{IRAS 17253-2831:} this object is classified as a possible pre-PN in HASH. It has no emission-lines and the stellar signature is that of a cool red star.

\emph{RZPM 25:} this object is listed as a possible PN in HASH, but carries a comment that it is unlikely to be a PN. It is a stellar H\,$\alpha$ source but perhaps at the unlikely extreme of those objects that should be classified as `possible'.

\emph{JaFu 1:} a definitive result for JaFu~1 would be exciting, as it is a member of the globular cluster Pal~6 and therefore suspected of originating through a binary process \citep{jacoby17}. The CS, though, having {\it V}$=$23.1, is too faint for a variable detection.

\emph{PN G359.6+01.9:} there is no optical signature at the coordinates for this object. It is listed in HASH as a possible PN, but there is no evidence of a nebula in the images, nor any notes or comments about it.


\section{The Variable PN Central Stars}
\label{sec:var-PN}

Table~\ref{tab:var-PN-list} summarizes those PN candidates having periodic variability as identified through the Python variable finder script, ordered similarly to Table~\ref{tab:nonvar-PN-list}.

For each object that shows periodic variability we have classified the type of light curve based on shape and period. Column 8 in Table~\ref{tab:var-PN-list} lists the classification based on its variability type: eclipses, irradiation, and ellipsoidal modulation due to tidal distortion. H~2-48 and Th~3-14 are special cases where stellar winds or internal gravity waves (IGWs) may be the variability driver. Here we describe each of these classifications in more detail, along with comments on the period, amplitude, and potential complications in the {\it K2} light curves, where appropriate.

We use four notations in Table~\ref{tab:var-PN-list} to denote uncertainty in column entries. A single colon, `:', denotes an uncertain entry that is our current best estimate. The periods given are the best-fit period {\it if} the variability is periodic, and their uncertainty is given by the number in `parentheses' which applies to the last digit. A double colon, `::', is used in the case of Th~3-14 to note that the reported amplitude is an approximate value for typical cyclical variability, rather than a minimum-to-maximum value. A question mark, `?', in Table~\ref{tab:var-PN-list} indicates not enough information is available to provide a companion type.

We have checked these candidates against the OGLE catalog of variables to find that Th~3-15 and H~2-13 were previously identified as variables. In addition, there is an OGLE identification within the bounds of the PHR~1738-2419 nebula, but it is not the central star identified by HASH. We discuss each of these matched identifications in the context of each object's description.

\begin{figure*}
        \begin{subfigure}[b]{0.47\textwidth}
         \centering
         \includegraphics[width=\textwidth]{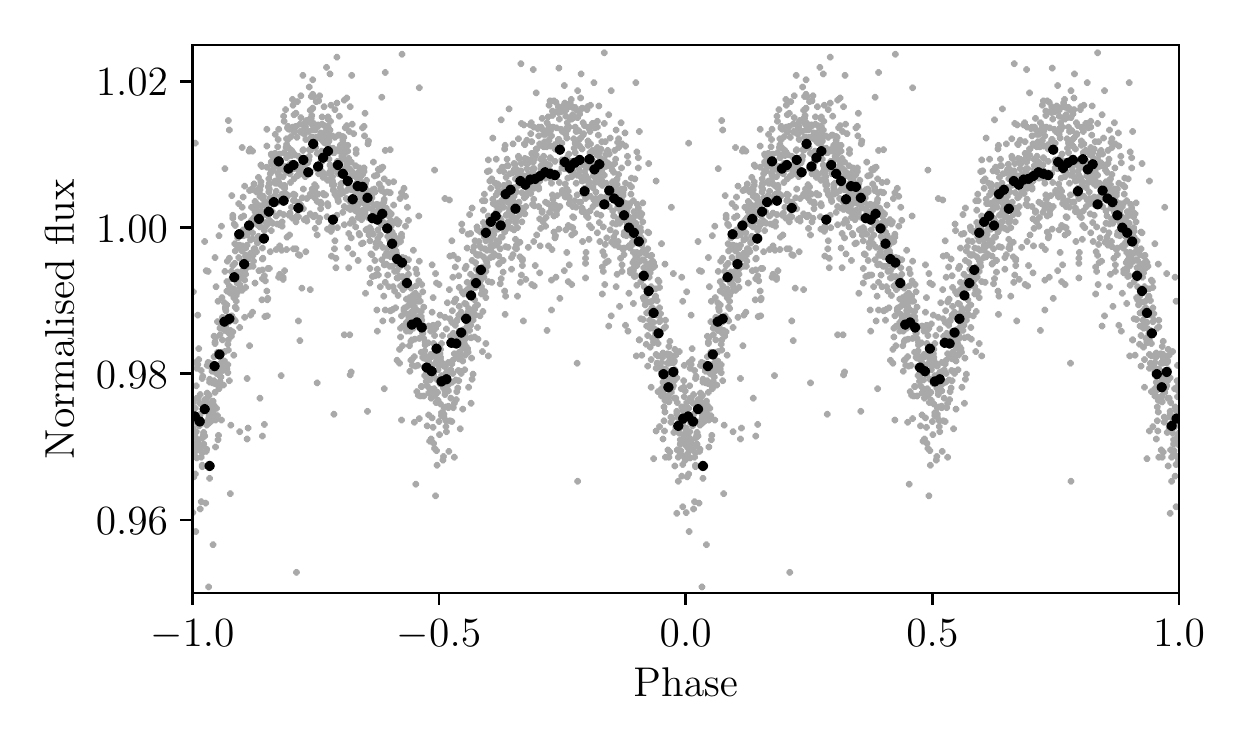}
         \caption{PC~12}
         \label{fig:PC12}
     \end{subfigure}
     \hfill
     \begin{subfigure}[b]{0.47\textwidth}
         \centering
         \includegraphics[width=\textwidth]{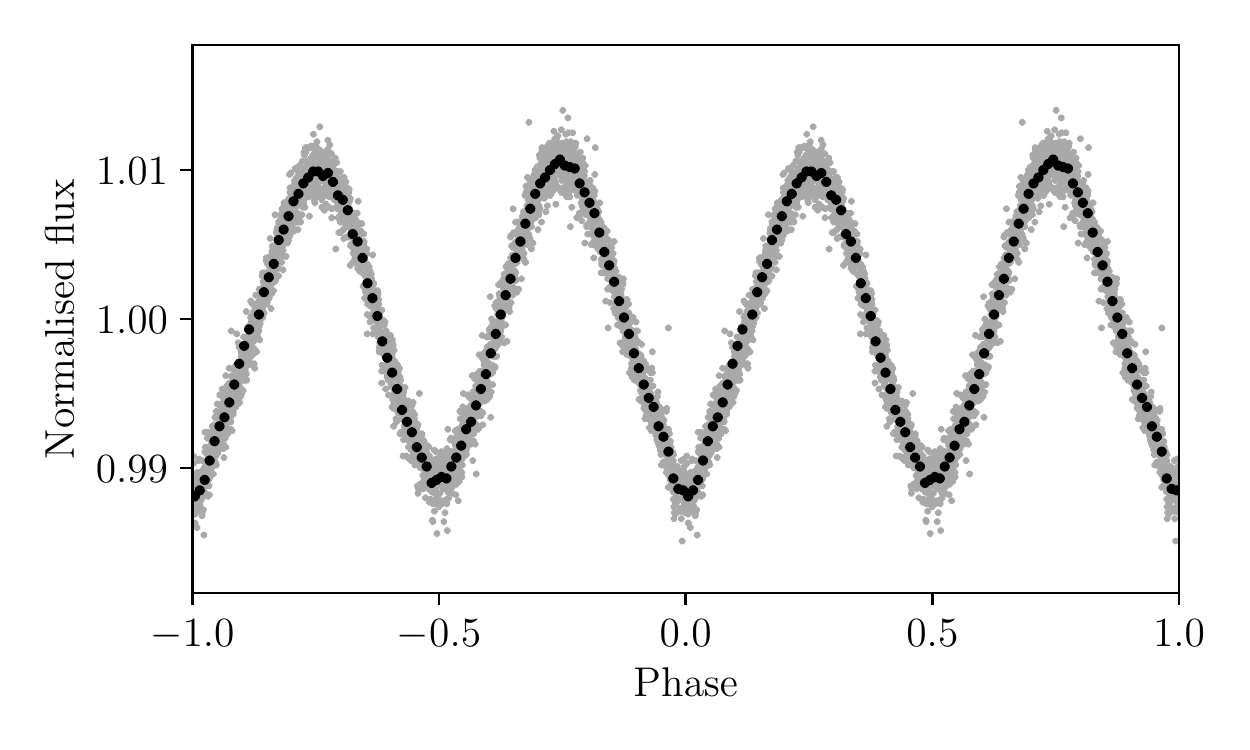}
         \caption{CGMW~4-3783}
         \label{fig:CGMW4-378}
     \end{subfigure}
     \\
     \begin{subfigure}[b]{0.47\textwidth}
         \centering
         \includegraphics[width=\textwidth]{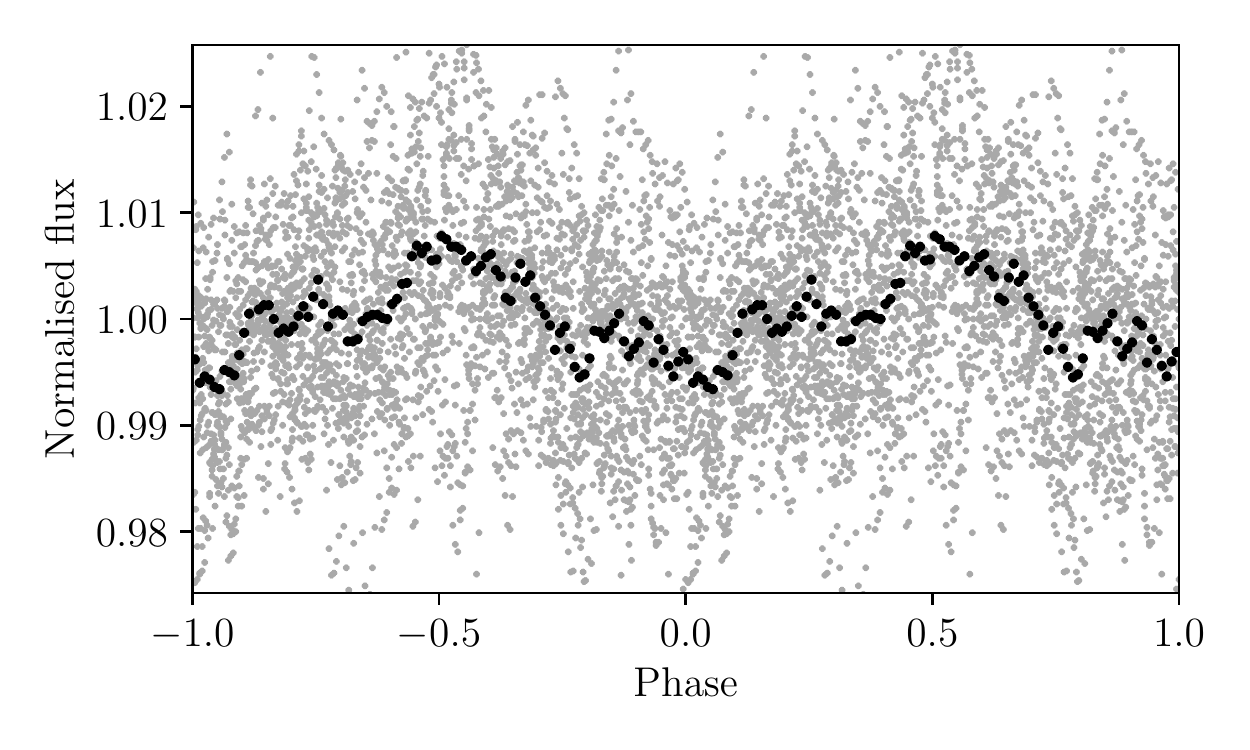}
         \caption{H~2-48}
         \label{fig:H2-48fold}
     \end{subfigure}
     \hfill
     \begin{subfigure}[b]{0.47\textwidth}
         \centering
         \includegraphics[width=\textwidth]{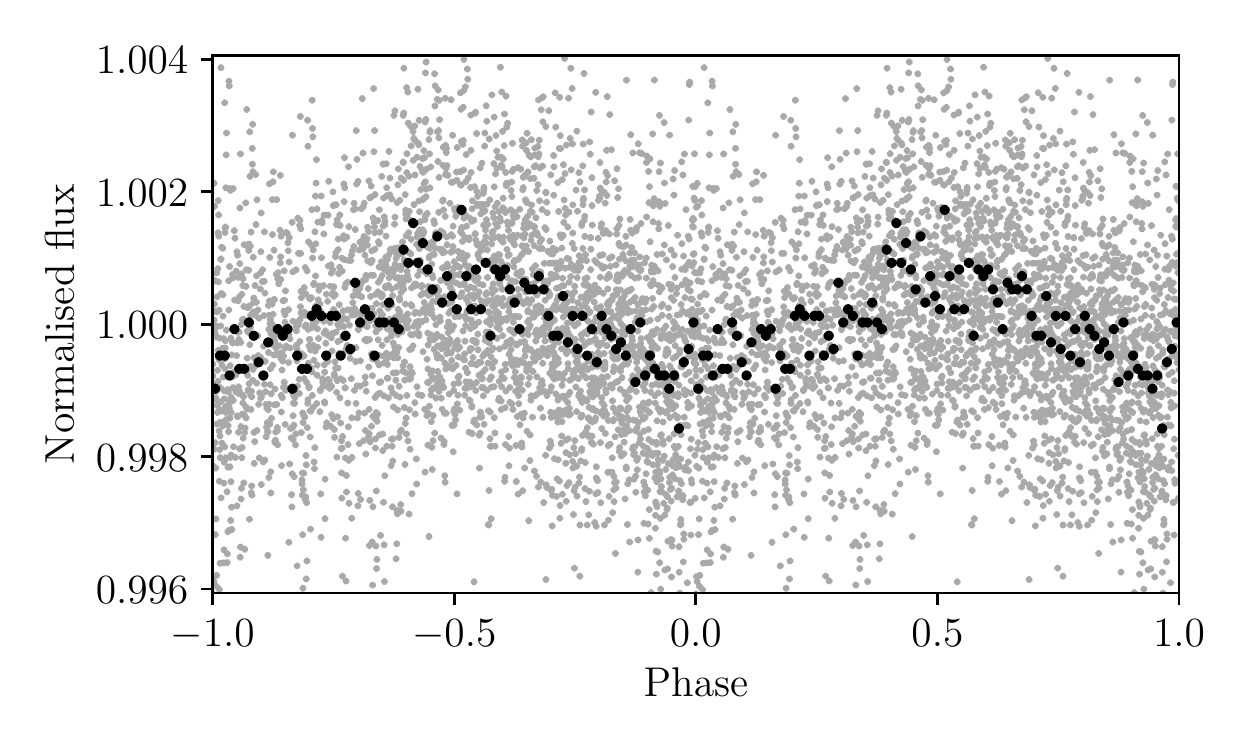}
         \caption{YC~2-32}
         \label{fig:YC2-32}    
     \end{subfigure}
     \\
      \begin{subfigure}[b]{0.47\textwidth}
         \centering
         \includegraphics[width=\textwidth]{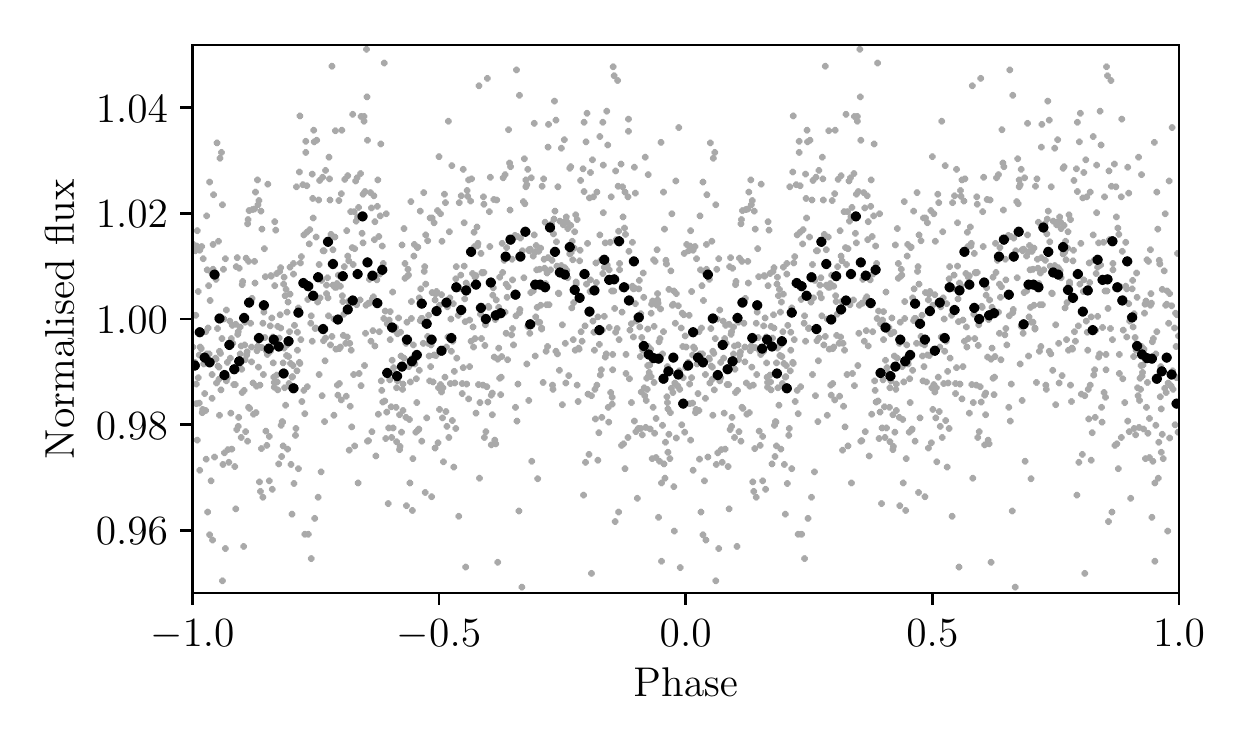}
         \caption{JaSt~52}
         \label{fig:JaSt52}
     \end{subfigure}
     \hfill
         \begin{subfigure}[b]{0.47\textwidth}
         \centering
         \includegraphics[width=\textwidth]{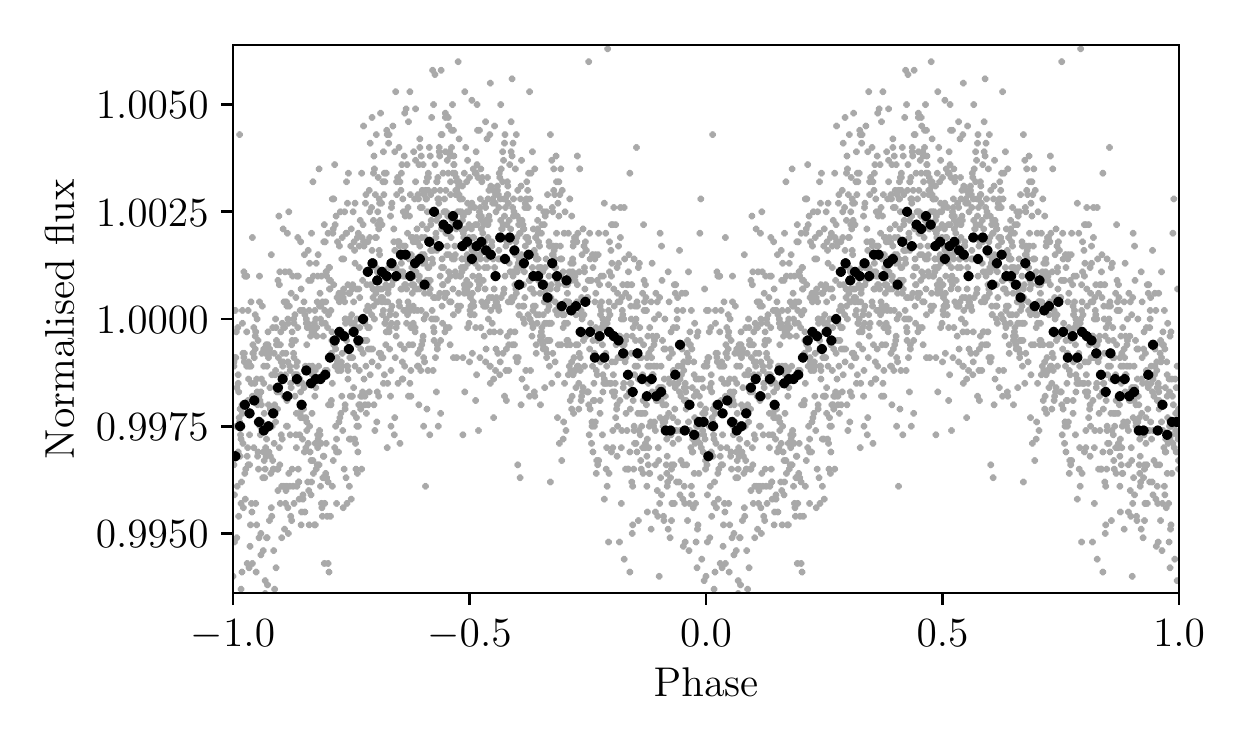}
         \caption{PHR~J1724$-$2302}
         \label{fig:PHRJ1724-2302}
     \end{subfigure}
     \\
      \begin{subfigure}[b]{0.47\textwidth}
         \centering
         \includegraphics[width=\textwidth]{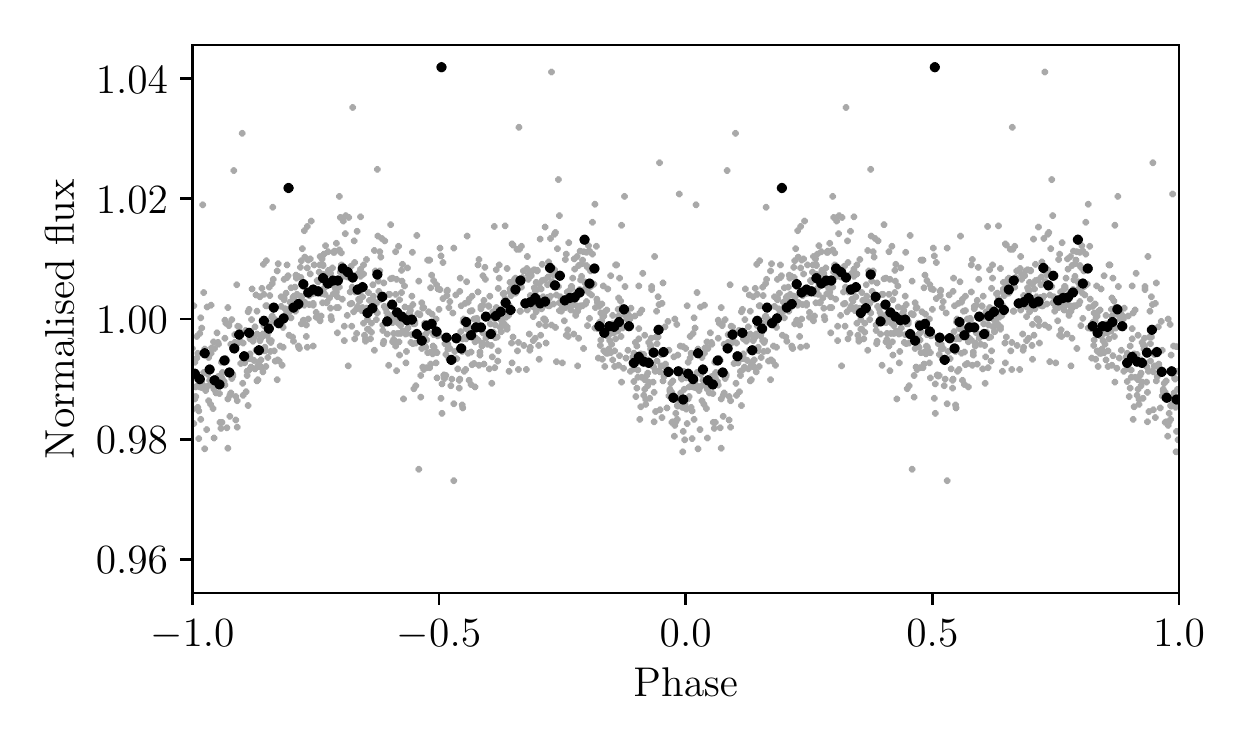}
         \caption{Th~3-12}
         \label{fig:Th3-12}
     \end{subfigure}
     \hfill
          \begin{subfigure}[b]{0.47\textwidth}
         \centering
         \includegraphics[width=\textwidth]{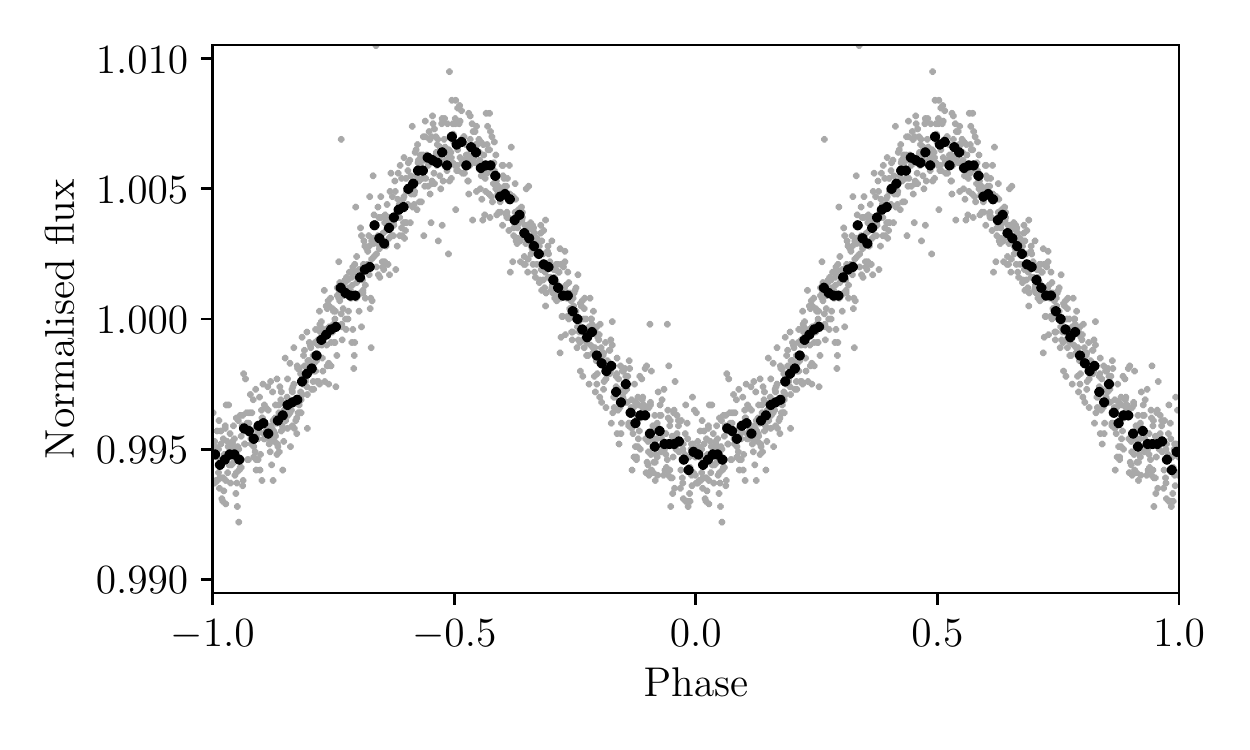}
         \caption{M~2-11}
         \label{fig:M2-11}
     \end{subfigure}
     \\
     \caption{K2 folded (grey) and binned (black) light curves for: (a) PC~12, (b) CGMW~4-3783, (c) H~2-48 (see also Figure\ \ref{fig:H2-48}), (d) YC~2-32, (e) JaSt~52, (f) PHR~J1724$-$2302, (g) Th~3-12, and (g) M~2-11. The binned light curves represent the average values for bins of 0.01 in phase for the folded light curve.}
        \label{fig:LCs1}
\end{figure*}

\begin{figure*}
      \begin{subfigure}[b]{0.47\textwidth}
         \centering
         \includegraphics[width=\textwidth]{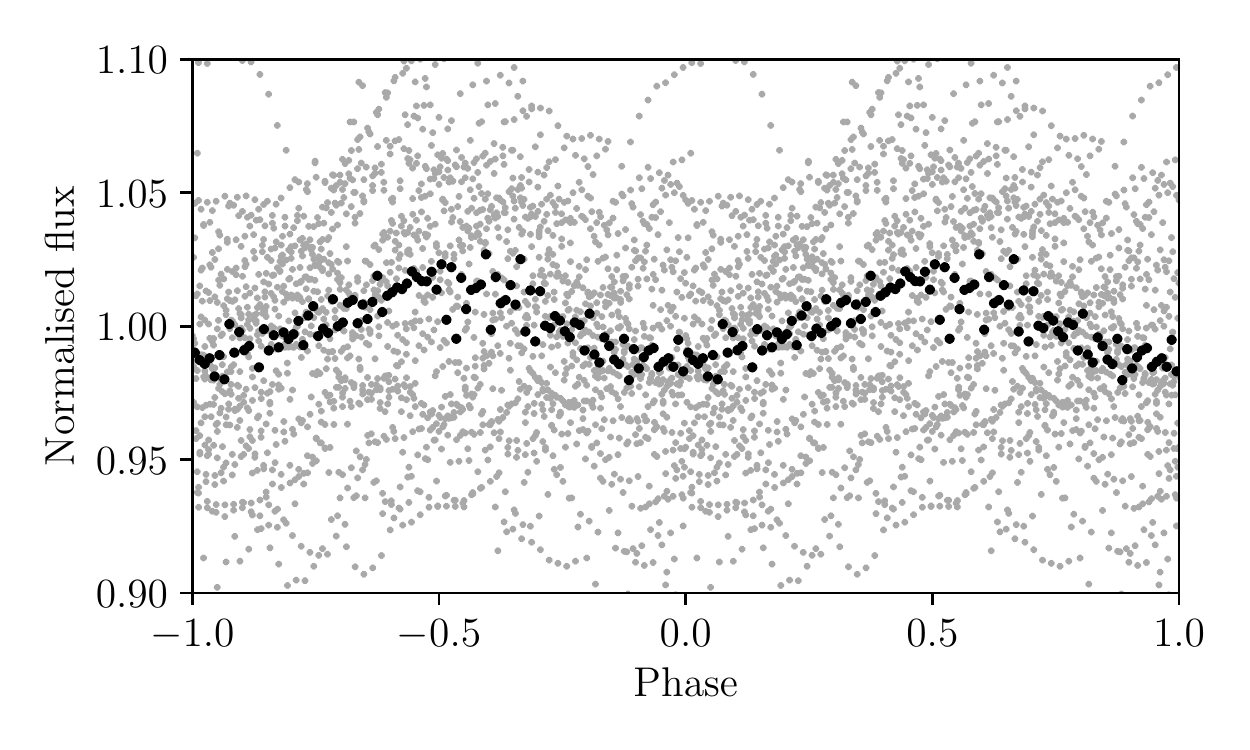}
         \caption{Th~3-14}
         \label{fig:Th3-14fold}
     \end{subfigure}
     \hfill
        \begin{subfigure}[b]{0.47\textwidth}
         \centering
         \includegraphics[width=\textwidth]{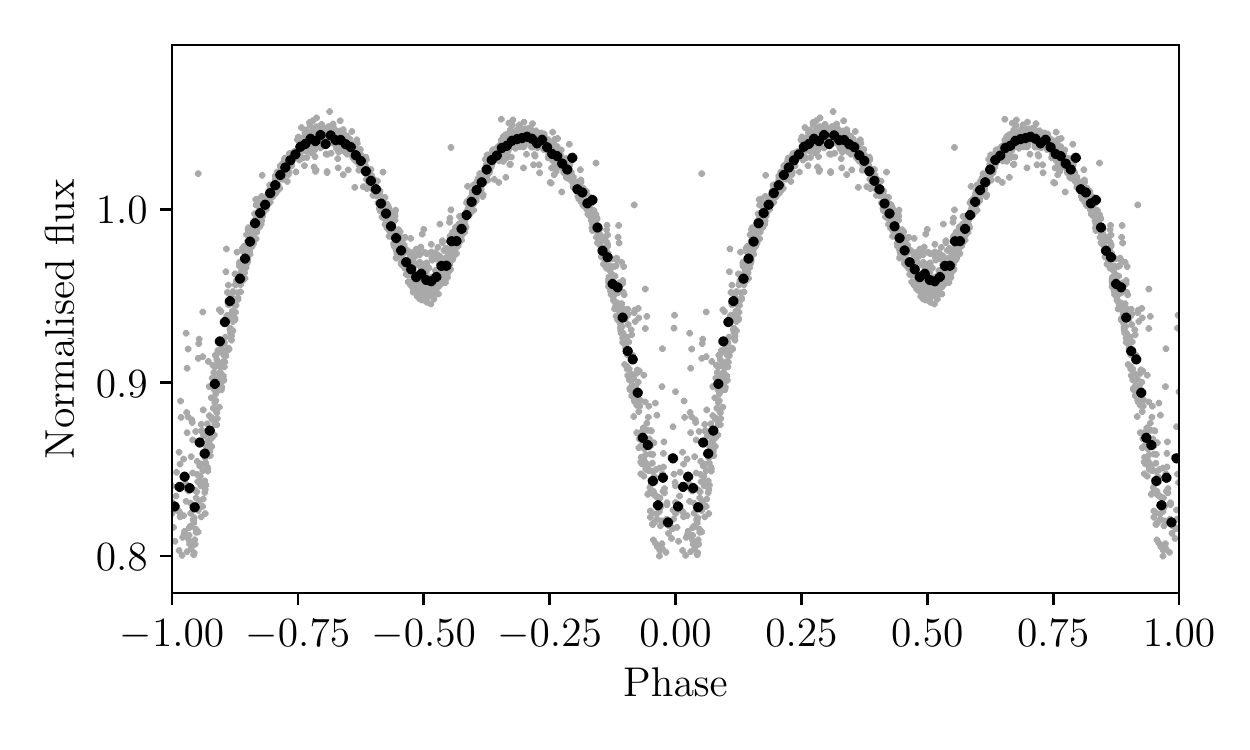}
         \caption{PTB~26}
         \label{fig:PTB26}
     \end{subfigure}
\\     
          \begin{subfigure}[b]{0.47\textwidth}
         \centering
         \includegraphics[width=\textwidth]{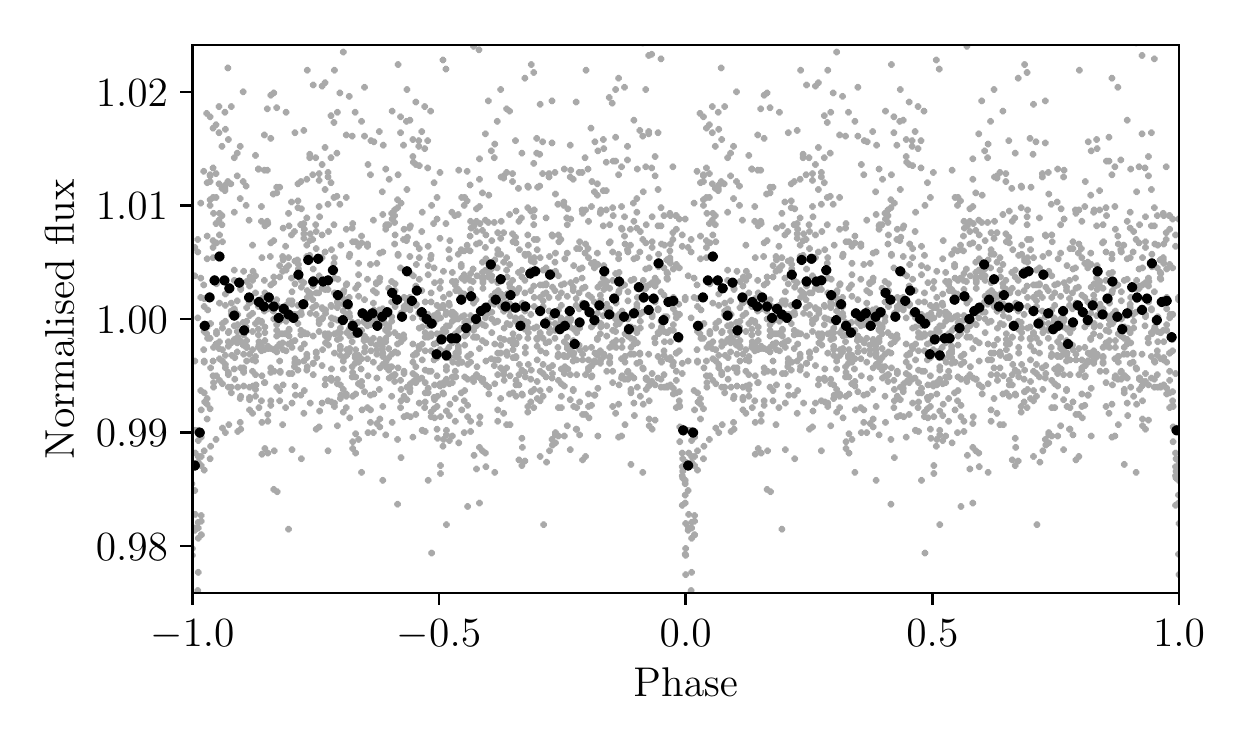}
         \caption{Th~3-35}
         \label{fig:Th3-35}
     \end{subfigure}
\hfill
     \begin{subfigure}[b]{0.47\textwidth}
         \centering
         \includegraphics[width=\textwidth]{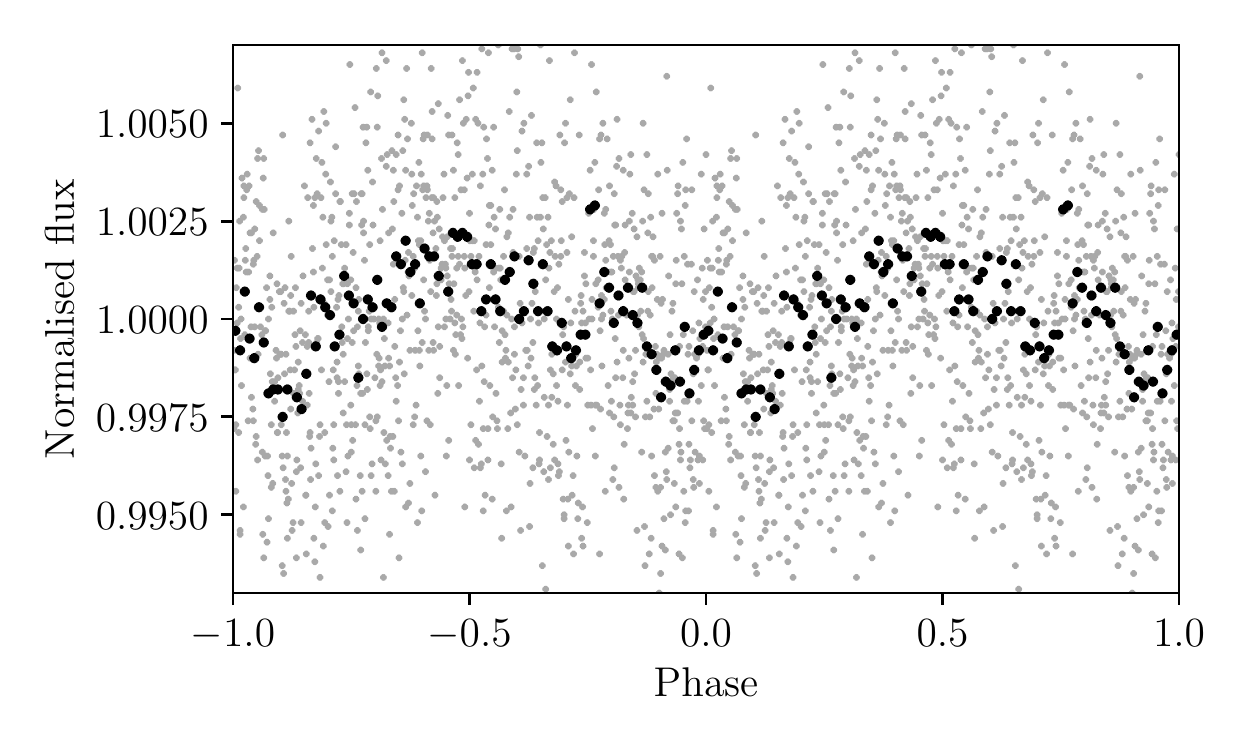}
         \caption{H~1-17}
         \label{fig:H1-17}
     \end{subfigure}
     \\
     \begin{subfigure}[b]{0.47\textwidth}
         \centering
         \includegraphics[width=\textwidth]{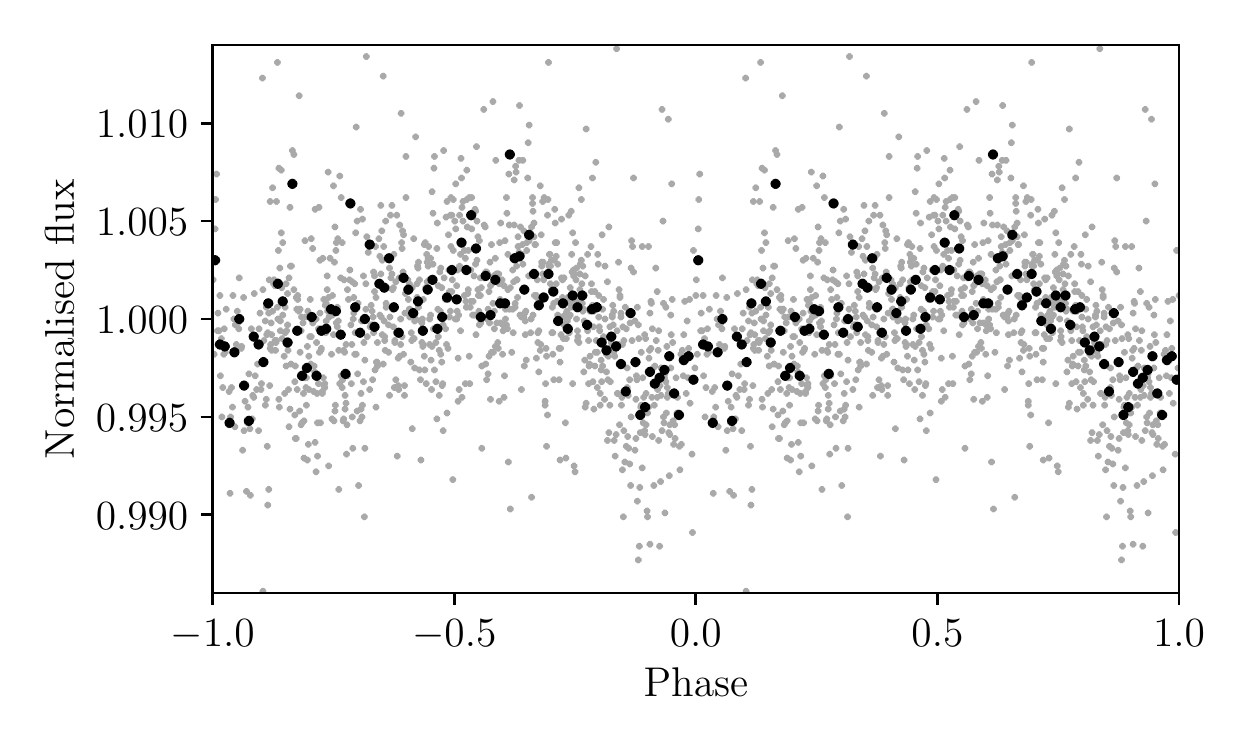}
         \caption{M~4-4}
         \label{fig:M4-4}
         \end{subfigure}
     \hfill
      \begin{subfigure}[b]{0.47\textwidth}
         \centering
         \includegraphics[width=\textwidth]{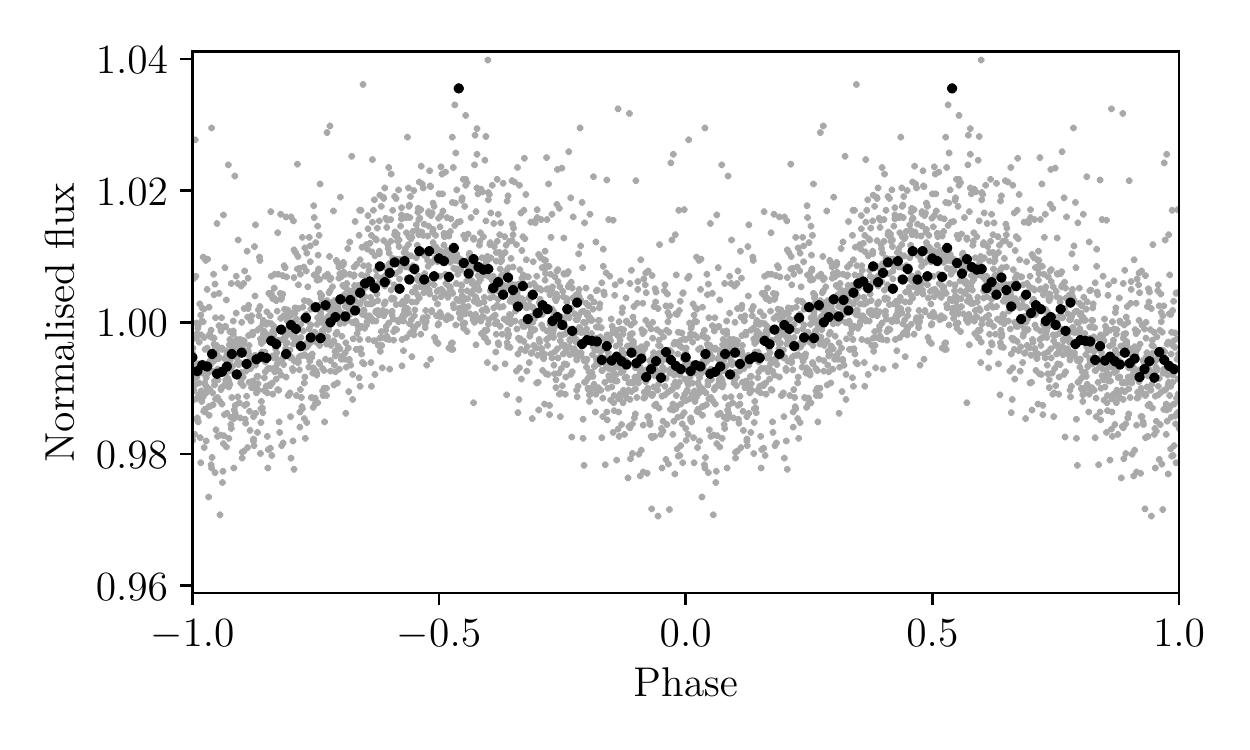}
         \caption{H~2-13}
         \label{fig:H2-13}
     \end{subfigure}
     \\
         \begin{subfigure}[b]{0.47\textwidth}
         \centering
         \includegraphics[width=\textwidth]{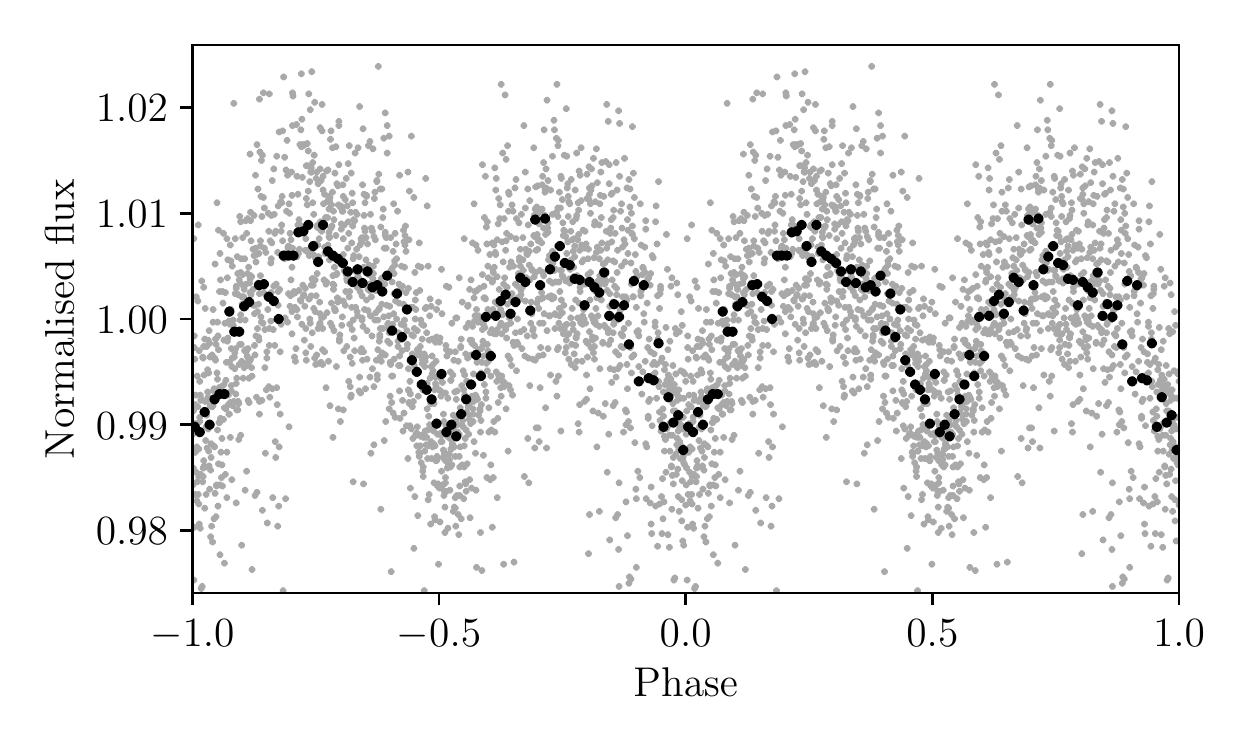}
         \caption{MPA~J1711$-$3112}
         \label{fig:MPAJ1711-3112}
     \end{subfigure}
     \hfill
      \begin{subfigure}[b]{0.47\textwidth}
         \centering
         \includegraphics[width=\textwidth]{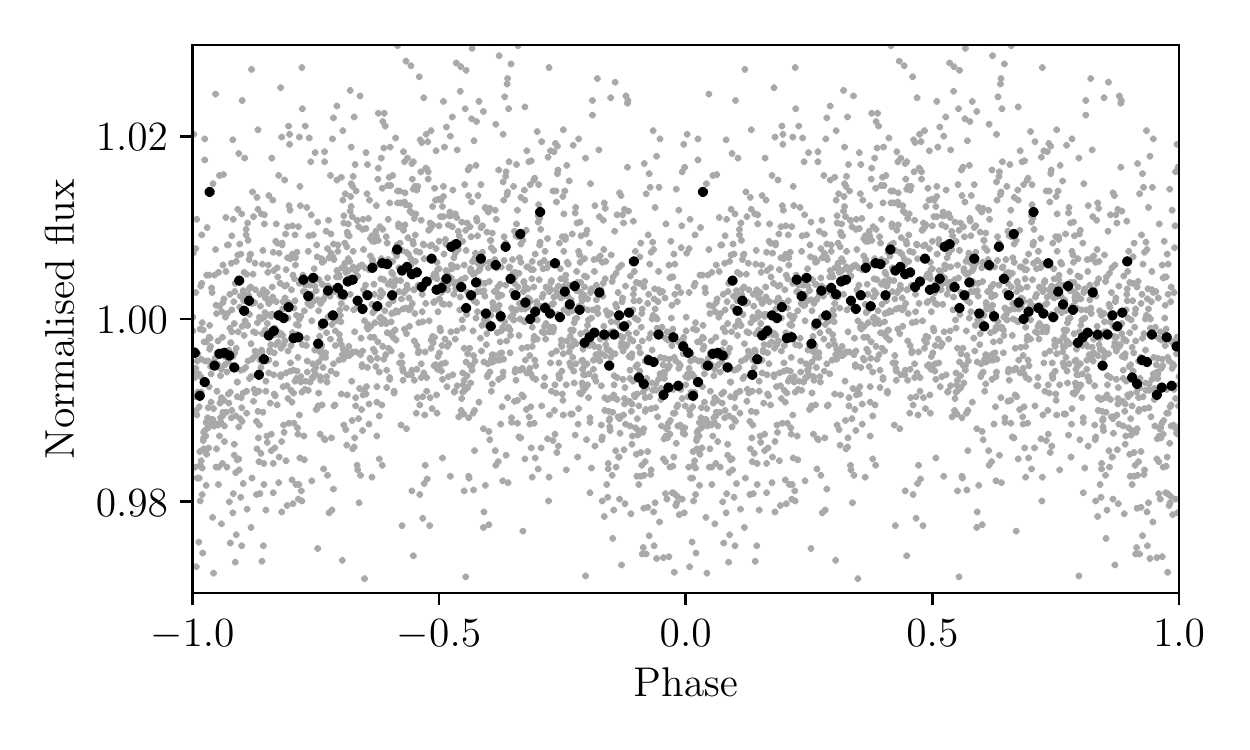}
         \caption{PHR~J1713$-$2957}
         \label{fig:PHRJ1713-2957}
     \end{subfigure}
     \\
     \caption{K2 folded (grey) and binned (black) light curves for: (a) Th~3-14 (see also Figure\ \ref{fig:Th3-14}) (b) PTB~26, (c) Th~3-35, (d) H~1-17, (e) M~4-4, (f) H~2-13, (g) MPA~J1711$-$3112, and (h) PHR~J1713$-$2957. The binned light curves represent the average values for bins of 0.01 in phase for the folded light curve.}
        \label{fig:LCs2}
\end{figure*}

\begin{figure*}
          \begin{subfigure}[b]{0.47\textwidth}
         \centering
         \includegraphics[width=\textwidth]{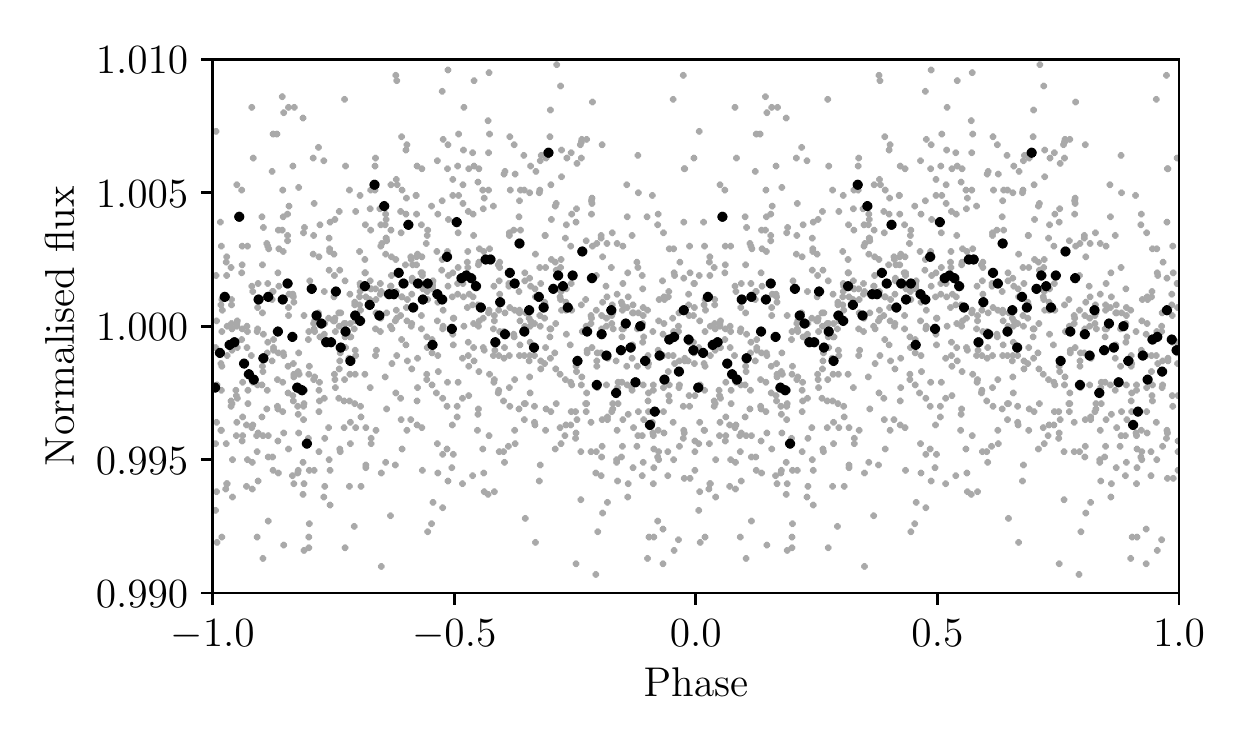}
         \caption{MPA~J1714$-$2946}
         \label{fig:MPAJ1714-2946}
     \end{subfigure}
     \hfill
        \begin{subfigure}[b]{0.47\textwidth}
         \centering
         \includegraphics[width=\textwidth]{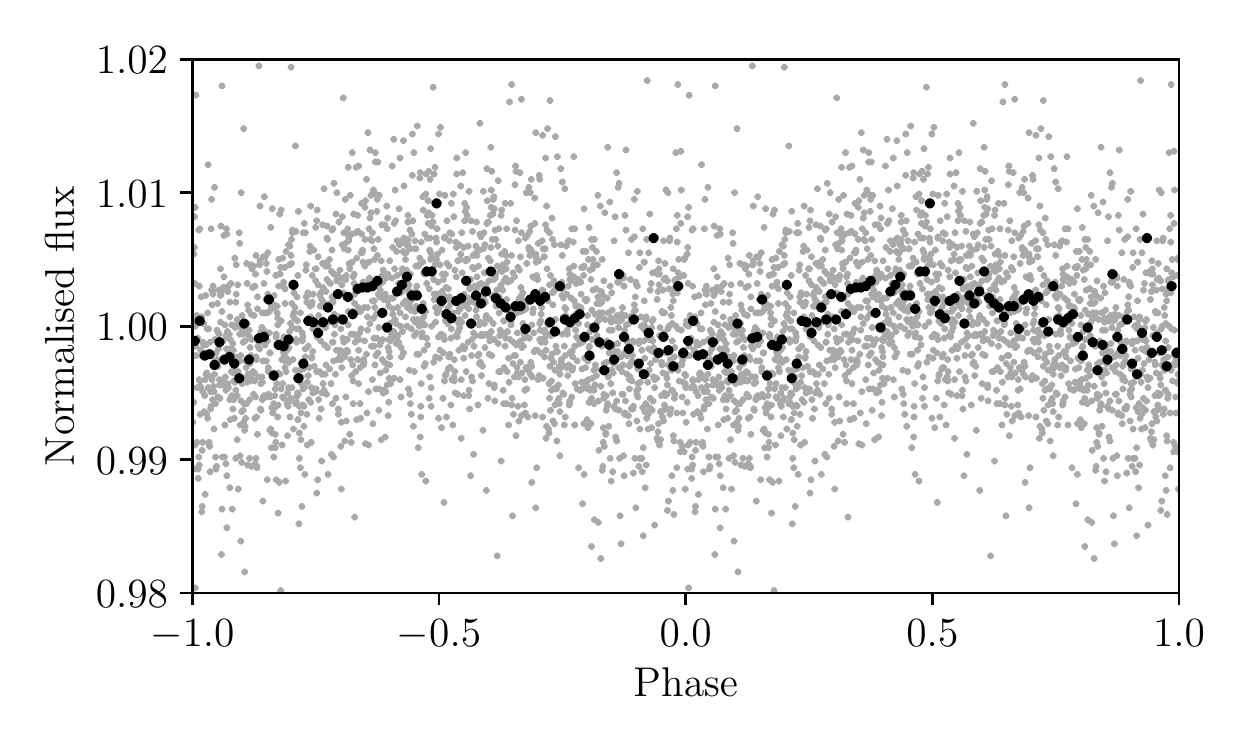}
         \caption{MPA~J1717$-$2356}
         \label{fig:MPAJ1717-2356}
     \end{subfigure}
     \\
     \begin{subfigure}[b]{0.47\textwidth}
         \centering
         \includegraphics[width=\textwidth]{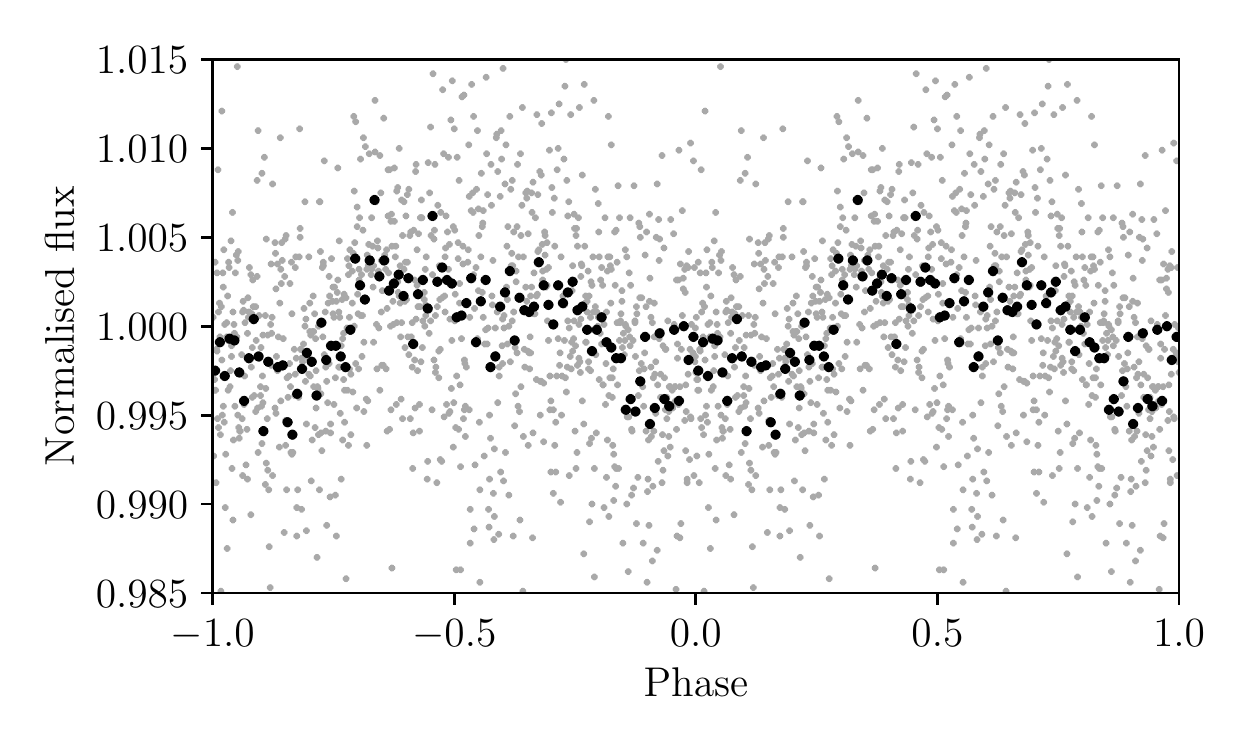}
         \caption{PHR~J1718$-$2441}
         \label{fig:PHRJ1718-2441}
     \end{subfigure}
     \hfill
     \begin{subfigure}[b]{0.47\textwidth}
         \centering
         \includegraphics[width=\textwidth]{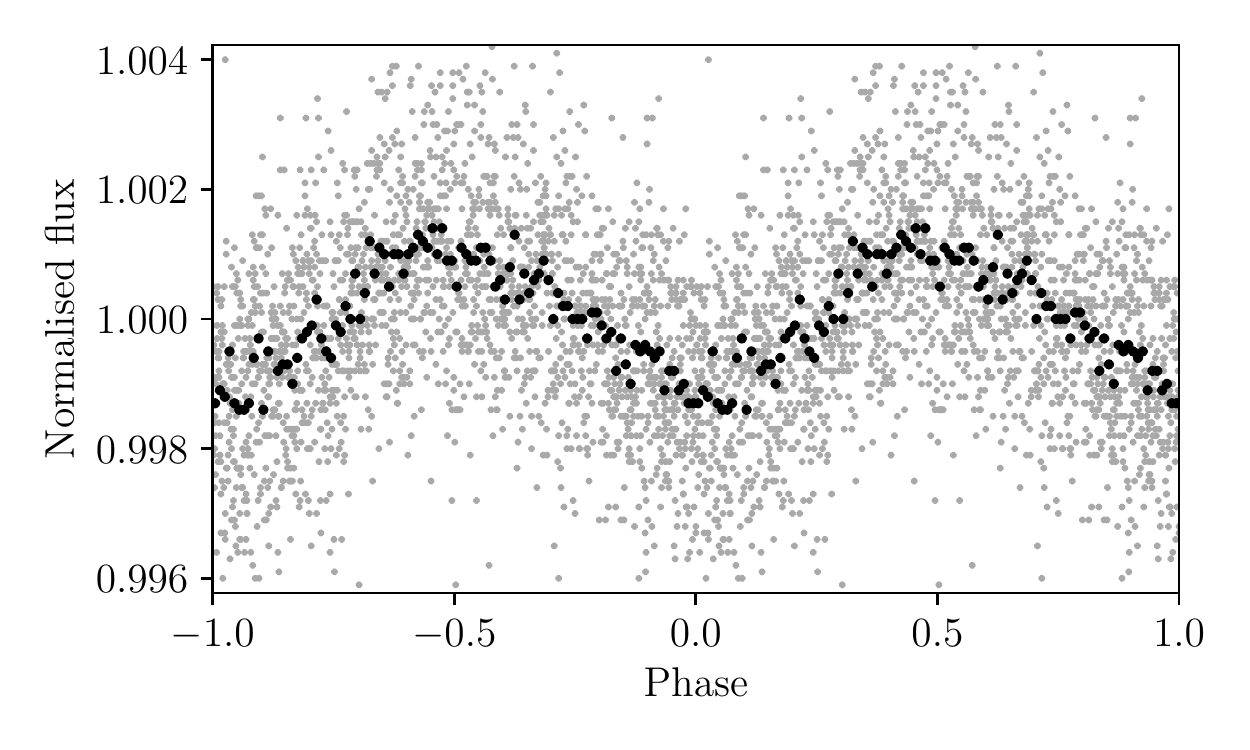}
         \caption{M~3-39}
         \label{fig:M3-39}
     \end{subfigure}
     \\
     \begin{subfigure}[b]{0.47\textwidth}
         \centering
         \includegraphics[width=\textwidth]{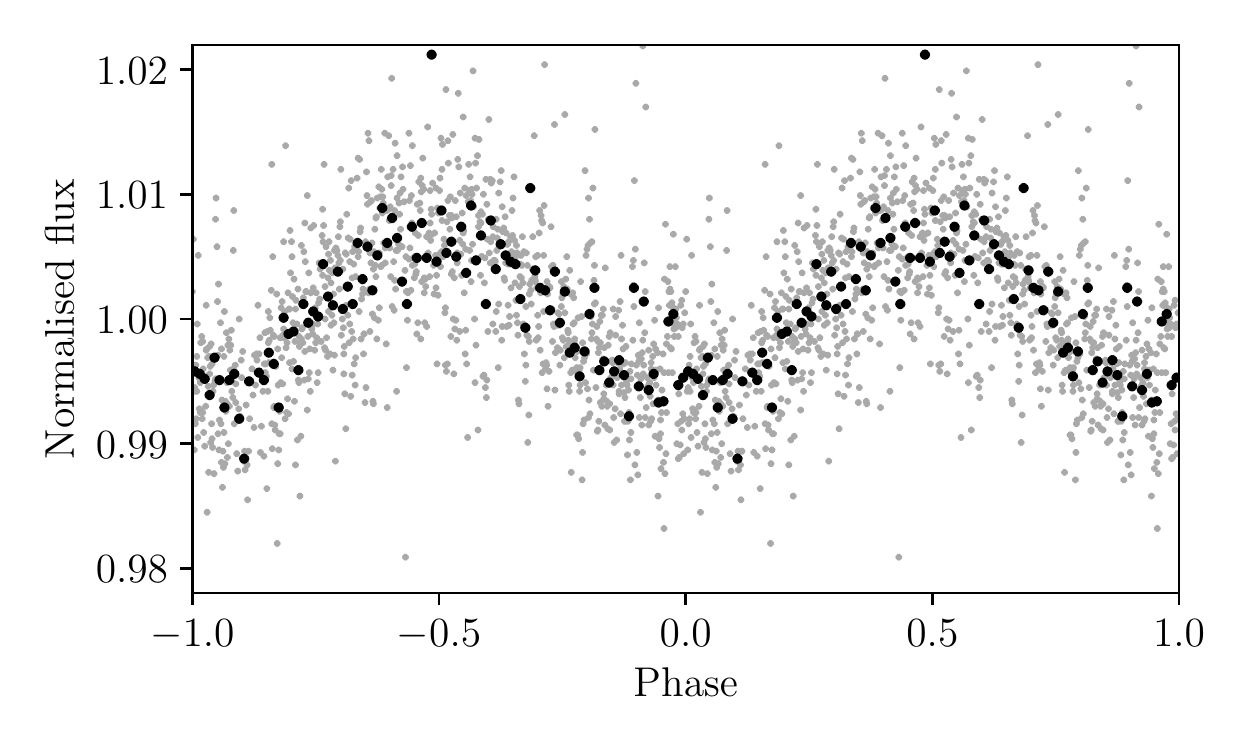}
         \caption{K~5-31}
         \label{fig:K5-31}
         \end{subfigure}
     \hfill
      \begin{subfigure}[b]{0.47\textwidth}
         \centering
         \includegraphics[width=\textwidth]{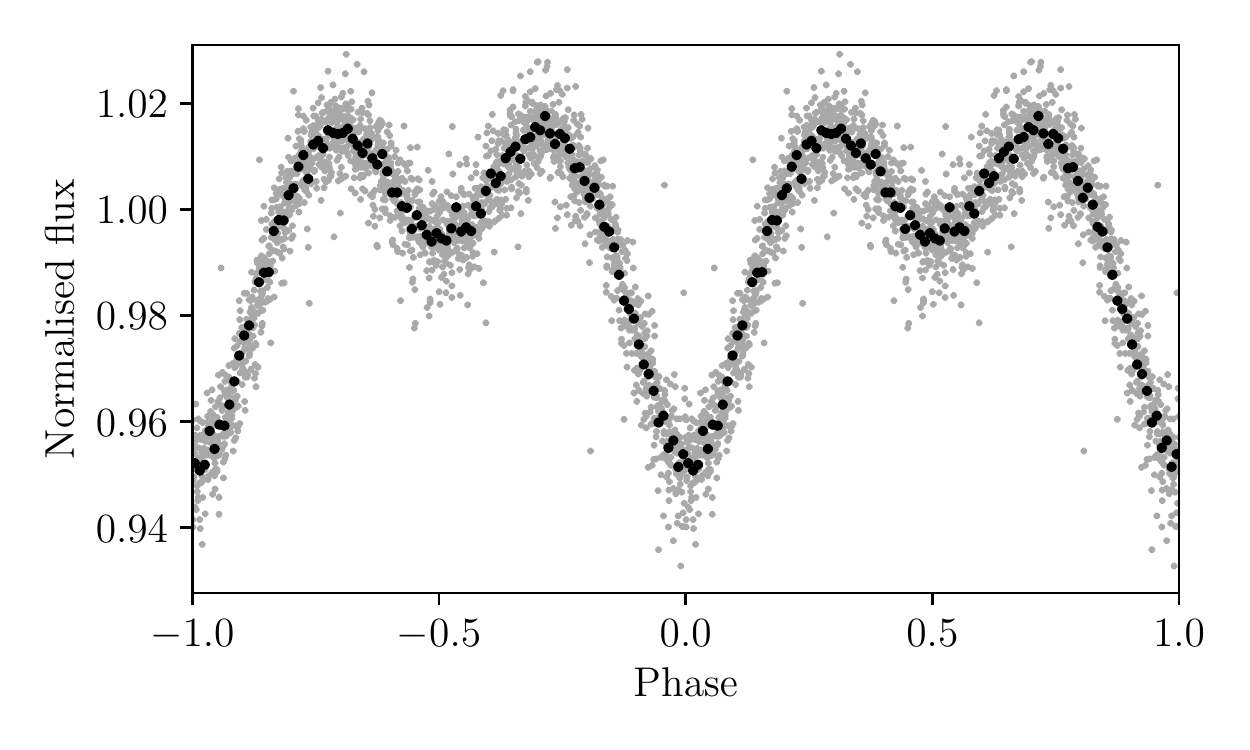}
         \caption{Terz~N19}
         \label{fig:TerzN19}
     \end{subfigure}
     \\
         \begin{subfigure}[b]{0.47\textwidth}
         \centering
         \includegraphics[width=\textwidth]{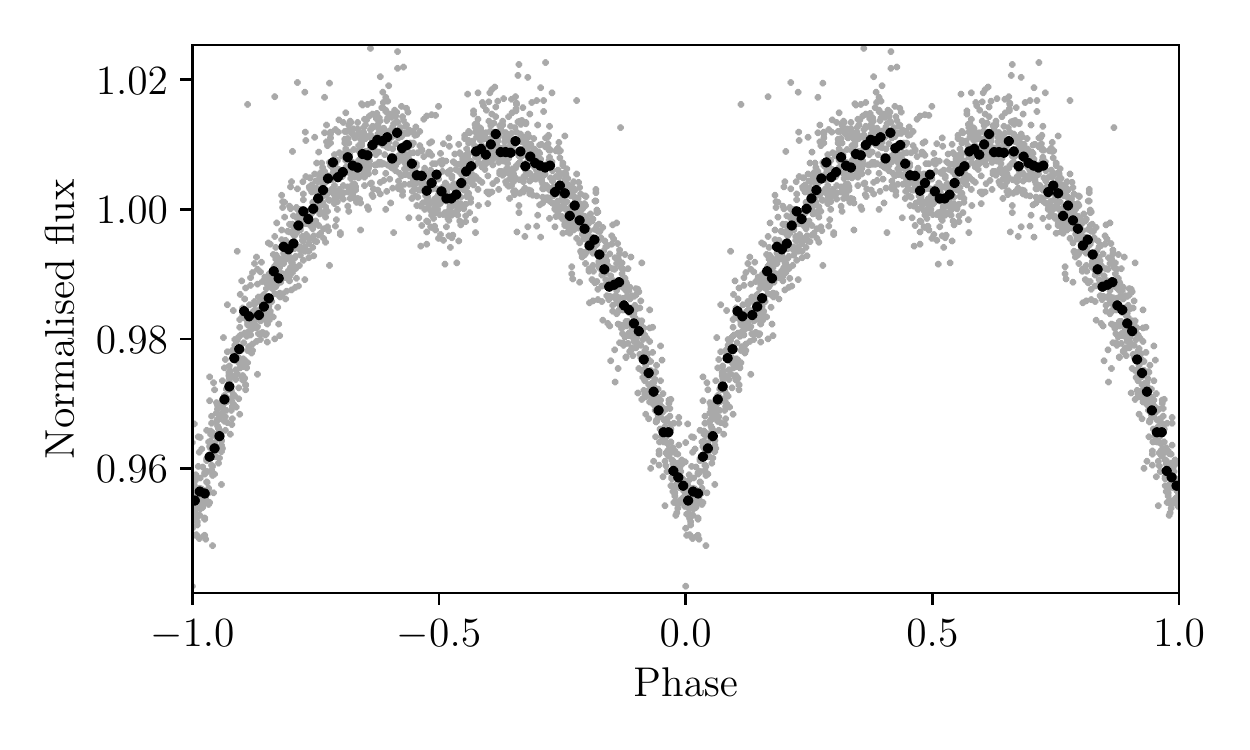}
         \caption{PHR~J1725$-$2338}
         \label{fig:PHRJ1725-2338}
     \end{subfigure}
     \hfill
      \begin{subfigure}[b]{0.47\textwidth}
         \centering
         \includegraphics[width=\textwidth]{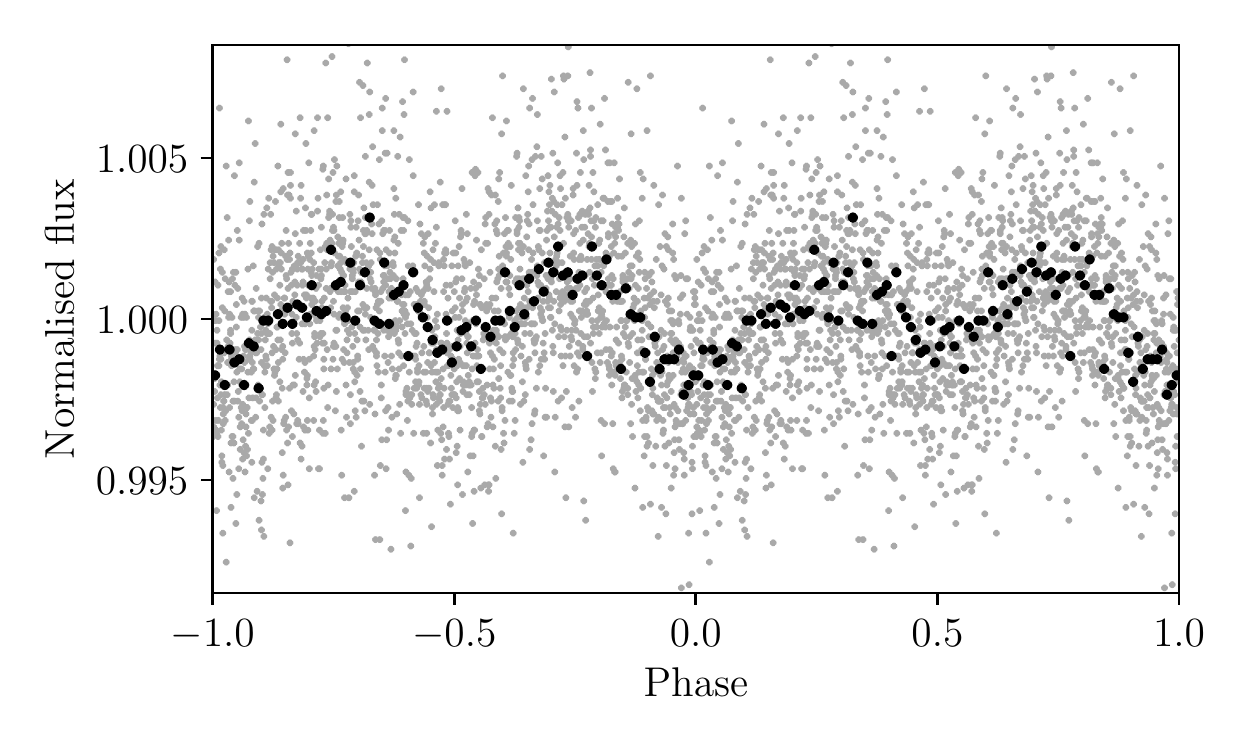}
         \caption{M~3-42}
         \label{fig:M3-42}
     \end{subfigure}
     \\
     \caption{K2 folded (grey) and binned (black) light curves for: (a) MPA~J1714$-$2946, (b) MPA~J1717$-$2356, (c) PHR~J1718$-$2441, (d) M~3-39, (e) K~5-31, (f) Terz~N19, (g) PHR~J1725$-$2338, and (h) M~3-42. The binned light curves represent the average values for bins of 0.01 in phase for the folded light curve.}
        \label{fig:LCs3}
\end{figure*}

The phased light curves for the candidate variables are shown in Figure~\ref{fig:LCs1} through Figure~\ref{fig:LCs3}, listed in the order of appearance in Table~\ref{tab:var-PN-list}, although we also present the light curves of H~2-48 (Figure~\ref{fig:H2-48}) and Th~3-14 (Figure~\ref{fig:Th3-14}) separately as a special cases.

\emph{PC~12:} Figure~\ref{fig:PC12} shows the phased light curve for PC~12, indicating a very high confidence detection for a period of 0.5915(5)~d. The ATLAS survey `c' filter data phases at 0.5916(2)~d in agreement with the {\it K2} period but at twice the amplitude (8.4 percent). The amplitude difference illustrates the effects of nebula dilution due to the large {\it K2} pixels. This PN (see image in Figure 15 of \citet{richer2017}) is very centrally compact ($\sim$2\arcsec) and because ATLAS also has a relatively large measuring aperture ($\sim$3\arcsec), there is likely dilution from the nebula affecting both {\it K2} and ATLAS.
The light curve shows ellipsoidal variability with two peaks and two troughs per orbit.  Ellipsoidal variability in a CSPN suggests a compact companion \citep{hillwig10,santander-garcia15}, such as a white dwarf, particularly when the minima are so similar. Therefore it appears that the CS is a double degenerate (DD) binary star in which the CS is tidally distorted. The minima may be different depths, either due to a very small irradiation effect, to differential gravity darkening of the inner and outer faces of the distorted star, or as a result of different temperatures of the two stars combined with grazing eclipses. 

\emph{CGMW~4-3783:} this CSPN exhibits a very clear periodic variability at the 2.0 percent level with a period of 0.6982~d (see Figure~\ref{fig:CGMW4-378}). We note that the amplitude derived from 10 epochs of PanSTARRS observations is 2.2 percent, but these sparsely spaced data may not have captured the full peak-to-valley amplitude.
As with the CS of PC~12, this light curve is dominated by an ellipsoidal effect. However in this case there is an eclipse at phase $\phi=0.0$ and a likely eclipse present at $\phi=0.5$ as well.  The shape of the curve is consistent with Doppler beaming, especially when combined with the observed amplitude. A similar effect was seen in the \emph{Kepler} target CSPN J19310888+4324577 \citep[the CS of AMU 1,][]{demarco15}. The light curve is also remarkably similar to that of the CS of NGC~6026, which is a close DD binary system \citep{hillwig10}.
We have obtained several epochs of spectra of the CS of CGMW~4-3783 with GMOS on Gemini North. Those spectra show clear radial velocity variability. Therefore we can confirm that the CS here is a short period binary and likely a DD. We will present a detailed study of this system in a future paper.

\emph{H~2-48:} this CSPN shows a marginal and questionable detection for a periodicity near 4.1~d with an amplitude of $\sim$3 percent. There is clear variability here, though the cause is difficult to identify. It is likely due to strong winds. The \ion{He}{i} Zanstra temperature of this CS is low, $32.0\pm10.2$kK \citep{gleizes89} and central stars in this temperature range have been shown to have strong wind variability \citep{prinja12a}. Figure~\ref{fig:H2-48fold} shows a potential underlying periodicity with noise and aliasing effects. It is possible that this is simply an aliasing of some of the larger excursions in brightness with the observing window (see Figure\ \ref{fig:H2-48}), or it may be evidence of a real periodic signal. If so, it is weak and we do not believe that it represents strong evidence of binarity. As such we do not include H~2-48 in our discussion of binary population statistics in Section \ref{sec:discuss} below.

\emph{Y-C~2-32:} this CSPN is an unusual detection because of the relatively long period seen in the {\it K2} observations. Few binary PN central stars have periods longer than $\sim$4~d, so the period that we find, 30~d, is rare. Periods nearly as long are occasionally seen, as with NGC~2346 and MyCn~18, which have periods of 16.0~d and 18.2~d, respectively \citep{miszalski18,brown19}. The amplitude of $\sim$0.1 percent is at the limit for PNe with {\it K2}, although the low amplitude would be consistent with a distant irradiated secondary. Given that the observations extend over only 2-3 cycles, the amplitude is very small, and the periodicity could be a consequence of a data reduction residual, we consider this a tentative detection that needs a follow-up photometric check. Detecting a variable at this amplitude level, though, will be difficult from the ground, even if there is a factor of 10 dilution effect. See Figure~\ref{fig:YC2-32}.


\begin{figure*}
          \begin{subfigure}[b]{0.47\textwidth}
         \centering
         \includegraphics[width=\textwidth]{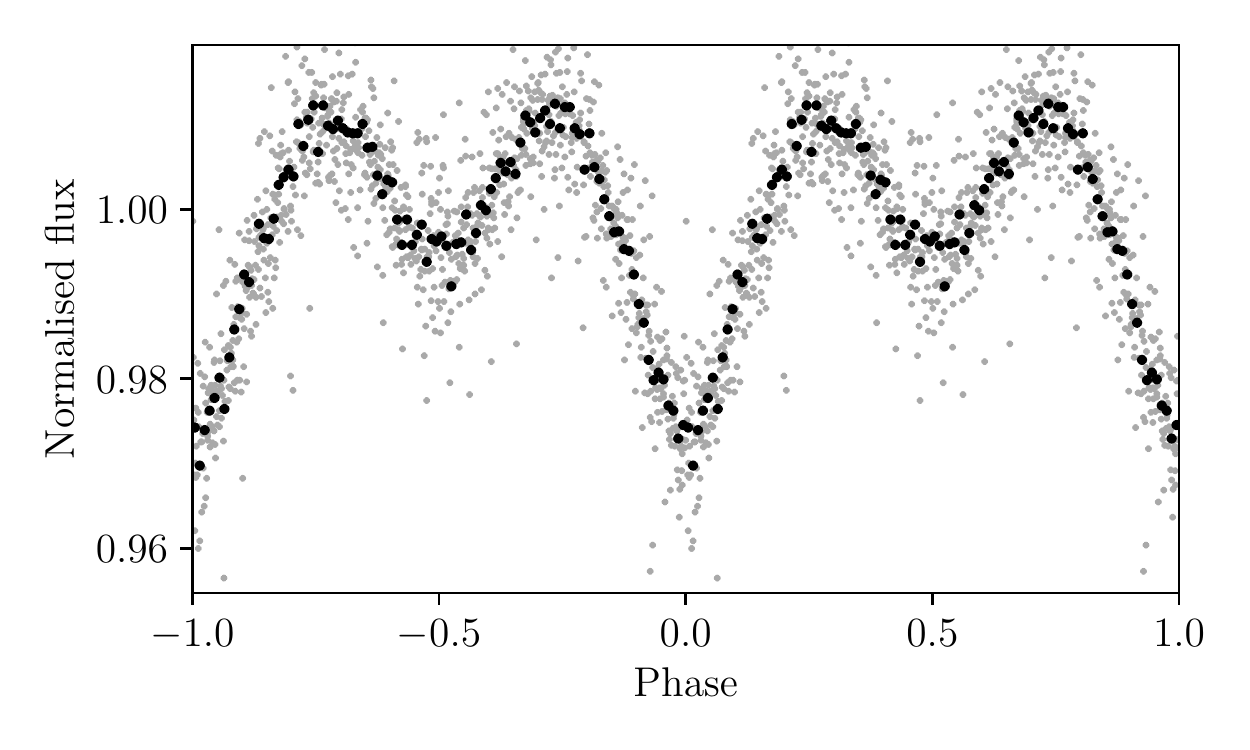}
         \caption{Th~3-15}
         \label{fig:Th3-15}
     \end{subfigure}
     \hfill
        \begin{subfigure}[b]{0.47\textwidth}
         \centering
         \includegraphics[width=\textwidth]{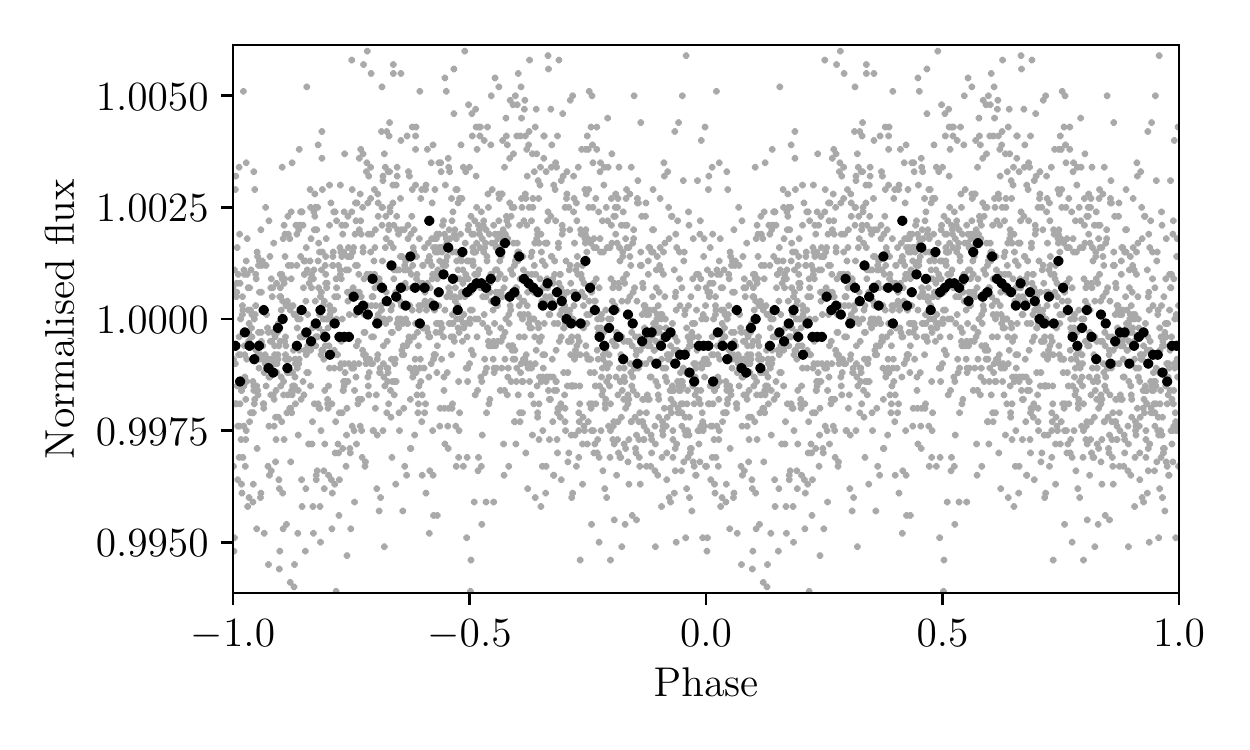}
         \caption{K~5-3}
         \label{fig:K5-3}
     \end{subfigure}
     \\
     \begin{subfigure}[b]{0.47\textwidth}
         \centering
         \includegraphics[width=\textwidth]{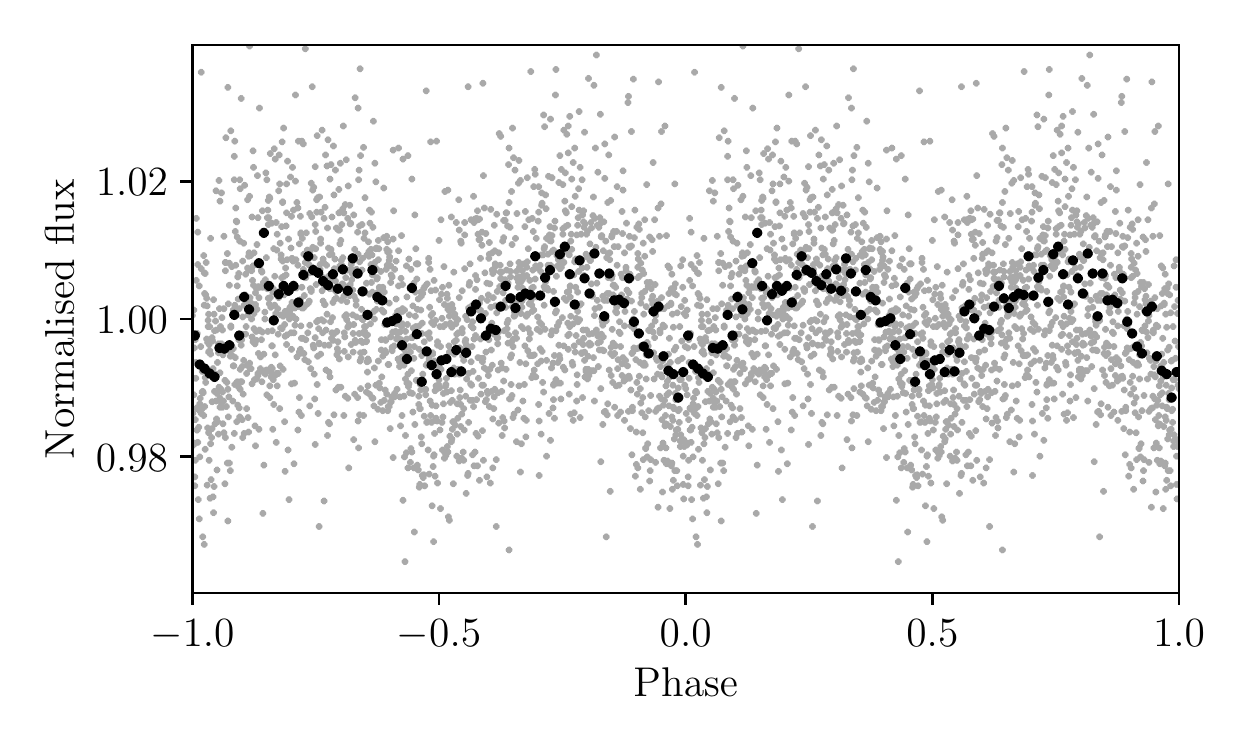}
         \caption{PHR~J1734$-$2000}
         \label{fig:PHRJ1734-2000}
     \end{subfigure}
     \hfill
     \begin{subfigure}[b]{0.47\textwidth}
         \centering
         \includegraphics[width=\textwidth]{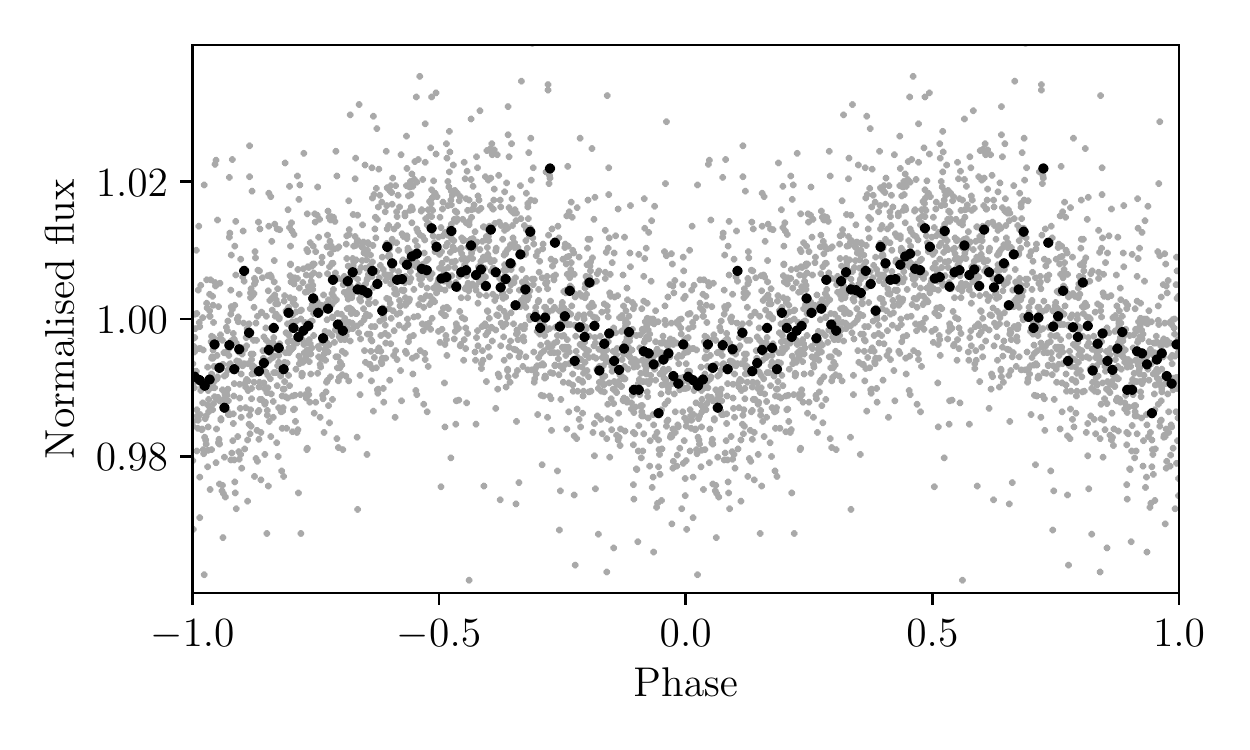}
         \caption{PPA~J1734$-$3015}
         \label{fig:PPAJ1734-3015}
     \end{subfigure}
     \\
     \begin{subfigure}[b]{0.47\textwidth}
         \centering
         \includegraphics[width=\textwidth]{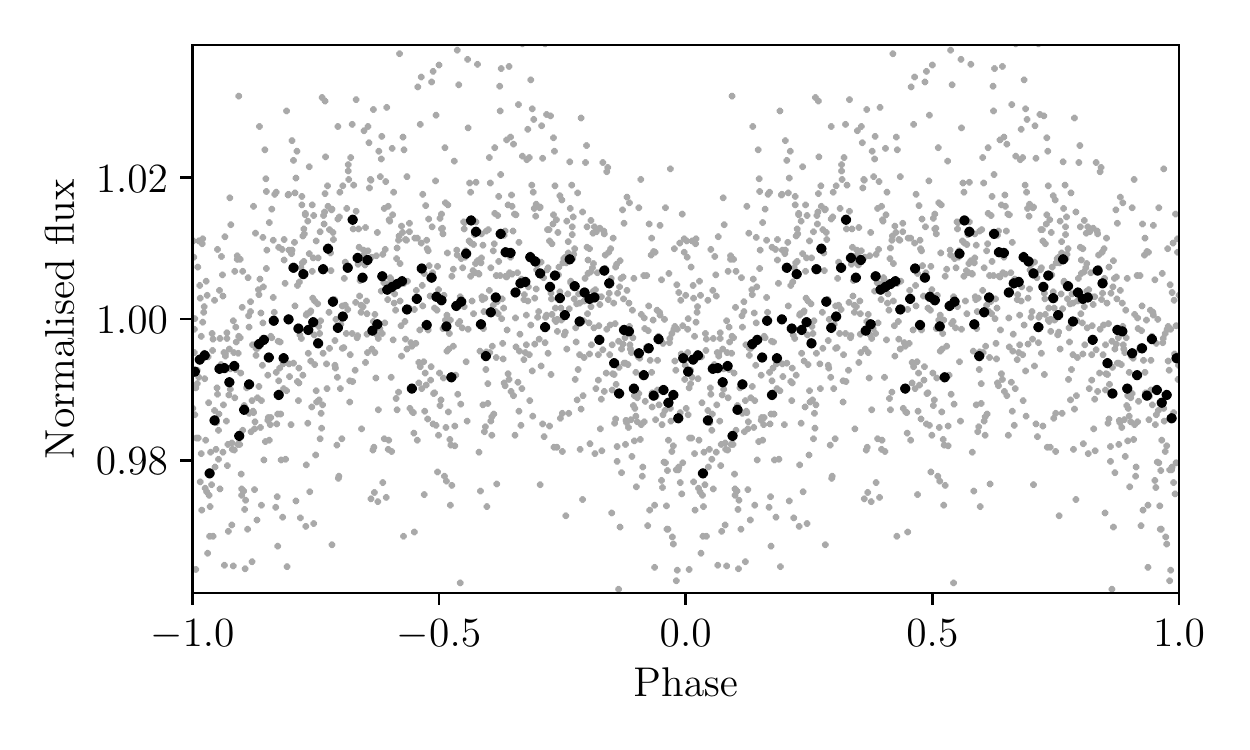}
         \caption{V972~Oph}
         \label{fig:V972Oph}
         \end{subfigure}
     \hfill
      \begin{subfigure}[b]{0.47\textwidth}
         \centering
         \includegraphics[width=\textwidth]{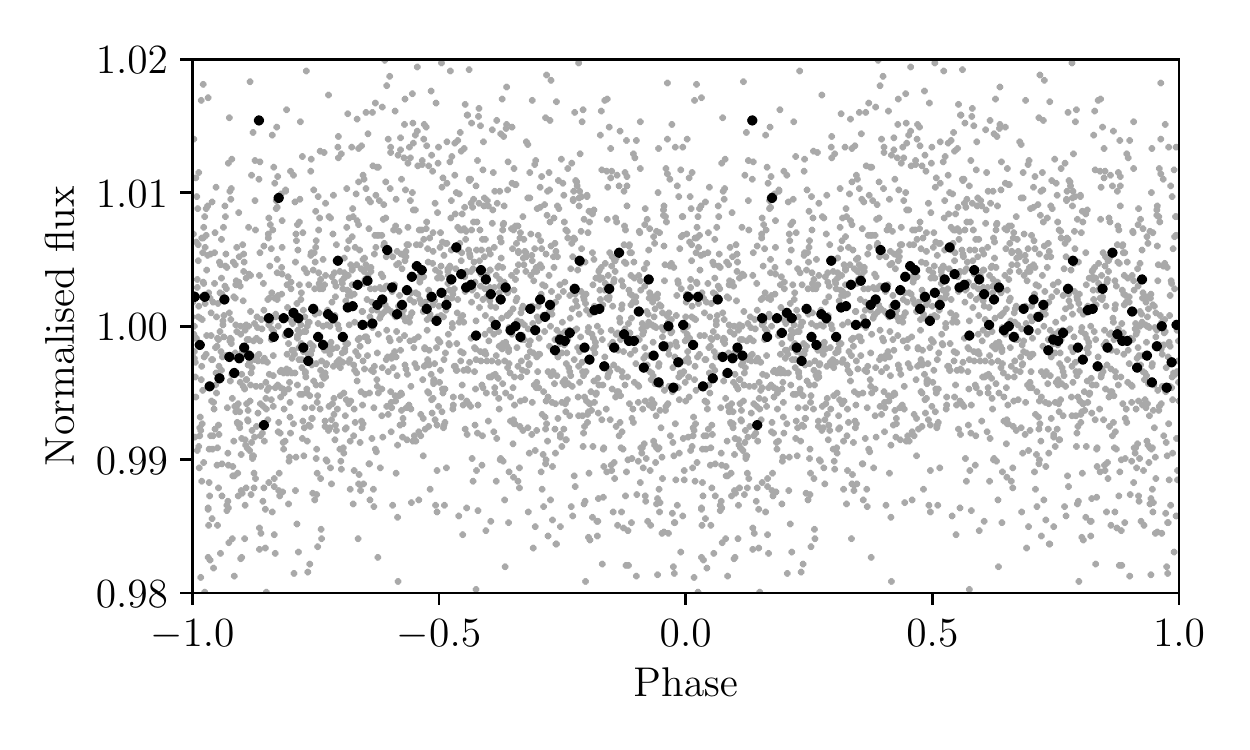}
         \caption{MGE~359.2412+02.3353}
         \label{fig:MGE359}
     \end{subfigure}
     \\
         \begin{subfigure}[b]{0.47\textwidth}
         \centering
         \includegraphics[width=\textwidth]{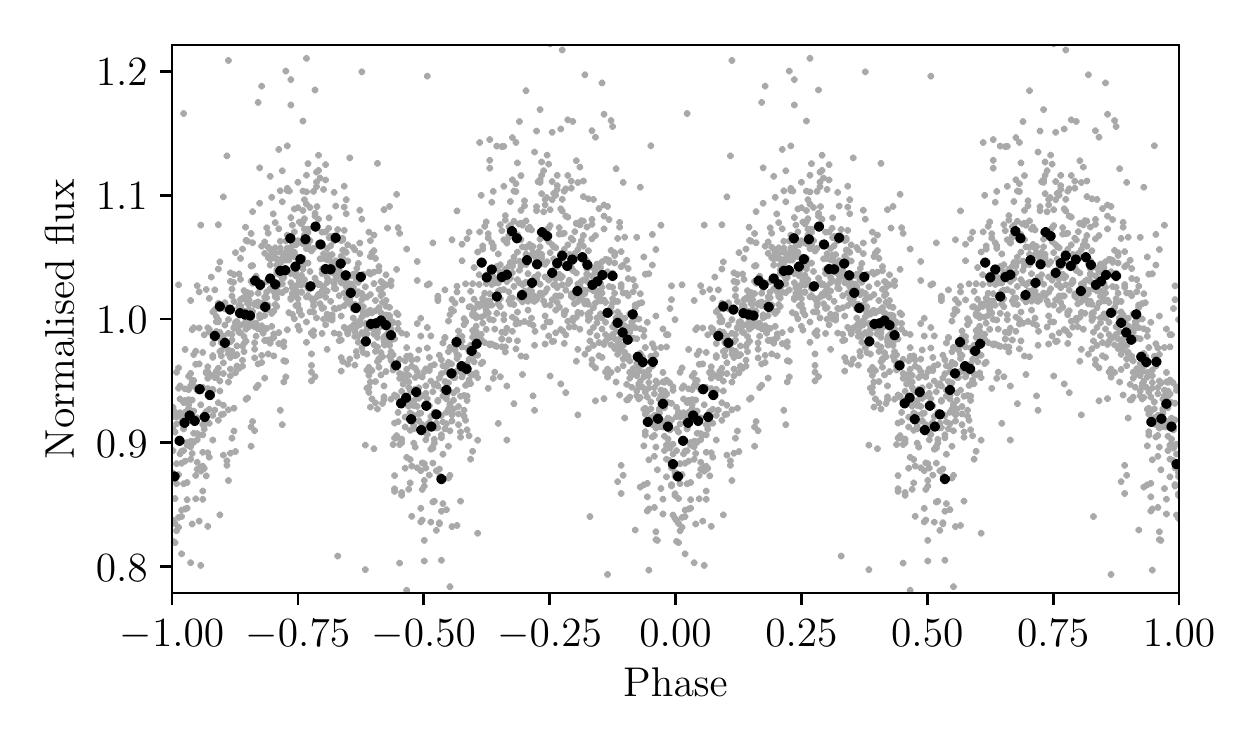}
         \caption{JaSt2~2}
         \label{fig:JaSt22}
     \end{subfigure}
     \hfill
      \begin{subfigure}[b]{0.47\textwidth}
         \centering
         \includegraphics[width=\textwidth]{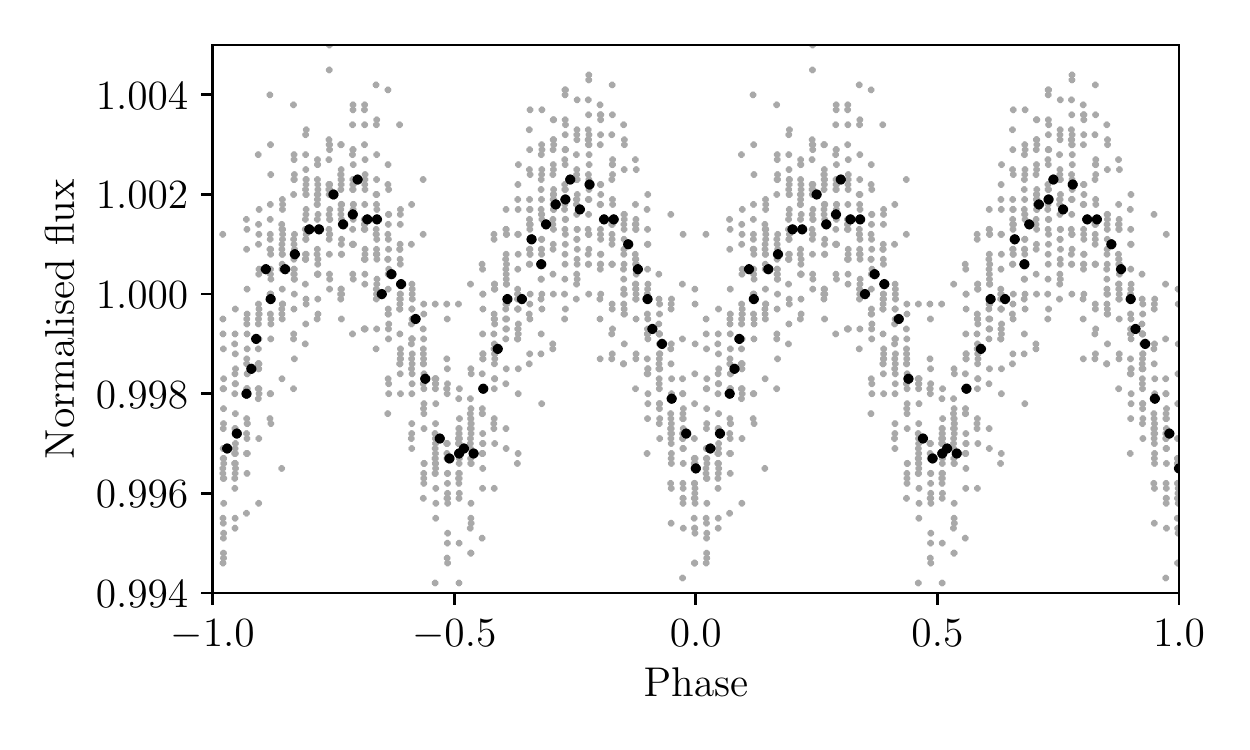}
         \caption{PHR~J1738$-$2419}
         \label{fig:PHRJ1738-2419}
     \end{subfigure}
     \\
     \caption{K2 folded (grey) and binned (black) light curves for:  (a) Th~3-15, (b) K~5-3, (c) PHR~J1734$-$2000, (d) PPA~J1734$-$3015, (e) V972~Oph, (f) MGE~359.2412+02.3353, (g) JaSt~2-2, and (h) PHR~J1738$-$2419. The binned light curves represent the average values for bins of 0.01 in phase for the folded light curve.}
        \label{fig:LCs4}
\end{figure*}

\begin{figure*}
          \begin{subfigure}[b]{0.47\textwidth}
         \centering
         \includegraphics[width=\textwidth]{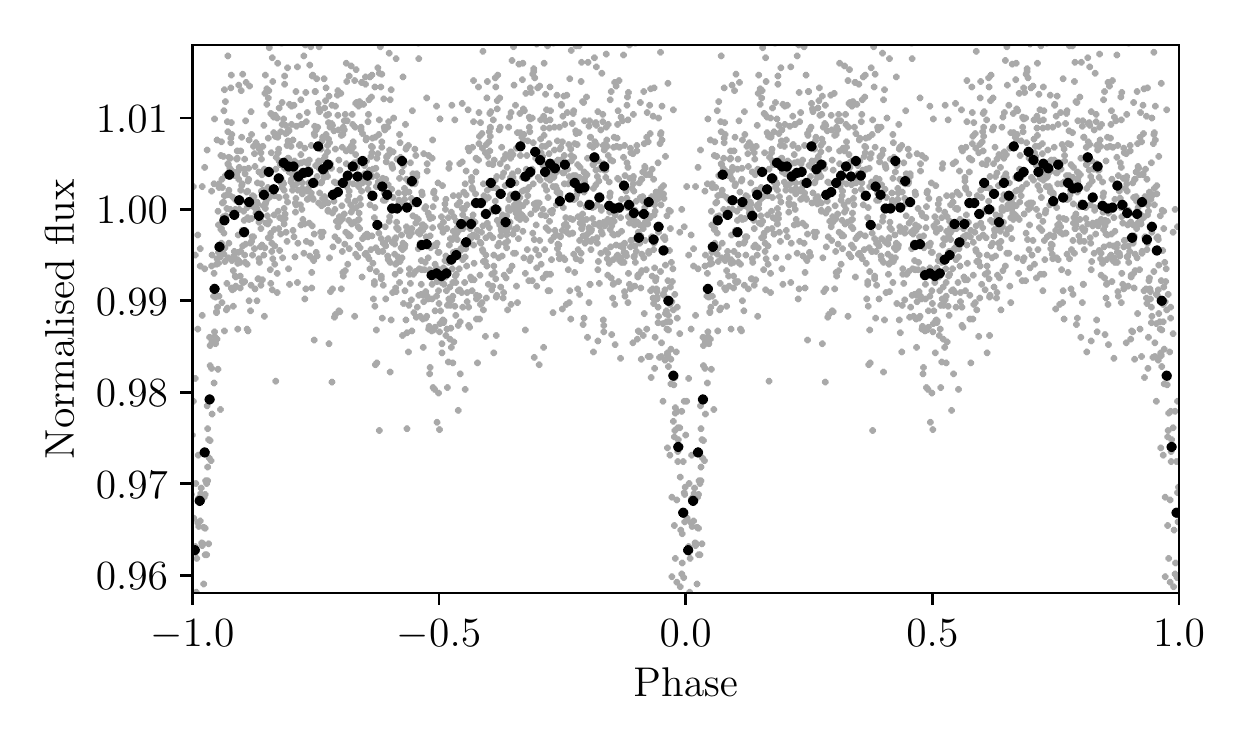}
         \caption{JaSt~12}
         \label{fig:JaSt12}
     \end{subfigure}
     \hfill
        \begin{subfigure}[b]{0.47\textwidth}
         \centering
         \includegraphics[width=\textwidth]{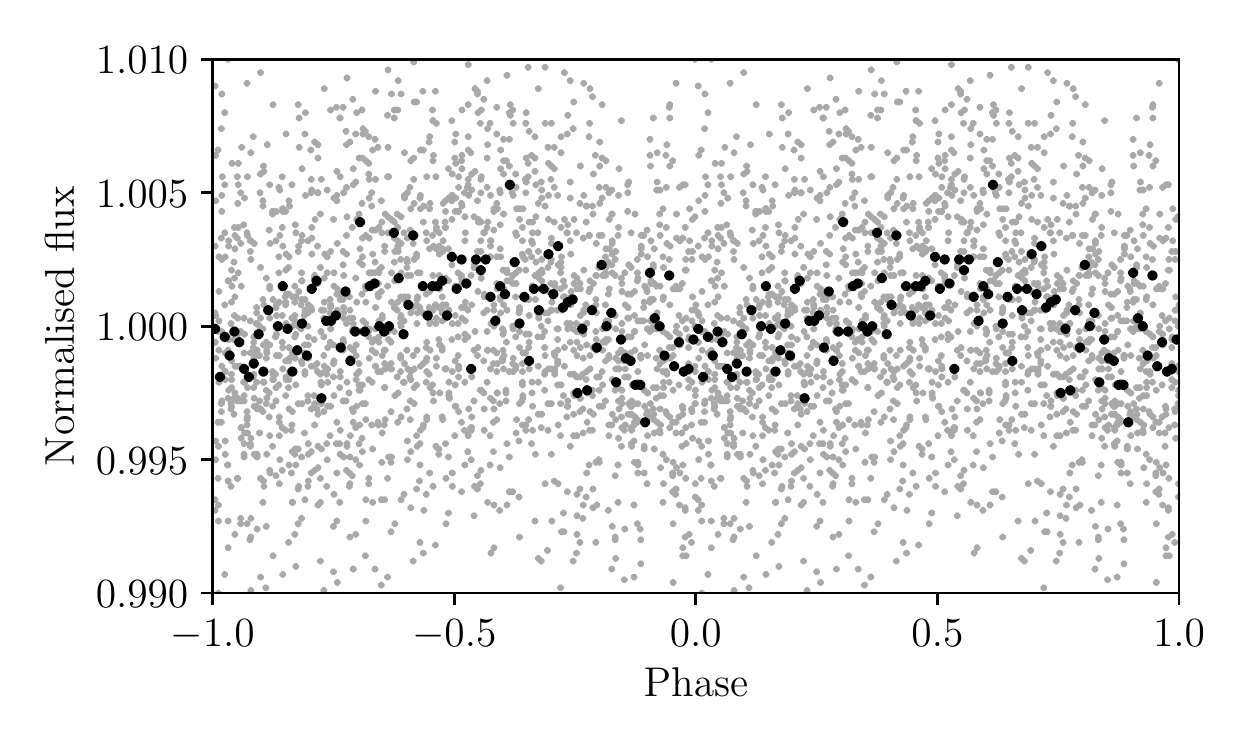}
         \caption{Terz~N1567}
         \label{fig:TerzN1567}
     \end{subfigure}
     \\
     \begin{subfigure}[b]{0.47\textwidth}
         \centering
         \includegraphics[width=\textwidth]{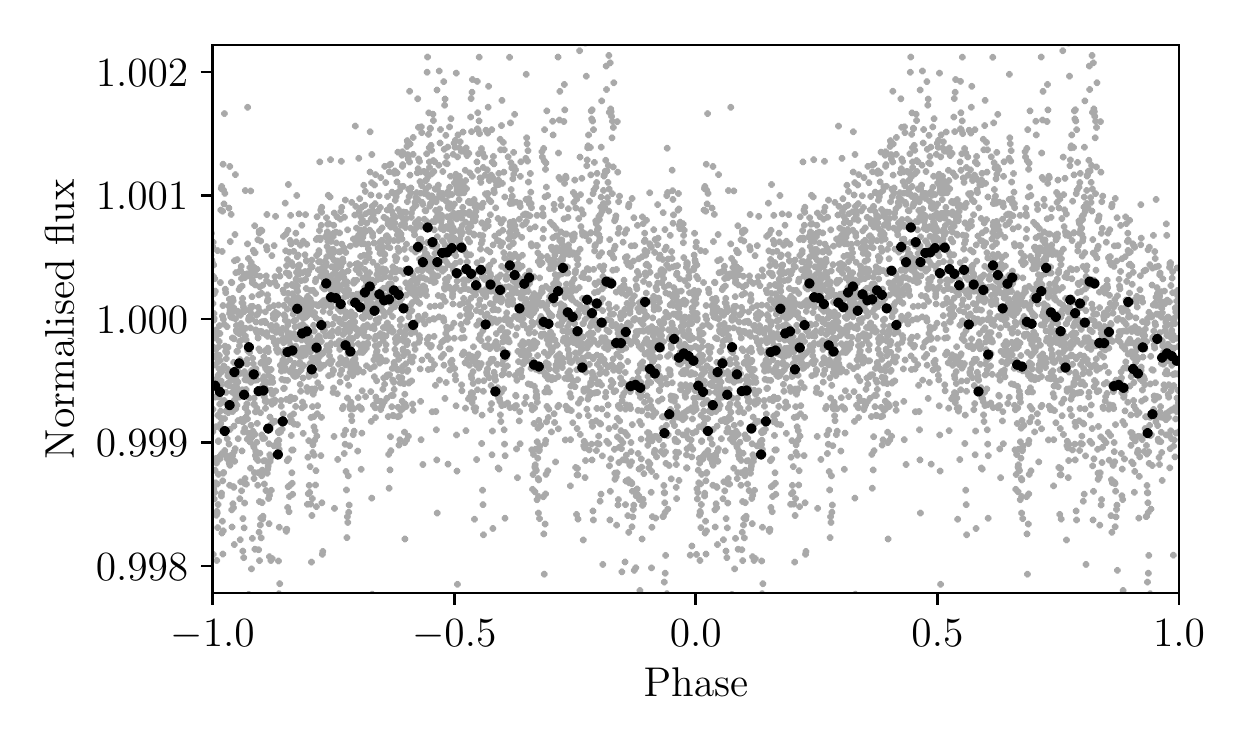}
         \caption{Me~2-1}
         \label{fig:Me2-1}
     \end{subfigure}
     \hfill
     \begin{subfigure}[b]{0.47\textwidth}
         \centering
         \includegraphics[width=\textwidth]{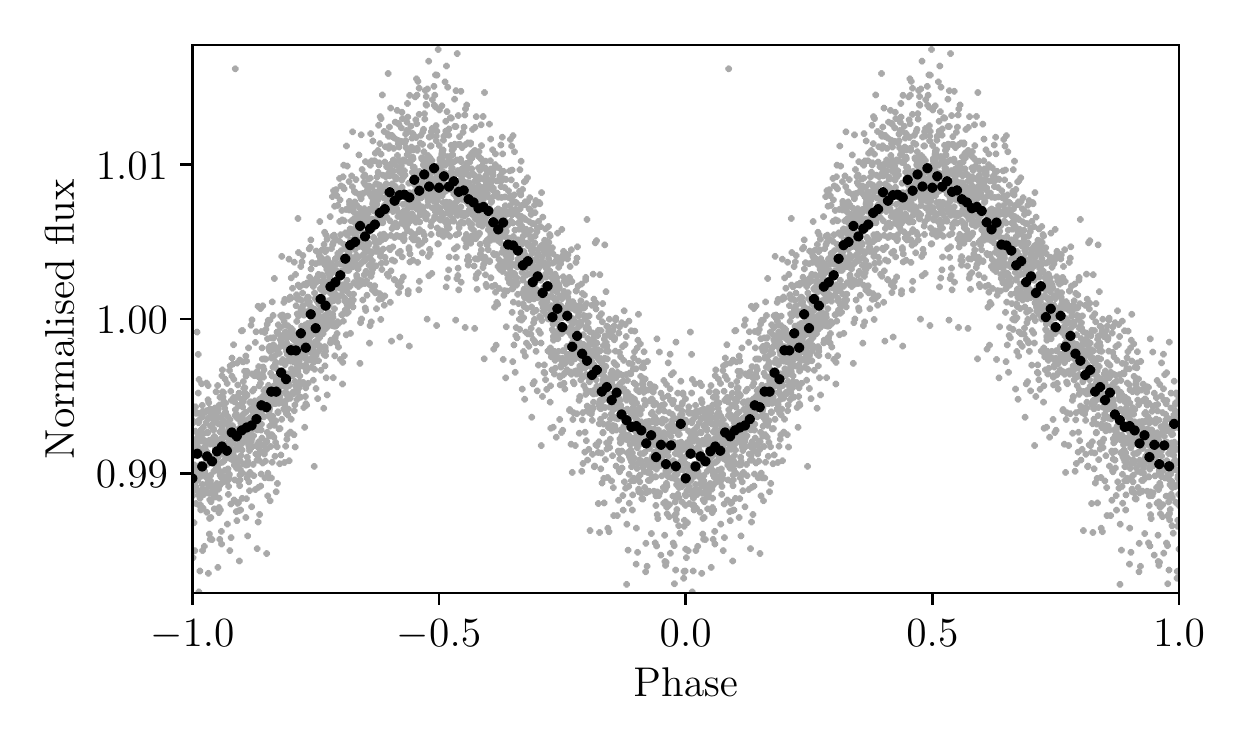}
         \caption{A~30}
         \label{fig:A30}
     \end{subfigure}
     \\
     \caption{K2 folded (grey) and binned (black) light curves for: (a) JaSt~12, (b) Terz~N1567, (c) Me~2-1, and (d) Abell~30 \citep[originally presented in][]{jacoby2020}. The binned light curves represent the average values for bins of 0.01 in phase for the folded light curve.}
        \label{fig:LCs5}
\end{figure*}

\begin{figure}
     \centering
     \begin{subfigure}[b]{0.47\textwidth}
         \centering
         \includegraphics[width=\textwidth]{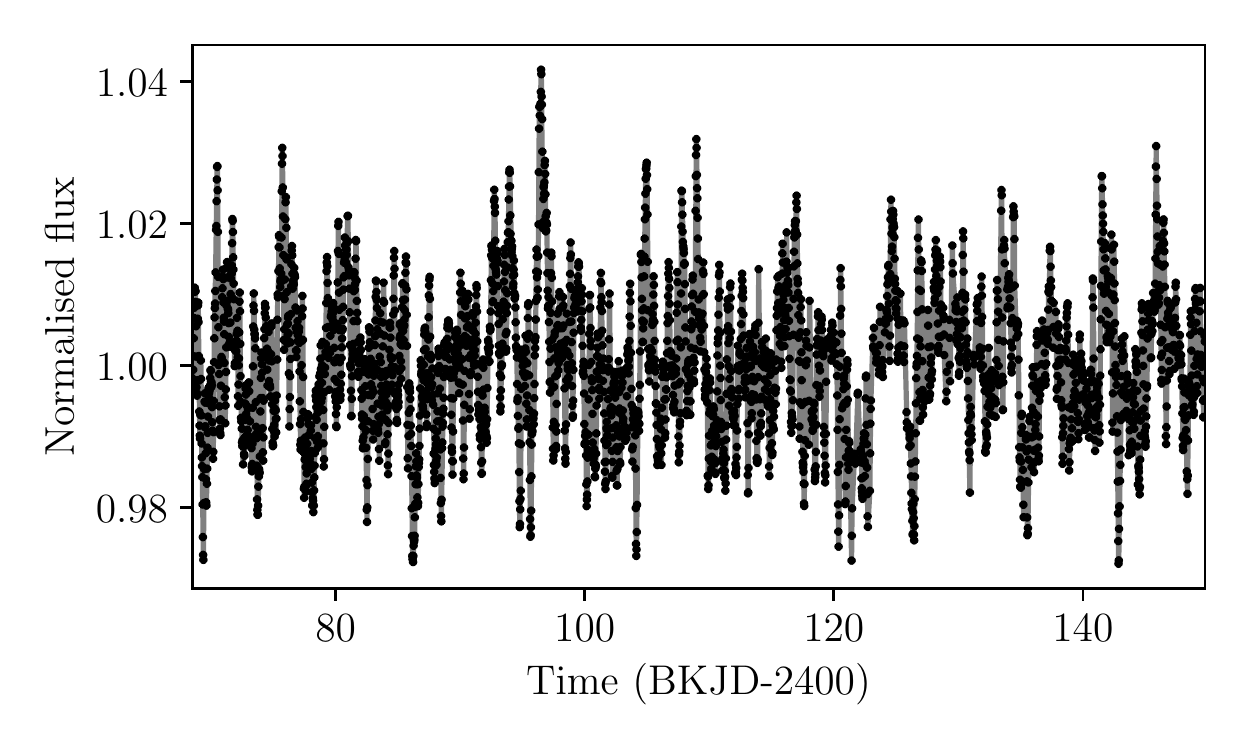}
         \caption{}
         \label{fig:H2-48full}
     \end{subfigure}
     \begin{subfigure}[b]{0.47\textwidth}
         \centering
         \includegraphics[width=\textwidth]{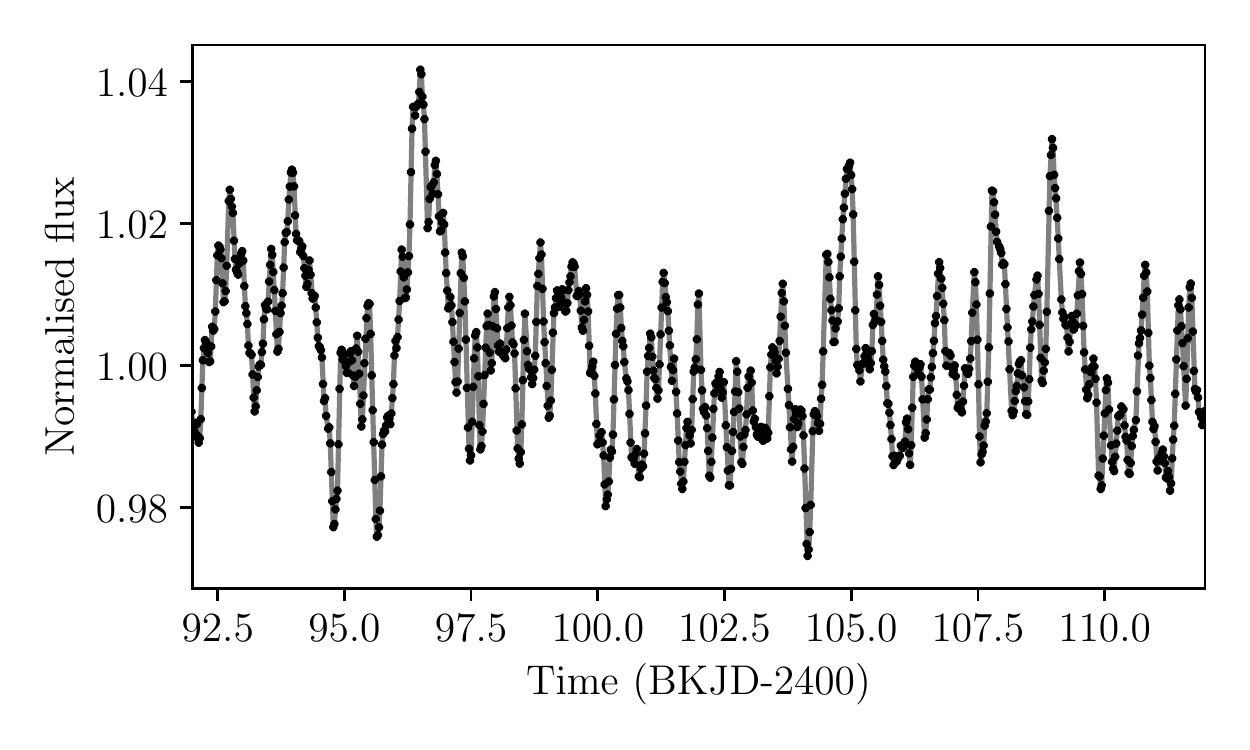}
         \caption{}
         \label{fig:H2-48blowup}
     \end{subfigure}
     \caption{The {\it K2} light curve of the Campaign 11 PN H~2-48 (a) and a zoomed sub-section (b). Solid points represent the {\it K2} data and the solid grey line connects those points.  The average uncertainty on the flux measurements is approximately 0.07 percent (much smaller than the observed variability).}
        \label{fig:H2-48}
\end{figure}

\emph{JaSt 52:} the CS appears to have some cyclical variability in C11b data that may be evident in C11a, but at a lower S/N. It is therefore difficult to determine if the variability lasts for the entire observing window. A period of 3.12~d produces a noisy phased light curve in the C11b data that shows evidence of periodic behavior. A long-term trend in the C11b data also contributes to the scatter as do apparent changes in the variability amplitude. A bright nearby star means that the observed amplitude (averaging 2.3 percent) is likely to be highly diluted. Ground-based follow-up observations should be performed on this object to produce a firm classification of the apparent variability. 

\emph{PHR J1724-2302:} a strong nearly sinusoidal variability is seen in this object with a period of 4.55~d. Ellipsoidal variability normally demonstrates broader maxima and narrower minima. Because that is not the case here, we classify this object's variability as most likely due to the irradiated hemisphere of a cool companion. In addition, if the variability were due to an ellipsoidal effect, the period would be doubled to over 9 days. For typical masses of a PN central star with companions, that would require a star far larger than would be expected in this situation. However, there are measurable changes in cycle-to-cycle amplitude which is not typically seen in irradiated binaries. Either there is some additional source of variability (from the CS or a nearby star), or the periodic variability is due to something other than typical close binary star variability.

\emph{Th 3-12:} this light curve shows two peaks per cycle (period of 0.2640~d) with markedly different minimum depths. The shape is suggestive of both ellipsoidal variability and broad eclipses from a tidally distorted semi-detached binary. While ellipsoidal-dominated systems in CSPNe are expected to be DD systems, as described above for PC~12, and to-date all modeled CSPN systems dominated by an ellipsoidal effect have been DDs, this type of curve in a CS, with two very different minima depths, has not been previously modeled. The closest example is that of M~3-1 \citep{jones19} which does not appear to be a DD. So here we defer a classification of the companion until follow-up observations and modeling can be completed. This central star is easily measured in the LCO data but shows no variation among four visits within the photometric errors of $\sim$0.5 percent. The {\it K2} amplitude of 1.4 percent might have been recovered at LCO if we had additional and/or more frequent visits.
 
\emph{M 2-11:} with a period of 0.5746~d, the maxima of this CS lightcurve are noticeably narrower than the minima, typical for an irradiation effect seen in a moderate to moderately-high inclination binary system. Thus we classify this as an irradiation effect.
 
 \emph{Th 3-14:} this CS has clear cyclical variations in both C11 datasets.  The periodogram produces a series of peaks with periods from about half a day out to about six days. This CS has a \ion{He}{i} Zanstra temperature of $27.5\pm10.2$ kK \citep{gleizes89}, even lower than the CS of H~2-48 above. In this case, the nebula is also known to be very young, with a derived expansion age of 974 years \citep{sahai11}. The variability is most likely due to strong winds and/or internal gravity waves (IGWs) similar to effects observed in hot O stars \citep[e.g.][]{bowman20}. The period of 1.15~d reported in Table~\ref{tab:var-PN-list} is the highest peak in the periodogram but neither the C11a nor C11b light curves fold particularly well on this period. While this is certainly a variable CSPN there is no clear evidence of binarity. We provide a plot of the C11b light curve for Th~3-14 in Figure~\ref{fig:Th3-14} along with a subsection of the same data to show in more detail the cyclical, and possible multi-periodic, variability. There is a weak detection of variability in the ATLAS data with a best-fit period of 0.9645(4)~d. As with H~2-48, we do not include Th~3-14 in our discussion of binary population statistics in Section \ref{sec:discuss}.

\emph{PTB 26:} the very strong double-peak light curve with different minimum depths exhibited in this CS is again strongly suggestive of an ellipsoidal effect with eclipses and one or both stars filling a significant fraction of their Roche lobe. This object shows one of the largest differences between minima depths with the primary (deepest) minimum nearly three times deeper than the secondary minimum. This CS was also observed with ATLAS, LCO, and SARA-CTIO. It is the only clear and strong ATLAS confirmation from Campaign~11. All datasets confirm that the CS is indeed the variable, as well as confirming the shape of the variability. The period given in Table \ref{tab:var-PN-list} of 0.152184~d is determined from the longer time baseline ATLAS observations. Although we measure an amplitude of 22 percent from {\it K2}, the ATLAS observations show an amplitude of 27 percent, our LCO observations suggest an amplitude of at least 37 percent based on only four visits randomly spaced across the variability cycle, and SARA $V$ and $I$ photometry show an amplitude of $48\pm5$ percent. The SARA observations (see Figure \ref{fig:PTB26SARA}) are the least likely to experience dilution. PTB~26 is also a good example of a very short period system that is still well sampled by the 30 min cadence observations. The shape of the \emph{K2} light curve is well reproduced by the shorter 10 minute SARA exposures. At least part of the dilution of the \emph{K2} light curve is likely due to the longer exposure times. This CS is also one of the 58 possible binary CSPNe identified by \citet{chornay21}. Again, with some ambiguity regarding what type of companion results in this light curve shape, we defer classifying the companion.

\emph{Th 3-35:} the light curve shows no clear variability other than a very narrow eclipse. There is perhaps marginal evidence for a secondary eclipse half a phase away from the primary. The period is 3.330~d.

\emph{H 1-17:} the C11a dataset for this CS shows noisy, but present variability with a period of 2.07~d. It is unclear from the light curve if the variability is truly periodic, semi-periodic, or multi-periodic. The C11b light curve has what appears to be instrumental noise in the first quarter of the observing window, and the remainder of the curve only has hints of low amplitude variability, too noisy to confirm the period. With an amplitude of only 0.3 percent and no bright nearby stars to be adding significant dilution it is unlikely that the variability can be observed with ground-based telescopes.

\emph{M 4-4:} the C11a data show a clear, low amplitude (0.6 percent) variability with a period of about 12.5 days. However, with coverage of only about two cycles, confirmation of a consistent variability is not possible. The C11b data are dominated by larger amplitude variability that does not show clear periodicity. It is unclear if the variability in C11b is from the star or is instrumental in nature.

\emph{H 2-13:} this object has been identified as OGLE-BLG-ECL-024233 with a period of 1.7924822~d and an {\it I}-band amplitude of 0.534 mag. However we find a period of 0.897~d (i.e., half the OGLE period), no evidence of eclipses, and an amplitude of only 0.015 mag, suggesting a highly diluted {\it K2 detection} (see Figure~\ref{fig:H2-13OGLE}). Folded on our shorter period, the phase averaged \emph {K2} light curve suggests a very slightly narrower maximum than minimum, thus an irradiation effect rather than ellipsoidal variability.  We therefore tentatively report the shorter period in Table \ref{tab:var-PN-list} that is found by combining both the OGLE and \emph{K2} data, although we cannot rule out the period being double and the source of variability being ellipsoidal modulation.

\emph{MPA J1711-3112:} We find two possible periods for this object, at 0.9825~d with two maxima per cycle, or half that, at 0.4912~d with one maximum per cycle. However, in a phase-averaged light curve (with the data averaged over bins of 0.01 in phase, as shown in Figure~\ref{fig:MPAJ1711-3112}) the maxima appear to be broader than minima, suggesting an ellipsoidal effect. We therefore adopt the longer of the two periods and tentatively identify the variability as due to an ellipsoidal effect and a DD. 

\emph{PHR J1713-2957:} this CS exhibits a noisy, but consistent sinusoidal light curve with a period of 0.1667~d, but only in the C11a data. The C11b data suffers from significantly lower S/N that swamps the variability. Even though no supporting period can be found in C11b data, we include this target as a periodic variable based on the C11a data. Though noisy, the shape of the light curve appears to be sinusoidal. Barring any clear indication of ellipsoidal variability we classify this as due to irradiation. The LCO photometry offers little insight. This nebula is faint and its center is poorly defined. Combined with the fact that the central region is heavily crowded with many extremely faint stars for which none is obviously blue, the central star identification may be incorrect. The crowding likely dilutes the {\it K2} amplitude significantly.

\emph{MPA J1714-2946:} this CS provides a limiting case for our detection of periodic variability at 1.94~d. Both C11a and C11b show indications of periodic variability with the same period. Both curves are very noisy and, on their own, might not result in classifying the system as a detection. However, together the light curves result in our designation of this as a binary candidate. The noisiness of the curves does not allow a clear variability-type classification.

\emph{MPA J1717-2356:} as with PHR1713-2957, we find a consistent variability in C11a data for this CS at 0.2073~d. The C11b data is once again noisier though here it does show a weak signal at the same period. Given that the phase-averaged curve is consistent with a sine curve we classify this as an irradiation dominated light curve. 

\emph{PHR J1718-2441:} we find a weak variability candidate if we offset the aperture mask 1 pixel south. This corresponds better to a star at RA=259$\fdg$67775, Dec$=-$24$\fdg$69122) than to the CS identified in HASH (RA=259$\fdg$67833, Dec$=-$24$\fdg$69028). Given the poorly defined nebula, it is not clear which is the true CS. The period is 2.07~d.

\emph{M 3-39:} here again, the C11a data has better S/N and shows a stronger variability than the C11b data. However, the consistent sinusoidal variability at 4.75~d is clearly present and we classify this CS as having an irradiation dominated light curve.

\emph{K 5-31:} the light curve for this CS is very similar to that for M~4-4; the C11a light curve shows a clear and coherent variability of just under 12~d, while the C11b data are dominated by variations that do not appear to be periodic or of constant amplitude. If the C11b data are dominated by instrumental noise, the 11.9~d variability seen in C11a may be indicative of an irradiation effect. We also note that the period range we find is consistent with being half of the C11a observing window, and thus could be an instrumental and aliasing effect. 

\begin{figure}
     \centering
     \begin{subfigure}[b]{0.47\textwidth}
         \centering
         \includegraphics[width=\textwidth]{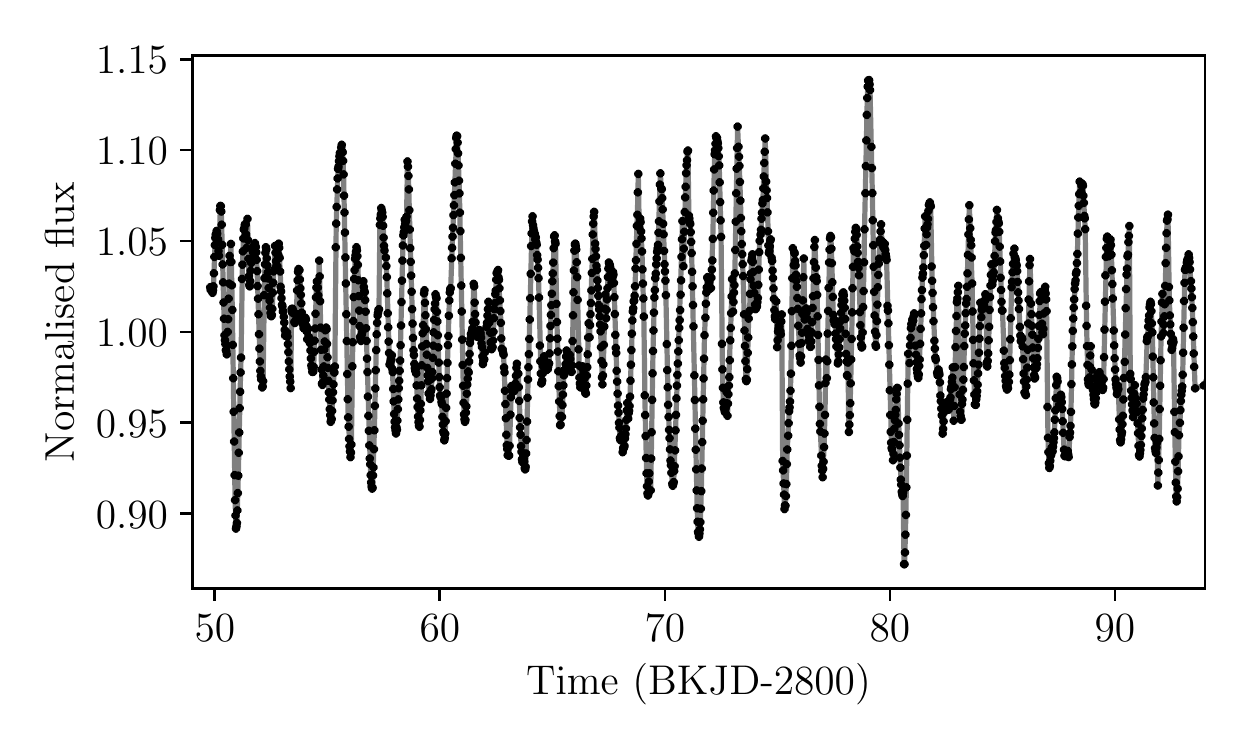}
         \caption{}
         \label{fig:Th3-14full}
     \end{subfigure}
     \begin{subfigure}[b]{0.47\textwidth}
         \centering
         \includegraphics[width=\textwidth]{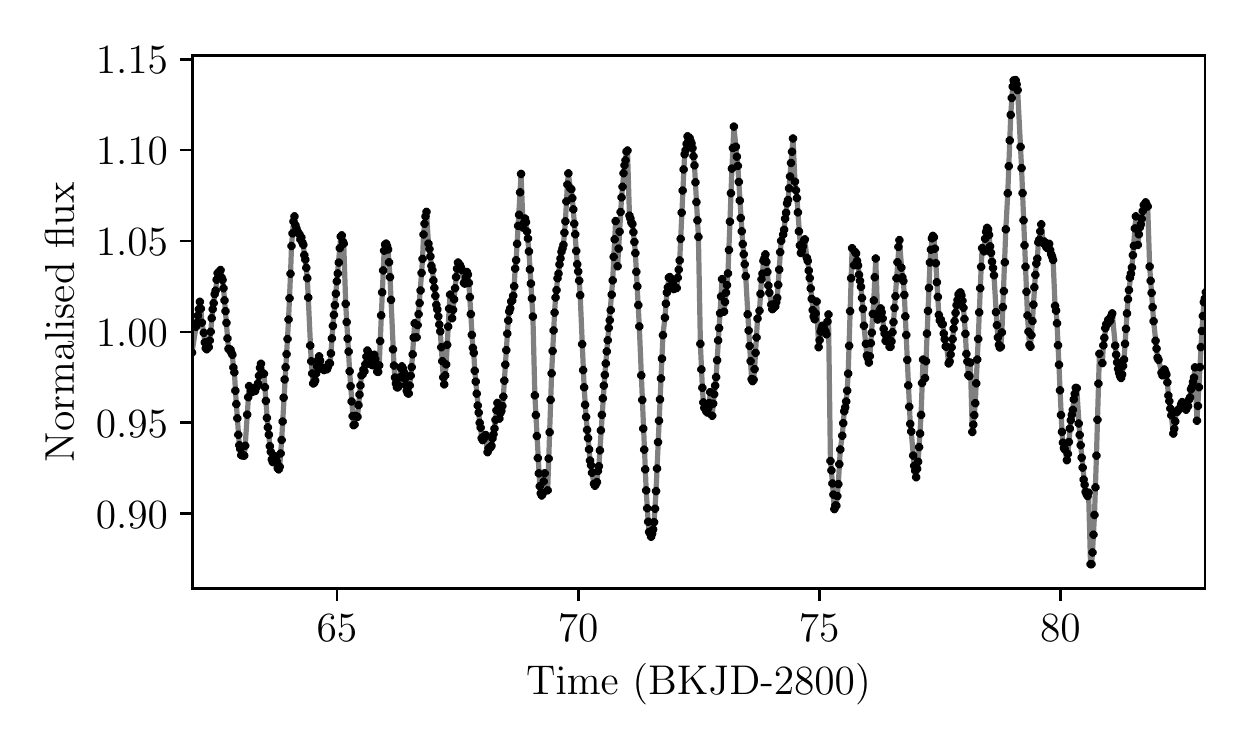}
         \caption{}
         \label{fig:Th3-14blowup}
     \end{subfigure}
     \caption{The {\it K2} light curve of the Campaign 11 PN Th~3-14 (a) and a zoomed sub-section (b). Solid points represent the {\it K2} data and the solid grey line connects those points. The average uncertainty on the flux measurements is approximately 0.04 percent (much smaller than the observed variability).}
        \label{fig:Th3-14}
\end{figure}

\emph{Terz N19:} our shortest period variable at 2.28~h, this CS shows another light curve apparently dominated by a strong ellipsoidal effect and eclipses in a binary with one or both stars filling all or most of their Roche lobe. We again defer classification of the companion in this situation. We note that as the shortest period variable discovered here, this light curve is likely the most affected by the 30 min cadence, and prior to any significant binary modeling, details of the light curve shape should be further confirmed with additional observations. However Terz~N19 was observed for 0.105~d (just over a full cycle) with the Kitt Peak Mayall 4-m telescope. Its amplitude was 37 percent, 6 times the amplitude in the {\it K2} measurements (Figure~\ref{fig:TerzN19Mayall}); yet, the data support the same basic light curve shape, demonstrating that even with 30-min cadence data we are able to recover short period binary CS variability in the \emph{K2} data. The period of 0.0950349(4)~d is derived from the combined {\it K2} and Mayall data.

\emph{PHR J1725-2338:} the light curve of this CS is dominated by an irradiation effect, but shows a primary eclipse at the irradiation minimum and a secondary eclipse at irradiation maximum.  The shape of the primary eclipse, however, is a strong indication that it is close to Roche-lobe filling, and so some ellipsoidal modulation is likely present \citep[at a lower amplitude than the evident irradiation effect, somewhat similar to M~3-1;][]{jones19}. The amplitude given in Table \ref{tab:var-PN-list} is from maximum to minimum light, including both irradiation and primary eclipse. The period is 0.20985~d.

\begin{figure*}
          \begin{subfigure}[b]{0.47\textwidth}
         \centering
         \includegraphics[width=\textwidth]{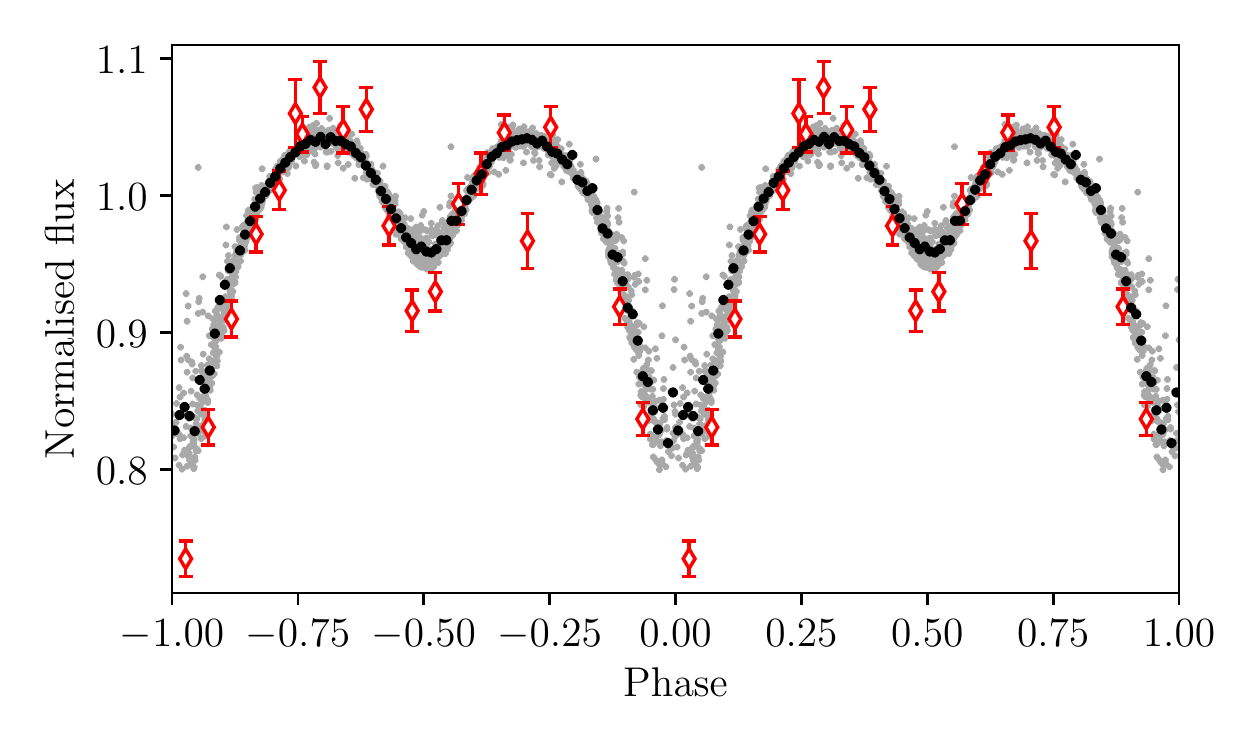}
         \caption{PTB~26}
         \label{fig:PTB26SARA}
     \end{subfigure}
     \hfill
        \begin{subfigure}[b]{0.47\textwidth}
         \centering
         \includegraphics[width=\textwidth]{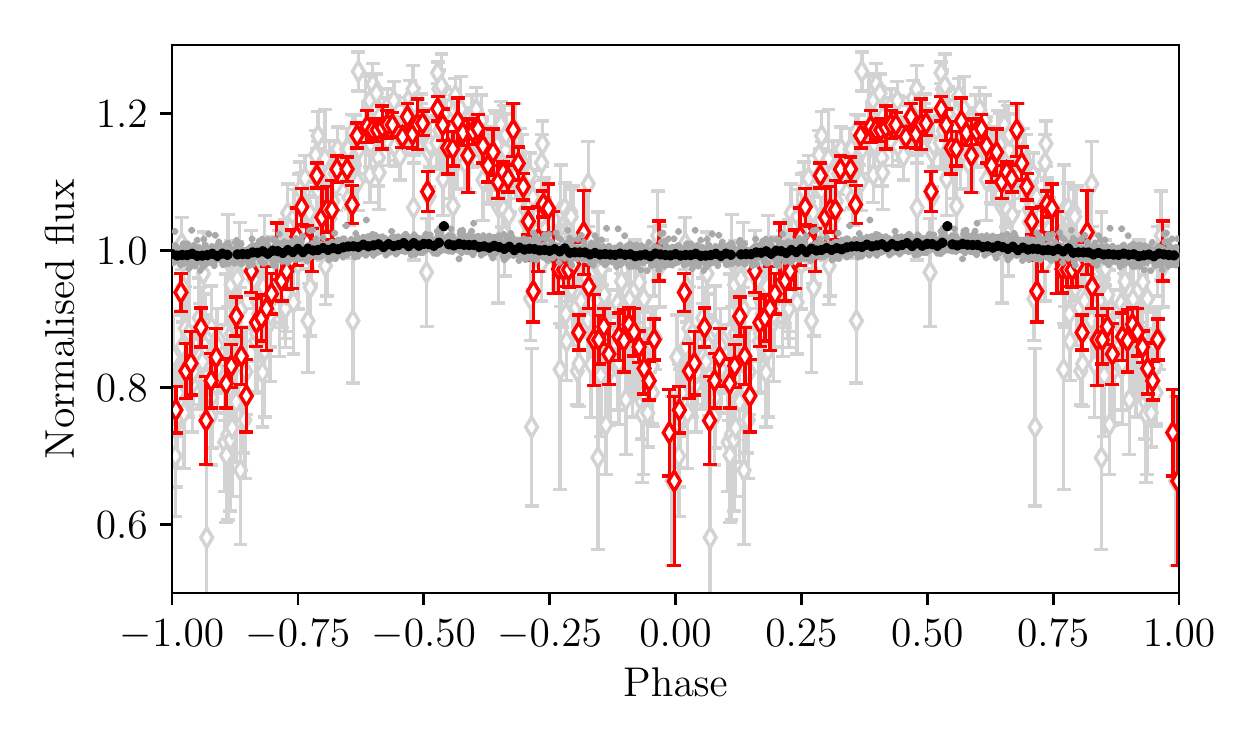}
         \caption{H~2-13}
         \label{fig:H2-13OGLE}
     \end{subfigure}
\\
     \begin{subfigure}[b]{0.47\textwidth}
         \centering
         \includegraphics[width=\textwidth]{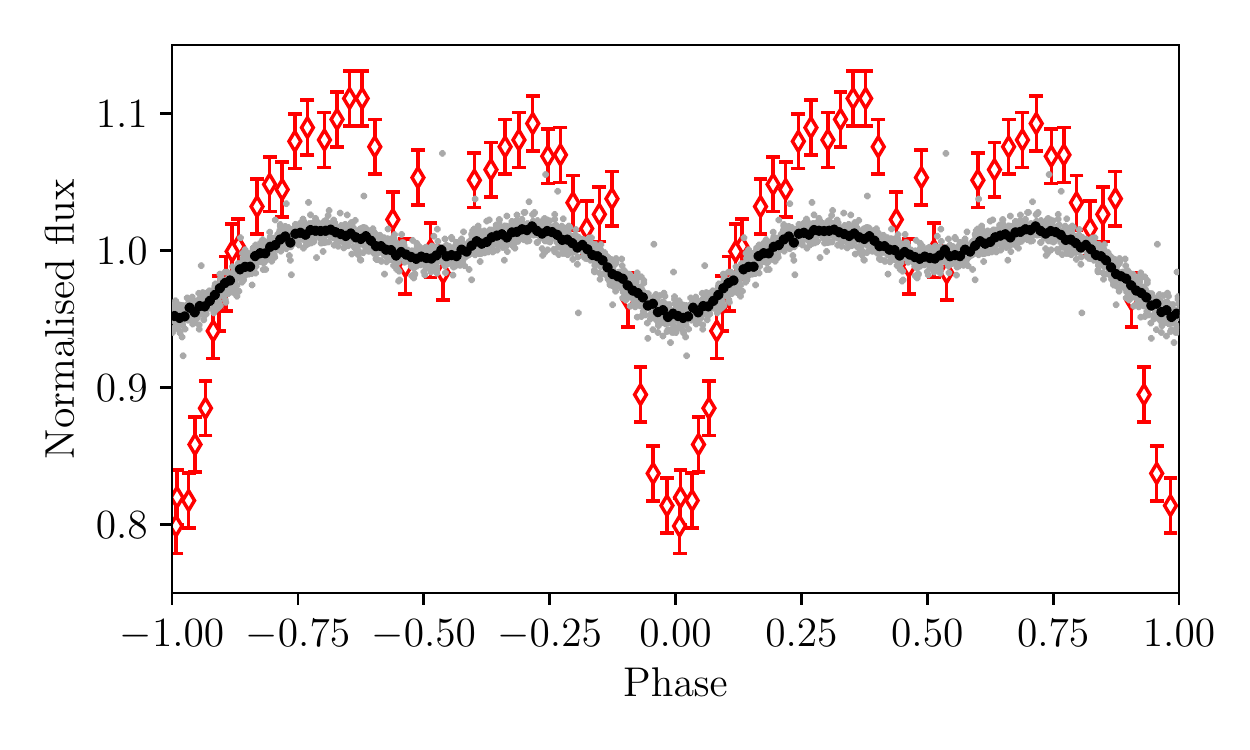}
         \caption{Terz~N19}
         \label{fig:TerzN19Mayall}
     \end{subfigure}
     \hfill
     \begin{subfigure}[b]{0.47\textwidth}
         \centering
         \includegraphics[width=\textwidth]{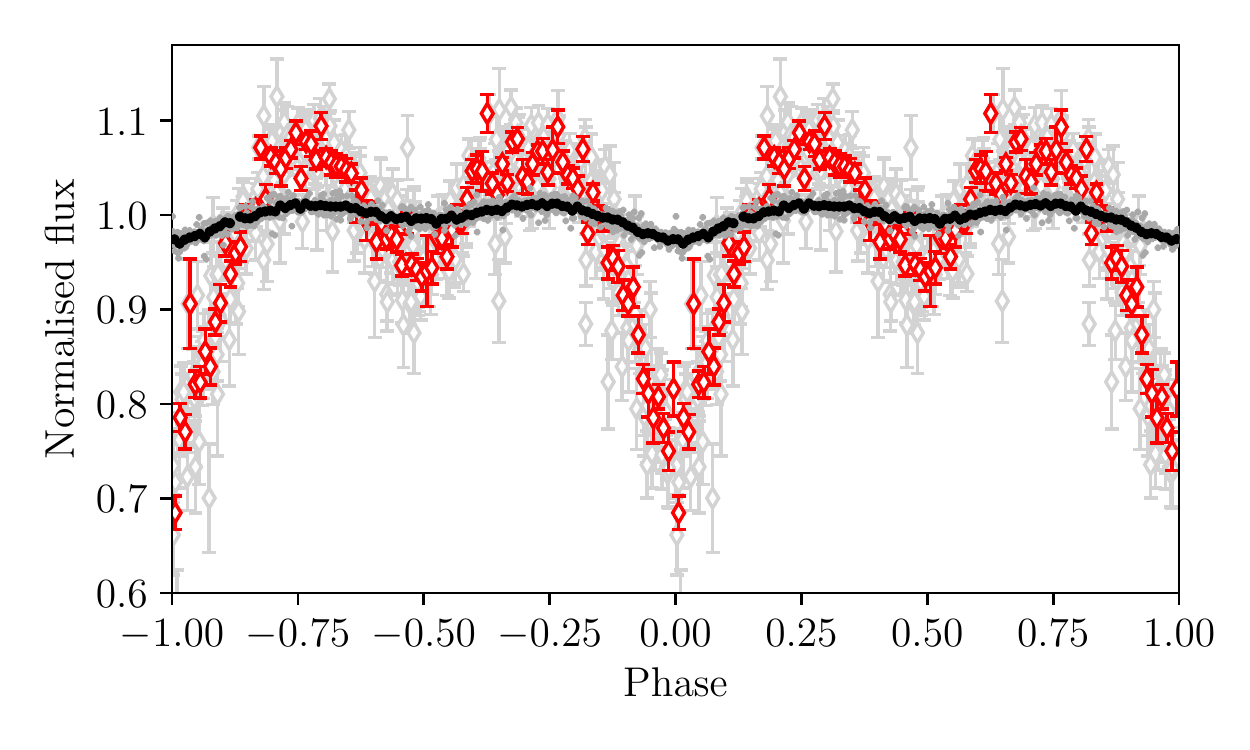}
         \caption{Th~3-15}
         \label{fig:Th3-15OGLE}
     \end{subfigure}
     \\
          \begin{subfigure}[b]{0.47\textwidth}
         \centering
         \includegraphics[width=\textwidth]{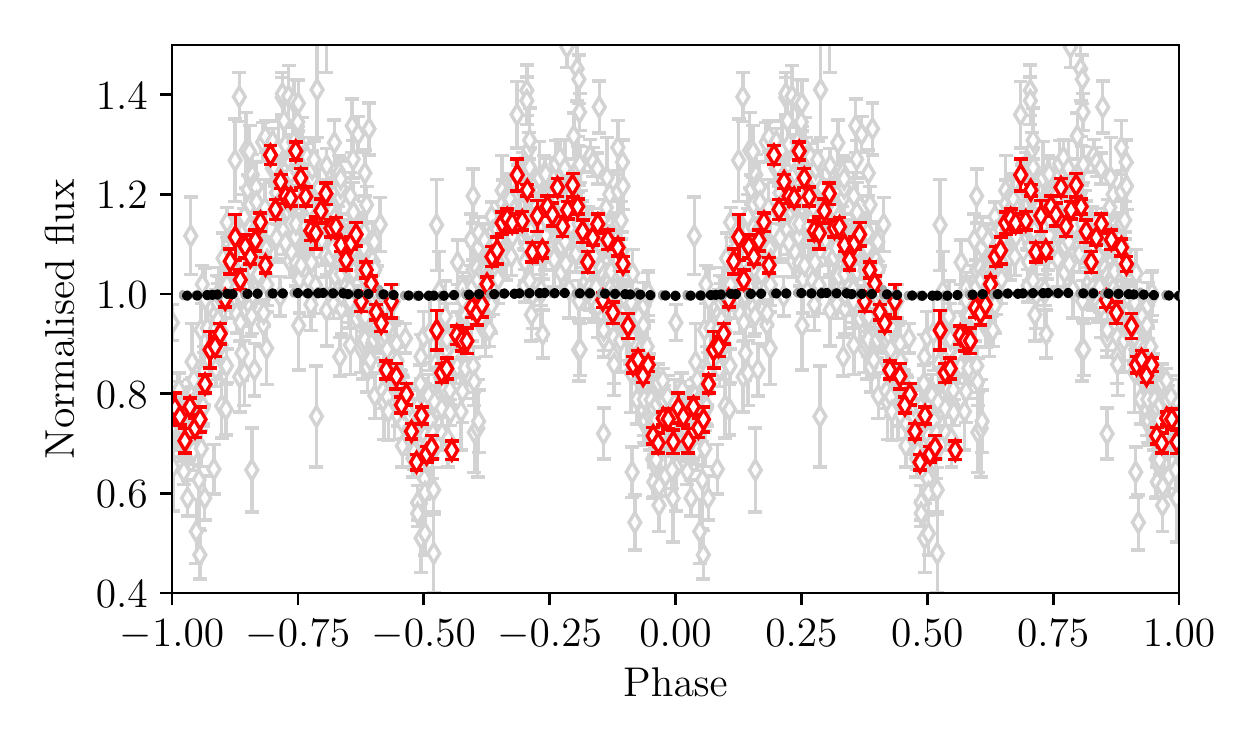}
         \caption{PHR~J1738$-$2419}
         \label{fig:PHRJ1738OGLE}
     \end{subfigure}
     \\
     \caption{K2 and ground-based folded light curves for: (a) PTB~26 , (b) H~2-13, (c) Terz~N19, (d) Th~3-15 and (e) PHR~J1738-2419.  The K2 data is plotted as in Figures~\ref{fig:PTB26}, \ref{fig:H2-13}, \ref{fig:TerzN19}, \ref{fig:Th3-15} and \ref{fig:PHRJ1738-2419}, respectively, while the available ground-based data is underlaid in red (except in the cases of H~2-13, Th~3-15 and PHR~J1738$-$2419 where the red points show the data averaged to bins of 0.01 in phase and the light gray points represent the data without binning).  See the text for details of the origin and passbands of the ground-based light curves. The effects of dilution are dramatically illustrated here for the CS of H~2-13, Terz~N19, Th~3-15, and especially PHR~J1738-2419.}
        \label{fig:dilution}
\end{figure*}

\emph{M 3-42:} the variability from this object shows consistent periodicity in both C11a and C11b datasets. Again we find two possible periods corresponding to either irradiation (0.1601~d) or ellipsoidal (0.3202~d) variability. Despite the low amplitude relative to the noise it does appear that when folded on a period of 0.3202~d the two minima are different depths. We therefore adopt the longer period and tentatively classify this CS as an ellipsoidal variable and a DD. The ATLAS data for this CS supports a period of 0.3075(2)~d which is at odds with the {\it K2} period. However the variability signal was only weakly detected. Nebular emission masks any evidence of a central star in the LCO data preventing any photometric measurements. Given that the star is not discernible in any of the PanSTARRS bands ({\it grizy}), ground-based photometry will have limited value. The bright nebula likely dilutes the central star variability amplitude from {\it K2} significantly.

\emph{Th 3-15:} the light curve from this CS indicates ellipsoidal effects with broad eclipses. This object was identified to be an eclipsing variable as OGLE-BLG-ECL-000009, with a period of 0.1507714(6)~d in excellent agreement with the {\it K2} period of 0.15078(6)~d, and an amplitude of 0.376 mag, much larger than the {\it K2} amplitude of 0.03 mag (Figure \ref{fig:Th3-15OGLE}). It was identified as a binary CSPN by \cite{hlabathe15} and \cite{soszynski15}. The final period given in Table \ref{tab:var-PN-list} was found using both the OGLE and \emph{K2} data.

\emph{K 5-3:} in the first quarter of C11b data, the CS for this PN shows what appears to be consistent variability at the 0.6 percent level with an approximate period of 2.568~d. However, the variability then either goes away, or is lost amidst larger non-periodic variations in the data.  These variations may be instrumental in nature. The C11a data show no clear evidence of variability to within 0.3 percent. It is unclear then what the source of variability may be, and if it is inherent to the CS. With a brighter, nearby neighbor 5\arcsec\  away, we expect significant dilution in the light curve. Therefore, ground-based follow-up observations are likely to help resolve the ambiguity in the light curve of this CS.

\emph{PHR J1734-2000:} as with several other central stars, this light curve is noisy but the shape, slightly broader maxima than minima, suggests ellipsoidal variability with a period of 0.3367~d. The minimum depths are approximately equal to within the scatter of the phase-averaged data. Thus we tentatively classify this as an ellipsoidal variable and a DD. The central star is faint leading to $>$10 percent photometric errors and no evident variability in the LCO data.

\emph{PPA J1734-3015:} the light curve for this CS is typical of an irradiation effect with slightly narrower maxima than minima. The ATLAS data suggest a period of 0.172397(6)~d, close to, but not quite in agreement with, the {\it K2} period of 0.1702(3)~d. Though again, the ATLAS detection of variability was weak.

\emph{V972 Oph:} classified as an old nova, this object has what may be an old PN shell around it and is classified in the HASH database as a possible PN. When folding the light curve on the known orbital period from spectroscopy of 0.281~d \citep{Tappert2013}, we find a noisy, but clear sinusoidal photometric variability. It is possible that the variability is disc or even outflow related. Given the similarity to our other light curves we adopt a classification of irradiation. We note that the variability signal peaks on the star about 6\arcsec\ east of the HASH coordinates, and so, is consistent with the SIMBAD coordinates. This object was also observed at the Kitt Peak 4-m Mayall, but for only about a third of a period, so we were not able to confirm the light curve shape. The weakly detected variability from ATLAS data suggest a period of 0.28436(1)~d, again disagreeing slightly with the {\it K2} period of 0.2800(13)~d and the spectroscopic period of 0.281~d. Data from the Mayall 4-m covering 0.086~d, corresponding to 0.307 of its phase, show a photometric variation of 11 percent.

\emph{MGE 359.2412+02.3353:} despite a small amplitude and noisy light curve, this CS shows sinusoidal variations of consistent amplitude with a period of 0.2682~d in both C11a and C11b. The variation appears to be sinusoidal and we classify the tentative cause as irradiation.

\emph{JaSt2 2:} we find this CS has consistent periodicity 0.3716~d and amplitude in both sets of C11 data, though with considerable noise.  We also find that the maxima appear broader than the minima and there may also be a slight deviation in the depths of the two minima. We therefore adopt a tentative classification of ellipsoidal and a DD.

\emph{PHR J1738-2419:} the light curve shows consistent periodic variability in C11a data and very weak variability at the same period in C11b data. Upon further checking, this signal is certainly from the nearby (10\arcsec\  distant) eclipsing binary OGLE-BLG-ECL-053461. As confirmation, we obtain a very good signal from {\it K2} by moving the aperture mask 2 pixels to the northwest. Our period and light curve shape also agree with that from OGLE. The phase-averaged light curves of OGLE-BLG-ECL-053461 appear to have an amplitude of roughly 65 percent and a broader maximum than minimum so we adopt a period of 0.418839~d with two maxima per cycle and classify the target as an ellipsoidal variable. The corresponding designation as a DD we include as tentative since, if this is not the CS, an ellipsoidal effect need not be due to a DD system. We use the longer baseline OGLE data to determine the period given in Table~\ref{tab:var-PN-list}. Because the PN is poorly defined, so is the identification of the CS; the OGLE identification may actually be the CS, but we suspect not. The CS as identified in HASH is unexpectedly bright for a CS in this sample ({\it i}=16.0 from PanSTARRS) and the OGLE variable is $\sim$3 mag fainter and off-center to the west. From our LCO data, there is a very faint star that appears more geometrically centered than either the OGLE or HASH stars and is $\sim$4\arcsec\  northeast of the HASH coordinates at: RA$=$264$\fdg$71595, Dec$=-$24$\fdg$32794. It is about 6 mag fainter than the HASH CS. 

\emph{JaSt 12:} the two eclipses visible in the light curve for this CS dominate the variability with a period of 4.357~d. However additional variability is also evident in the light curve continuum. With the existing S/N, it is difficult to classify this system with high confidence, but appears to be an ellipsoidal component, and so we tentatively classify this as a DD.

\emph{Terz N 1567:} the nebula shows a tight waist at high inclination with polar lobes that have not yet broken out. The equatorial nebulosity is bright and it is difficult to detect the CS. Thus we expect a large level of dilution to any variability of the CS. The light curve for the CS exhibits a periodic variability of 0.1714~d just above the noise level in both sets of C11 data. However, the amplitudes and periods are consistent, so we include it in our list of probable binary systems.  The light curve shape is roughly sinusoidal and we classify it tentatively as an irradiated system.

\emph{Me~2-1:} this was the only PN observed in Campaign~15.  The nebula is small ($\sim$8\arcsec) and the CS is relatively bright ({\it g}$\sim$14.0). The observing duration of 90~d is relatively long for {\it K2}, so there is reason to have confidence in the period that we find here of 22~d since the photometric coverage extends over four cycles. However, given the small amplitude of 0.1 percent, further confirmation, while difficult to achieve at this low amplitude, is needed as the long period could be an artifact of the pipeline processing. See Figure~\ref{fig:Me2-1}.

\emph{Abell~30:} in a recent paper, \cite{jacoby2020} discussed the variability of the central star of Abell~30. As one of a handful of `born-again' PNe, Abell~30's unique status argued for a quick report that was published separately. In brief, the central star of Abell~30 has a periodic variability of 1.060~d that is very likely due to the presence of an irradiated companion. Binarity had been suspected among the born-again PNe for some time \citep{corradi15, wesson18} due to their high abundance discrepancy factors. Because Abell~30 has an extreme abundance discrepancy factor \citep{wesson03}, the strong case for binarity is consistent with prior expectations.  The light curve for Abell~30 is shown in Figure~\ref{fig:A30}, folded on the period found by \citet{jacoby2020}.


\section{Discussion}
\label{sec:discuss}

\subsection{Statistics of the Binary Population}
\label{subsec:stats}

For the purposes of this section, we adopt the notion that the periodic variability we see in the central stars is a consequence of being a binary star. Those with eclipses are definitely so, and those arising from ellipsoidal effects also must be binary. Variability that is attributed to irradiation, under some circumstances, may originate from other binary-related mechanisms such as Doppler beaming, or from non-binary mechanisms. See \citet{jacoby2020} for a discussion of the alternatives and why they may be unlikely, but not fully refutable. These non-binary alternatives include pulsations of the CS and star spots on its surface. It may also be possible that the cool companions to the hot CS exhibits pulsations or spots, such as in the cases of K~1-6 \citep{frew2011} or LoTr 5 \citep{aller18}. Regardless, it is possible that the periods in Table~\ref{tab:var-PN-list} do not represent orbital periods for every case, but for our discussion here we treat them as such. When follow-up spectroscopy can be obtained, this assumption should be revisited.

The total number of binary candidate central stars for all of the {\it K2} campaigns is 34 (after removing H~2-48 and Th~3-14 because the variability is highly likely to be due to causes other than binarity) 
and the total number of PNe targeted is 204, as summarized in Table~\ref{tab:campaigndata}. These numbers correspond to a 16.7 percent fraction of targets having a signature of periodic variability that likely originates from a binary central star.

Given the current uncertainties, one could explore how this fraction changes by bringing in additional considerations. For example, one could remove all targets that are not `true' PNe from the sample. There are five binary candidates in the `true' sample, from Campaigns 0, 2, 7, 15, and 16, and 22 binary candidates from Campaign~11. The fraction then becomes 27 binary candidates out of 145 `true' PNe, or 18.6 percent. Whether we take 16.7 or 18.6 percent, we show below that, due to a number of additional considerations, this percentage should be considered a lower limit.

\subsubsection{Incompleteness Due to Faint Targets in Campaign 11}

Many Campaign~11 targets are extremely faint, having a  signal that is in the noise in the {\it K2} pixel images. The `noise' is primarily due to the flux from numerous nearby stars. This effect represents an incompleteness due to a combination of the limiting magnitude for {\it K2} with the crowding from neighboring stars. We  attempt to correct for this effect below.
    
\citet{gilliland2011} discussed the details of photometry precision for {\it Kepler}. The {\it Kepler Instrument Handbook} \citep{vancleve2016} also provides valuable guidance in estimating the noise of {\it K2} observations. However, due to the extreme crowding and reddening for Campaign~11 targets and the object-specific conditions, we found that an empirical approach is more tractable and provides a similar result.
    
We derived an estimate of the incompleteness in the following way. First, we assumed that the faintest CS (that of JaSt~12) in the sample of Campaign~11 variables represents the magnitude for completeness. That CS has {\it r}=20.10 from PanSTARRS. PHR~1718-2441 has a nearly identical value with {\it r}=20.07. This is similar to the {\it I }$<20.5$ Campaign~9 targets selected by \citet{henderson2016} under similar conditions. 

Next, we identified the fraction of the non-variable targets in Table~\ref{tab:nonvar-PN-list} that are below some limiting magnitude, using PanSTARRS as the magnitude source. Of those 153 Campaign~11 targets, there are 27 that are too far south for PanSTARRS. There are also four variables from Table~\ref{tab:var-PN-list} that are too far south for PanSTARRS. That yields a sample of 126 non-variable targets whose magnitudes we can consider, plus 26 variables (including one that is not a binary).
    
Of these, PanSTARRS fails to report an {\it r}-band magnitude for 24 objects, either due to coordinate errors, extreme reddening, or because the CS is extremely faint. For whatever reason, these would not have been detectable by {\it K2}. 

In addition, there are those targets with {\it r}-band magnitudes that are fainter than some ultimate detection limit. For those targets having {\it r}-band magnitudes, there are 25, 13, 7, and 4 objects fainter than magnitude 20.1, 21, 22, and 23, respectively.

As noted by Poleski (priv. comm. 2021) the likelihood of measuring a variable that is fainter than {\it r}=23 is close to zero. Thus, we adopt a conservative estimate for our incompleteness based on this magnitude limit. As shown in Table~\ref{tab:bfsummary}, which summarizes the numbers at various steps in the faint limit correction, the close binary fractions with these assumptions are 20.7 percent for all targets, and 23.5 percent for all targets if limited to the `true' PNe. Since we don't know the exact magnitude limit, we note that these fractions rise to 22.1 percent (`all' PNe) and 25.3 percent (`true' PNe only) if the magnitude limit was 21 rather than 23. Thus, the results are not a strong function of the completeness point.

\begin{table*}
\caption{Summary of Binary Fractions With and Without Faint Limit Correction in Campaign 11 (C11)}
\label{tab:bfsummary}
\begin{tabular}{l c c c c c c c}
\hline
 & &  Full Sample & & & & True PN Only & \\
 \cline{2-4}
\cline{6-8}
Subsample & Number & Sample & Fraction & & Number & Sample & Fraction \\
 & Binaries & Size & Percent & & Binaries & Size & Percent \\
\hline
Campaigns 0, 2, 7, 15, 16  & 5  &  21 & 23.8 & & 5  & 17  & 29.4 \\
C11 all                             & 29 & 183 & 15.8 & & 22 & 128 & 17.2 \\
C11 with DEC$>-$30                  & 25 & 152 & 16.4 & & 18 & 101 & 17.8 \\
C11 with DEC$>-$30 and {\it r}$<$23 & 25 & 124 & 20.2 & & 18 & 86  & 20.9 \\
All campaigns (with C11 corrected)  & 30 & 145 & 20.7 & & 23 & 98  & 23.5 \\
\hline
\end{tabular}
\end{table*}

\subsubsection{Incorrect Central Star Identification}

 For older, faint PNe, the CS may no longer be near the geometrical center of the nebula. This can lead to an incorrect 
 identification in these crowded and reddened fields, and consequently, incorrect {\it K2} targeting of the true central star. Targeting the wrong central star could both detract or increment the binary fraction. We do not attempt to make any correction for this effect, although the incompleteness factor above should account for some targets that fall into this category.

\subsubsection{Unreliable Variability and Limited Observing Duration}

Because useful data during Campaign~0 was limited to only 36 days and Campaign~11 was divided into two short observing sequences, the sensitivity to all periods, and especially long periods, was degraded for these campaigns. For Campaign~11, one can, in principle, combine the two light curves to recover a near full-length campaign duration. In practice, the systematics of the repositioning and the reduction pipeline produce two data sets with differing quality, limiting the value in combining them. On the plus side, we effectively have two largely independent data sets to compare results. 

To assess the severity of the shortened observing runs on our targets, we reviewed the {\it Kepler} study from \citet{demarco15} where targets were observed for 17 quarters of data. Typically a `quarter' represented 85-95 days, however, `quarters' 1 and 17 were each only 33 days. That duration is more-or-less representative of the 23 and 48 day durations for Campaigns 11a and 11b and the 36 day useful period of Campaign~0. 

Of the four PNe that \citet{demarco15} found to be variables, three (NGC~6826, Pa~5, and J19311) were included in the short quarters 1 and 17. In quarter 1, NGC~6826 showed no variability, J19311 showed a marginal signal, and Pa~5 had a good periodic signal detection. In quarter 17, NGC~6826 had a good signal, and the other two had compromised, but detectable, signals. 

The fourth PN, Kn~61, exhibited semi-regular variability and was observed only in four normal length quarters. Its apparent variability goes largely undetectable in any single quarter. Because this object is quite unusual \citep{hermes2015}, we don't consider it representative for our purposes. Similarly, NGC~6826 did not exhibit a stable pattern, and was only well-detected as variable in 9 of the 17 quarters. We don't consider this object representative either.

Thus, there are two objects from the original Kepler study, J~19311 and Pa~5, that exhibited clear variability only 75 percent of the time during the short quarters (1 and 17). These may be indicative of a similar effect during Campaign~0 and Campaign~11.

Although we may be missing some candidates having periodic variability signatures, we expect irradiation and ellipsoidal binary systems to exhibit reliable photometric effects. Consequently, we make no attempt to correct for the reduced variability detection either due to the shorter observing duration or to the possible existence of unusual PNe like those analysed by \citet{demarco15}.

\subsubsection{Dilution}
\label{subsub:duration}

Dilution factors of 10 or higher relative to ground-based observations are not unusual: Terz~N19 (observed with the Mayall 4m with 37 percent amplitude), H2-13 (OGLE-BLG-ECL-024233 with 0.534 mag amplitude (63 percent)), and Th~3-15 (OGLE-BLG-ECL-000009 with 0.376 mag amplitude (41 percent)). PTB 26, on the other hand, was less affected; the {\it K2} amplitude is only reduced by a factor 2 relative to the ground-based measurement with data from the SARA telescope at CTIO with a 0.42 mag amplitude (47 percent) and the LCO Swope telescope (38 percent). The 30  min cadence observations are likely the cause of some of this dilution, and it is possible that there are some variables with very sharp features that are not detected due to the 30 min cadence. For example, a short period system with partial eclipses and no significant additional variability may not be identified.
    
Although dilution is likely the main reason why we may not detect a binary CS, we make no correction for this effect because each target requires a unique and generally unknown factor.

\subsubsection{Astrophysical Factors}

Despite the excellent sensitivity of K2, there are several astrophysical situations where a binary signature will go undetected. These are (1) an unfavorable orbital plane inclination, (2) a companion that is smaller than some threshold to yield a measurable irradiation or elliptical signal, 
and (3) a companion that is so distant that its irradiation or elliptical signal is too small for detection. Like other surveys, our sensitivity for detecting binary stars decreases with longer periods (more than approximately two weeks) in systems having small secondaries, and with unfavorably aligned systems, though the latter parameter is relatively minor. See, e.g., discussions by \citet{bond00, demarco08, demarco15,jones17,boffin19}.

\subsubsection{Merged Pairs}

Some fraction of main sequence close binaries will result in a common envelope merger, usually due to a lower companion mass \citep{iaconi19}. Such mergers can happen on the red giant branch after which the, now single, star evolves through the asymptotic giant branch phase and on to become a single PN CS with a slightly larger mass and more angular momentum, something that may affect its shape. The common envelope merger can also happen on the asymptotic giant branch and affect the shape of the PN, because of the injection of orbital angular momentum. While common envelope mergers are likely one of the reason for non-spherical PNe, they result in single CSs and it is almost impossible to demonstrate that those single PN CSs come from mergers.Nevertheless, an example from the Kepler sample is NGC~6826 which was found to vary due to fast rotation \citep{prinja2012} showing it to be a merger remnant.  A fast rotation like this cannot be explained without a companion that adds its own orbital angular momentum to the primary star \citep{Dan2014}.

\subsubsection{Limits of the Analysis Tools}

In some cases, the exact positioning of the measuring aperture is critical. If it is displaced by a single pixel, the variability signal may decrease dramatically and be undetectable. Of our 30 variables in Campaign~11, we were able to improve the signals for six of these with a small pixel offset. We did not attempt to search for an offset variable among those targets in the non-variable list.

\subsubsection{Summary}
\label{subsub:stat_summary}

While our close binary fraction is well below the 80--85 percent fraction of PNe that are non-spherical, our value is a lower limit. Perhaps more importantly, the range of fractions that we derive, 20--24 percent, is indicative of values that are slightly larger than seen in the literature \citep[][see also Section~\ref{ssec:ComparisonofTheCloseBinaryFractionwithOtherVariabilitySurveys}]{bond00,miszalski09}. Thus, even with the drawbacks of the current survey, {\it K2} is a powerful tool in the detection of close binary CSs. 

\subsection{Comparison of The Close Binary Fraction with Other Variability Surveys}
\label{ssec:ComparisonofTheCloseBinaryFractionwithOtherVariabilitySurveys}

\citet{miszalski09} summarized the motivations for measuring the fraction of binary central stars in PNe. The principal drivers are (1) to assess whether most PNe form through a binary pathway, as proposed, for example by \citet{moe2006} and \citet{paczynski85}, and (2) to ascertain the impact, or requirement, that a binary CS has on shaping the PN. 

\citet{bond00} derived an estimate of the binary fraction at 10--15 percent, although one of the 13 variables in his list has since been rejected as a PN \citep[Abell~35; see discussion in][]{tyndall13}, and another has been shown to be a chance alignment rather than the true CS \citep[SuWt~2;][]{jones17b}. \citet{miszalski09}, using OGLE results for the Galactic bulge for a much more refined sample, derived a value of 12--21 percent, the high end value based on the subset of `true' PN only.

A similar close binary fraction is found by \citet{chornay21}, who arrived at 18 percent from the Gaia sample. However, as they point out, this value suffers from significant biases. 
So their result has significant uncertainty, but is consistent with previous values. 

In this study, we find close binary fractions (20--24 percent) that are somewhat higher than earlier surveys. We attribute the higher number to the advantages of {\it K2} relative to ground-based surveys. 

We emphasize that our close binary fractions are lower limits. To ensure a conservative result, we have not applied corrections for several systematic effects described above other than incompleteness for very faint CS. On the other hand, if we adopt the Poisson statistical error of approximately $\pm5$ percent on the binary fraction, then our results reasonably overlap other estimates.

\subsection{Understanding the Close Binary Central Star Fractions}

To understand whether non-spherical PNe come from binary interactions we need to reconcile the fraction of non-spherical PNe (80--85 percent) with the fraction of central stars that either have a companion close enough to have interacted, or can be shown to be the product of a merger.

As we have discussed, our binary fraction is a strict lower limit for technical as well as astrophysical reasons. One of the astrophysical limits is the correlation of the variability amplitude with the orbital period: this detection technique, even if not plagued by any technical limits is unlikely to find binaries with periods much in excess of 2 weeks \citep{demarco08}.

Other techniques are sensitive to binary systems with any separation or orientation, for example, the infrared excess method \citep{DeMarco13, Douchin15, Barker18}. Fractions of 50$\pm$24 percent were measured but this is also a lower limit as it cannot detect mergers or faint companions or compact companions such as WDs. Accounting for fainter and compact companions (but not for mergers) would bring the binary fraction to 83 percent but with an error of order 30 percent, leaving the value poorly constrained. We conclude that it is entirely plausible that the fraction of PNe with a binary CS that has interacted in the past, or with a CS that has merged in the past, could be of the order of 80-85 percent, although this cannot be confirmed at this time.

There is a second issue raised by the discovered fraction of close, post-common envelope binary CSs. This value at $>$20 percent, is already far greater than the 2.5 percent expectation derived from stellar population models and binary interaction scenarios. These assume that all stars in the $\sim1-8$~M$_\odot$ range produce a PN. Of these, given the main sequence binary fraction and period distribution, only a small fraction of systems would go through a common envelope on the AGB \citep{madappatt16}.
The reason why we might observe many more short period, post-common-envelope binaries than predicted by population models can be explained if  not all 1--8~$M_{\sun}$ stars produce PNe. \citet{moe2006} made a similar argument but based on a very different premise: they predicted the number of PNe that should exist in the Galaxy using a population synthesis model, but found that number to be about 6 times larger than observationally-based values.

Clearly, these percentages and factors either have large uncertainties or they tend to be limits. Nevertheless, the body of evidence strongly argues that observable PNe form more easily via a close-binary pathway than from single star evolution.

Further detailed observations and analyses are needed to place the Binary Hypothesis \citep{demarco09} on firmer ground. While the known sample of post-common-envelope central stars exhibits some  trends towards, for example, morphological symmetries \citep{miszalski09b, hillwig16} and high abundance discrepancies \citep{wesson18}, we still require a clearer understanding of the exact role that the central binary has played in producing the observed properties.  This understanding will require painstaking characterisation of a statistically significant fraction of nebulae (morphologies, kinematics, and chemistry) and central stars (masses, temperatures, radii, and chemistry), in order to define the underlying trends needed to inform theory.

\subsection{Comparison of the {\it Kepler/K2} Period Distribution with that Determined from Ground-based Observations}

\begin{figure}
\centering
\includegraphics[width=0.475\textwidth]{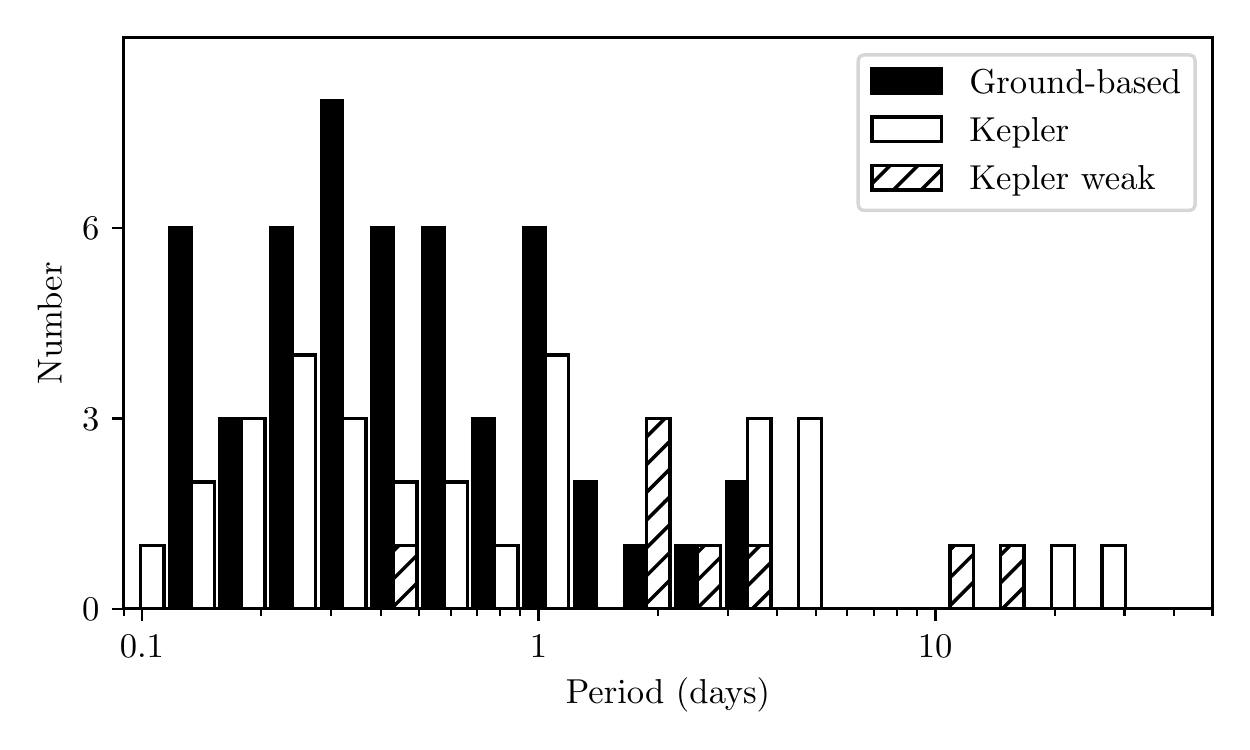}
\caption{The period distribution of ground-based photometrically-detected binary central stars (black bars) versus that of the Kepler sample (white bars).  The contribution from the weak candidates to the Kepler distribution is highlighted by the hatch regions.}
\label{fig:period_dist}
\end{figure}

We have examined the possible difference in the period distributions of the binary central star sample discovered by {\it Kepler/K2} and the ground-based sample found in the literature. We expected that the exceptional precision and stability of {\it Kepler/K2} should be more effective in unveiling any hidden population of binaries with periods longer than the few days typically discovered from the ground.  The period distribution of the ground-based sample\footnote{Taken from the compiliation hosted at http://drdjones.net/bcspn and restricted only to the systems that have been found to be photometrically variable from the ground (i.e. excluding systems discovered through their radial-velocity variability).} is plotted alongside the full {\it Kepler/K2} sample \citep[including systems from][]{demarco15} in Figure~\ref{fig:period_dist}.  

Intriguingly, the statistical correlation between the ground-based and {\it Kepler/K2} samples hinges on the inclusion or exclusion of the weaker candidates identified in this paper, which, as expected have longer periods and shallower amplitudes.  These include: JaSt~52, H~1-17, M~4-4, MPA~J1714-2946, PHR~J1718-2441, K~5-31, K~5-3, and PHR~J1738-2419.  An Anderson-Darling K-sample test that includes the weak candidates shows that the null hypothesis that the two samples are randomly drawn from the same parent population can be rejected with a significance (p-value) of $\sim$1 percent (i.e., one can be approximately 99 percent certain that the two are not drawn from one population).  However, if we restrict the calculation to only the strong candidates, the significance is roughly 15 percent, and thus the null hypothesis cannot be excluded.  The difference is due to the weak candidates exhibiting longer periods and fewer short periods than those typically identified with ground based observations.  A similar inference -- that space-based observations reveal a population of longer period post-common-envelope binaries -- can be drawn from the sample of binary candidates revealed by {\it TESS}, where all have periods longer than 1 day \citep[][]{aller20}.

The common envelope interaction naturally reduces orbital periods, but there is no expectation that the periods will be systematically reduced to below a day. \citet{iaconi19} showed that while hydrodynamic simulations cannot yet determine the post-common-envelope period distribution, there are several conditions under which separations of tens of days are readily achieved. For example, when the mass ratio at the time of common envelope is close to unity, the orbital separation tends to be wider. Further, population synthesis models \citep[e.g.,][]{Yungelson1993,Toonen2013}, based on the alpha or gamma formalisms, have never predicted a gap in periods above one-to-a-few days.

Hence we revise here the conclusion, typically drawn from ground-based samples, that  if binaries with wider separations than a couple of days existed, they would have been detected \citep{demarco08,jones2017}. While ground-based observations are, in principle, able to detect variability amplitudes smaller than 0.1~mag and binaries with periods as long as 1-2 weeks, in practice they tend not to because of the issues we have described. With space based searches such as {\it Kepler/K2} and {\it TESS} we are slowly uncovering those longer period, post-common envelope binaries that are predicted to exist. 

\section{Conclusions}

We have observed a total of 204 targets in {\it K2} campaigns 0, 2, 7, 11a/b, 15 and 16. Of these, 36 were found to have periodic variability, although two of them were complex and contained multiple periodicities. Five of those were previously known variables. Th~3-15 \citep{hlabathe15,soszynski15}, H~2-13 \citep{hlabathe15}, PHR~J1738-2419 \citep{soszynski15}, and V972~Oph \citep{Tappert2013} were previously identified as binary systems while PTB~26 was identified as a probable variable in the \emph{Gaia} data \citep{chornay21}. The remaining 31 objects are new discoveries, 29 of which are possible or likely new binary CSPNe.

We applied an approximate correction for incompleteness beyond the faint limit of {\it K2}, but we did not correct for other factors that likely impacted our ability to identify variability in some objects. From the sample statistics, we estimate that the fraction of PNe with a close binary CS is 20.7 percent when PNe of all status groups (`true', `likely', or `possible') are considered and 23.5 percent if only `true' PNe are considered. We believe that these are lower limits to the actual fraction for reasons discussed in Section~\ref{subsec:stats}.

The {\it K2} mission clearly provided us with excellent results for many PNe. The benefits of continuous time coverage for many days dramatically improves the sensitivity to very low levels of variability. Many of our targets yielded textbook-like phased light curves, especially after binning. For the detection of binary stars using photometric methods, a space-based facility offers clear advantages. 

While {\it K2} allowed us to detect many new binary candidates and recover several others, it has technical limitations. These require that additional imaging data be obtained to confirm the variability of many of the candidates, and to derive more accurate measures of the amplitudes required to analyse the systems. Unfortunately, this won't be possible for some very low amplitude objects. 

Ultimately, radial velocity measurements are needed for each candidate to confirm that there is a companion and to derive the parameters of the binary system. In some cases, deep infrared spectroscopy will be helpful to search for the signatures of cool companions.

As suggested by \citet{miszalski09b}, we would also like to compare the morphologies of those PNe having a binary CS to those without. This may require space facilities or ground-based adaptive optics imagers since many nebulae are very small and have ambiguous morphologies when observed with conventional ground-based facilities.

In conclusion, this {\it K2} survey was highly successful, yielding many excellent light curves and a binary fraction that is comparable to, or larger than, previous studies.

We reiterate that close binary central stars of PNe are only a part of the story of binary interactions generating PNe. Longer period systems, but still able to interact on the asymptotic giant branch, cannot be identified even by the best photometric variability technique. Finally there must be mergers that contributed to the PN shape, but that are today single stars and that can only be identified as past mergers by difficult observations of the rotational speed.

The close binary fraction, even as a lower limit, is already almost 10 times larger than stellar population models where we expect a low close binary fraction ($\sim$2.5 percent) of CSPN. This discrepancy is attributed to the small fraction of the stellar population that actually succeeds in forming a PN, as initially suggested by \citet{moe2006} using population synthesis models.

Despite some of the debilitating characteristics of the {\it K2} survey for CS variability, we have increased the close binary fraction. Our lower limit values of 20--24$\pm5$ percent exceed previous ground-based  estimates (13 percent, \citet{bond00}, 12--21 percent, \citet{miszalski09}), demonstrating that the space-based sensitivity yields a net gain. Indeed, {\it K2} is sensitive to longer periods, with the period distribution of the {\it K2} sample being statistically distinct from the ground-based sample compiled from the literature.

However, we cannot yet be certain that the variables are truly binary stars based on the {\it K2} light curves alone. To be certain, it is essential to obtain follow-up observations, namely photometry and radial velocity variability measurements. In addition, now that the number of close binary CSs is large enough, it would be interesting to determine their PN morphologies via an imaging campaign to associate the common envelope interaction that must have taken place with the shapes of the resulting nebulae.

\section*{Acknowledgements}

This paper includes data collected by the {\it Kepler Telescope} missions and obtained from the MAST data archive at the Space Telescope Science Institute (STScI). Funding for the {\it Kepler} mission is provided by the NASA Science Mission Directorate. STScI is operated by the Association of Universities for Research in Astronomy, Inc., under NASA contract NAS 5–26555. OD and KM acknowledge Macquarie University's ASTR3810 Professional and Community Engagement undergraduate programme (PACE), as well as funding from Macquarie's research centre for Astronomy Astrophysics and Astrophotonics Vacation programme that enabled the participation of Kayla Martin.

This research made use of \textsc{lightkurve}, a Python package for {\it Kepler} and {\it TESS} data analysis \citep{lightkurve}. 

This research has made use of the HASH PN database at hashpn.space \citep{parker16}.  

The Pan-STARRS1 Surveys have been made possible through contributions of the Institute for Astronomy, the University of Hawaii, the Pan-STARRS Project Office, the Max-Planck Society and its participating institutes, the Max Planck Institute for Astronomy, Heidelberg and the Max Planck Institute for Extraterrestrial Physics, Garching, The Johns Hopkins University, Durham University, the University of Edinburgh, Queen's University Belfast, the Harvard-Smithsonian Center for Astrophysics, the Las Cumbres Observatory Global Telescope Network Incorporated, the National Central University of Taiwan, the Space Telescope Science Institute, the National Aeronautics and Space Administration under Grant No. NNX08AR22G issued through the Planetary Science Division of the NASA Science Mission Directorate, the National Science Foundation under Grant No. AST-1238877, the University of Maryland, and Eotvos Lorand University (ELTE). 

This research has made use of the SIMBAD database, operated at CDS, Strasbourg, France.

This research has made use of the NASA Exoplanet Archive, which is operated by the California Institute of Technology, under contract with the National Aeronautics and Space Administration under the Exoplanet Exploration Program.

GHJ acknowledges support from {\it Kepler/K2} NASA grants NNX16AE57G and NNX17AE64G and wishes to thank Carnegie Observatories for access to the Las Campanas Observatory 1~m Swope telescope.
GHJ thanks Dana Patchick for his assistance is defining the {\it K2} target lists, Christina Hedges, Benjamin Pope for their assistance with Python/\textsc{LightKurve} details, and Steve Howell for suggesting to go down this rabbit hole. John Tonry kindly provided ATLAS data for many of our targets in advance of public access. Di Harmer provided valuable assistance with our Mayall 4~m observations. 

DJ acknowledges support from the Erasmus+ programme of the European Union under grant number 2020-1-CZ01-KA203-078200. DJ also acknowledges support from the State Research Agency (AEI) of the Spanish Ministry of Science, Innovation and Universities (MCIU) and the European Regional Development Fund (FEDER) under grant AYA2017-83383-P.

We further thank Gerald Handler and Paulina Sowicka for valuable discussions on the  nature of the Th~3-14 and H~2-48 variability. We greatly appreciate the contribution of the anonymous referee for providing very helpful and detailed suggestions that significantly improved the presentation and content of the paper.

\section*{Data availability}

All key data and most other supporting data, are available from public archives. Some secondary data sets, which are not publicly archived, can be made available upon request.



\bibliographystyle{mnras}
\bibliography{K2_PN} 

\begin{thebibliography}{}
\makeatletter
\relax
\def\mn@urlcharsother{\let\do\@makeother \do\$\do\&\do\#\do\^\do\_\do\%\do\~}
\def\mn@doi{\begingroup\mn@urlcharsother \@ifnextchar [ {\mn@doi@}
  {\mn@doi@[]}}
\def\mn@doi@[#1]#2{\def\@tempa{#1}\ifx\@tempa\@empty \href
  {http://dx.doi.org/#2} {doi:#2}\else \href {http://dx.doi.org/#2} {#1}\fi
  \endgroup}
\def\mn@eprint#1#2{\mn@eprint@#1:#2::\@nil}
\def\mn@eprint@arXiv#1{\href {http://arxiv.org/abs/#1} {{\tt arXiv:#1}}}
\def\mn@eprint@dblp#1{\href {http://dblp.uni-trier.de/rec/bibtex/#1.xml}
  {dblp:#1}}
\def\mn@eprint@#1:#2:#3:#4\@nil{\def\@tempa {#1}\def\@tempb {#2}\def\@tempc
  {#3}\ifx \@tempc \@empty \let \@tempc \@tempb \let \@tempb \@tempa \fi \ifx
  \@tempb \@empty \def\@tempb {arXiv}\fi \@ifundefined
  {mn@eprint@\@tempb}{\@tempb:\@tempc}{\expandafter \expandafter \csname
  mn@eprint@\@tempb\endcsname \expandafter{\@tempc}}}

\bibitem[\protect\citeauthoryear{{Aigrain}, {Parviainen}  \& {Pope}}{{Aigrain}
  et~al.}{2016}]{aigrain16}
{Aigrain} S.,  {Parviainen} H.,   {Pope} B.~J.~S.,  2016, \mn@doi [\mnras]
  {10.1093/mnras/stw706}, \href
  {https://ui.adsabs.harvard.edu/abs/2016MNRAS.459.2408A} {459, 2408}

\bibitem[\protect\citeauthoryear{{Aller}, {Lillo-Box}, {Vu{\v{c}}kovi{\'c}},
  {Van Winckel}, {Jones}, {Montesinos}, {Zorotovic}  \& {Miranda}}{{Aller}
  et~al.}{2018}]{aller18}
{Aller} A.,  {Lillo-Box} J.,  {Vu{\v{c}}kovi{\'c}} M.,  {Van Winckel} H.,
  {Jones} D.,  {Montesinos} B.,  {Zorotovic} M.,   {Miranda} L.~F.,  2018,
  \mn@doi [\mnras] {10.1093/mnras/sty174}, \href
  {https://ui.adsabs.harvard.edu/abs/2018MNRAS.476.1140A} {476, 1140}

\bibitem[\protect\citeauthoryear{{Aller}, {Lillo-Box}, {Jones}, {Miranda}  \&
  {Barcel{\'o} Forteza}}{{Aller} et~al.}{2020}]{aller20}
{Aller} A.,  {Lillo-Box} J.,  {Jones} D.,  {Miranda} L.~F.,   {Barcel{\'o}
  Forteza} S.,  2020, \mn@doi [\aap] {10.1051/0004-6361/201937118}, \href
  {https://ui.adsabs.harvard.edu/abs/2020A&A...635A.128A} {635, A128}

\bibitem[\protect\citeauthoryear{{Barker}, {Zijlstra}, {De Marco}, {Frew},
  {Drew}, {Corradi}, {Eisl{\"o}ffel}  \& {Parker}}{{Barker}
  et~al.}{2018}]{Barker18}
{Barker} H.,  {Zijlstra} A.,  {De Marco} O.,  {Frew} D.~J.,  {Drew} J.~E.,
  {Corradi} R. L.~M.,  {Eisl{\"o}ffel} J.,   {Parker} Q.~A.,  2018, \mn@doi
  [\mnras] {10.1093/mnras/stx3240}, \href
  {https://ui.adsabs.harvard.edu/abs/2018MNRAS.475.4504B} {475, 4504}

\bibitem[\protect\citeauthoryear{{Boffin} \& {Jones}}{{Boffin} \&
  {Jones}}{2019}]{boffin19}
{Boffin} H. M.~J.,  {Jones} D.,  2019, {The Importance of Binaries in the
  Formation and Evolution of Planetary Nebulae}.
{Springer Nature}, \mn@doi{10.1007/978-3-030-25059-1}

\bibitem[\protect\citeauthoryear{{Bond}}{{Bond}}{2000}]{bond00}
{Bond} H.~E.,  2000, in {Kastner} J.~H.,  {Soker} N.,   {Rappaport} S.,  eds,
  Astronomical Society of the Pacific Conference Series Vol. 199, Asymmetrical
  Planetary Nebulae II: From Origins to Microstructures. p.~115 (\mn@eprint
  {arXiv} {astro-ph/9909516})

\bibitem[\protect\citeauthoryear{{Bowman}, {Burssens}, {Sim{\'o}n-D{\'\i}az},
  {Edelmann}, {Rogers}, {Horst}, {R{\"o}pke}  \& {Aerts}}{{Bowman}
  et~al.}{2020}]{bowman20}
{Bowman} D.~M.,  {Burssens} S.,  {Sim{\'o}n-D{\'\i}az} S.,  {Edelmann}
  P.~V.~F.,  {Rogers} T.~M.,  {Horst} L.,  {R{\"o}pke} F.~K.,   {Aerts} C.,
  2020, \mn@doi [\aap] {10.1051/0004-6361/202038224}, \href
  {https://ui.adsabs.harvard.edu/abs/2020A&A...640A..36B} {640, A36}

\bibitem[\protect\citeauthoryear{{Brown}, {Jones}, {Boffin}  \& {Van
  Winckel}}{{Brown} et~al.}{2019}]{brown19}
{Brown} A.~J.,  {Jones} D.,  {Boffin} H. M.~J.,   {Van Winckel} H.,  2019,
  \mn@doi [\mnras] {10.1093/mnras/sty2986}, \href
  {https://ui.adsabs.harvard.edu/abs/2019MNRAS.482.4951B} {482, 4951}

\bibitem[\protect\citeauthoryear{{Chornay}, {Walton}, {Jones}, {Boffin},
  {Rejkuba}  \& {Wesson}}{{Chornay} et~al.}{2021}]{chornay21}
{Chornay} N.,  {Walton} N.~A.,  {Jones} D.,  {Boffin} H.~M.~J.,  {Rejkuba} M.,
   {Wesson} R.,  2021, \mn@doi [\aap] {10.1051/0004-6361/202140288}, \href
  {https://ui.adsabs.harvard.edu/abs/2021A&A...648A..95C} {648, A95}

\bibitem[\protect\citeauthoryear{{Coccato} et~al.,}{{Coccato}
  et~al.}{2009}]{coccato09}
{Coccato} L.,  et~al., 2009, \mn@doi [\mnras]
  {10.1111/j.1365-2966.2009.14417.x}, \href
  {https://ui.adsabs.harvard.edu/abs/2009MNRAS.394.1249C} {394, 1249}

\bibitem[\protect\citeauthoryear{{Corradi}, {Garc{\'\i}a-Rojas}, {Jones}  \&
  {Rodr{\'\i}guez-Gil}}{{Corradi} et~al.}{2015a}]{corradi15}
{Corradi} R. L.~M.,  {Garc{\'\i}a-Rojas} J.,  {Jones} D.,
  {Rodr{\'\i}guez-Gil} P.,  2015a, \mn@doi [\apj] {10.1088/0004-637X/803/2/99},
  \href {https://ui.adsabs.harvard.edu/abs/2015ApJ...803...99C} {803, 99}

\bibitem[\protect\citeauthoryear{{Corradi}, {Kwitter}, {Balick}, {Henry}  \&
  {Hensley}}{{Corradi} et~al.}{2015b}]{corradi15b}
{Corradi} R.~L.~M.,  {Kwitter} K.~B.,  {Balick} B.,  {Henry} R.~B.~C.,
  {Hensley} K.,  2015b, \mn@doi [\apj] {10.1088/0004-637X/807/2/181}, \href
  {https://ui.adsabs.harvard.edu/abs/2015ApJ...807..181C} {807, 181}

\bibitem[\protect\citeauthoryear{{Crowther}, {De Marco}  \&
  {Barlow}}{{Crowther} et~al.}{1998}]{Crowther1998}
{Crowther} P.~A.,  {De Marco} O.,   {Barlow} M.~J.,  1998, \mn@doi [\mnras]
  {10.1046/j.1365-8711.1998.01360.x}, \href
  {https://ui.adsabs.harvard.edu/abs/1998MNRAS.296..367C} {296, 367}

\bibitem[\protect\citeauthoryear{{Dan}, {Rosswog}, {Br{\"u}ggen}  \&
  {Podsiadlowski}}{{Dan} et~al.}{2014}]{Dan2014}
{Dan} M.,  {Rosswog} S.,  {Br{\"u}ggen} M.,   {Podsiadlowski} P.,  2014,
  \mn@doi [\mnras] {10.1093/mnras/stt1766}, \href
  {https://ui.adsabs.harvard.edu/abs/2014MNRAS.438...14D} {438, 14}

\bibitem[\protect\citeauthoryear{{Davis}, {Ciardullo}, {Jacoby}, {Feldmeier}
  \& {Indahl}}{{Davis} et~al.}{2018}]{davis18}
{Davis} B.~D.,  {Ciardullo} R.,  {Jacoby} G.~H.,  {Feldmeier} J.~J.,   {Indahl}
  B.~L.,  2018, \mn@doi [\apj] {10.3847/1538-4357/aad3c4}, \href
  {https://ui.adsabs.harvard.edu/abs/2018ApJ...863..189D} {863, 189}

\bibitem[\protect\citeauthoryear{{De~Marco}}{{De~Marco}}{2009}]{demarco09}
{De~Marco} O.,  2009, \mn@doi [\pasp] {10.1086/597765}, \href
  {https://ui.adsabs.harvard.edu/abs/2009PASP..121..316D} {121, 316}

\bibitem[\protect\citeauthoryear{{De Marco}, {Clayton}, {Herwig}, {Pollacco},
  {Clark}  \& {Kilkenny}}{{De Marco} et~al.}{2002}]{DeMarco2002}
{De Marco} O.,  {Clayton} G.~C.,  {Herwig} F.,  {Pollacco} D.~L.,  {Clark}
  J.~S.,   {Kilkenny} D.,  2002, \mn@doi [\aj] {10.1086/340569}, \href
  {https://ui.adsabs.harvard.edu/abs/2002AJ....123.3387D} {123, 3387}

\bibitem[\protect\citeauthoryear{{De Marco}, {Hillwig}  \& {Smith}}{{De Marco}
  et~al.}{2008}]{demarco08}
{De Marco} O.,  {Hillwig} T.~C.,   {Smith} A.~J.,  2008, \mn@doi [\aj]
  {10.1088/0004-6256/136/1/323}, \href
  {https://ui.adsabs.harvard.edu/abs/2008AJ....136..323D} {136, 323}

\bibitem[\protect\citeauthoryear{{De Marco}, {Passy}, {Frew}, {Moe}  \&
  {Jacoby}}{{De Marco} et~al.}{2013}]{DeMarco13}
{De Marco} O.,  {Passy} J.-C.,  {Frew} D.~J.,  {Moe} M.,   {Jacoby} G.~H.,
  2013, \mn@doi [\mnras] {10.1093/mnras/sts180}, \href
  {https://ui.adsabs.harvard.edu/abs/2013MNRAS.428.2118D} {428, 2118}

\bibitem[\protect\citeauthoryear{{De Marco}, {Long}, {Jacoby}, {Hillwig},
  {Kronberger}, {Howell}, {Reindl}  \& {Margheim}}{{De Marco}
  et~al.}{2015}]{demarco15}
{De Marco} O.,  {Long} J.,  {Jacoby} G.~H.,  {Hillwig} T.,  {Kronberger} M.,
  {Howell} S.~B.,  {Reindl} N.,   {Margheim} S.,  2015, \mn@doi [\mnras]
  {10.1093/mnras/stv249}, \href
  {https://ui.adsabs.harvard.edu/abs/2015MNRAS.448.3587D} {448, 3587}

\bibitem[\protect\citeauthoryear{{Decin} et~al.,}{{Decin}
  et~al.}{2020}]{decin20}
{Decin} L.,  et~al., 2020, \mn@doi [Science] {10.1126/science.abb1229}, \href
  {https://ui.adsabs.harvard.edu/abs/2020Sci...369.1497D} {369, 1497}

\bibitem[\protect\citeauthoryear{{Douchin} et~al.,}{{Douchin}
  et~al.}{2015}]{Douchin15}
{Douchin} D.,  et~al., 2015, \mn@doi [\mnras] {10.1093/mnras/stu2700}, \href
  {https://ui.adsabs.harvard.edu/abs/2015MNRAS.448.3132D} {448, 3132}

\bibitem[\protect\citeauthoryear{{Fabricius} et~al.,}{{Fabricius}
  et~al.}{2021}]{gaia_eDR3}
{Fabricius} C.,  et~al., 2021, \mn@doi [\aap] {10.1051/0004-6361/202039834},
  \href {https://ui.adsabs.harvard.edu/abs/2021A&A...649A...5F} {649, A5}

\bibitem[\protect\citeauthoryear{{Frew} et~al.,}{{Frew}
  et~al.}{2011}]{frew2011}
{Frew} D.~J.,  et~al., 2011, \mn@doi [\pasa] {10.1071/AS10017}, \href
  {https://ui.adsabs.harvard.edu/abs/2011PASA...28...83F} {28, 83}

\bibitem[\protect\citeauthoryear{{Gilliland} et~al.,}{{Gilliland}
  et~al.}{2011}]{gilliland2011}
{Gilliland} R.~L.,  et~al., 2011, \mn@doi [\apjs] {10.1088/0067-0049/197/1/6},
  \href {https://ui.adsabs.harvard.edu/abs/2011ApJS..197....6G} {197, 6}

\bibitem[\protect\citeauthoryear{{Gleizes}, {Acker}  \& {Stenholm}}{{Gleizes}
  et~al.}{1989}]{gleizes89}
{Gleizes} F.,  {Acker} A.,   {Stenholm} B.,  1989, \aap, \href
  {https://ui.adsabs.harvard.edu/abs/1989A&A...222..237G} {222, 237}

\bibitem[\protect\citeauthoryear{{Hajduk}, {Zijlstra}  \& {Gesicki}}{{Hajduk}
  et~al.}{2010}]{hajduk10}
{Hajduk} M.,  {Zijlstra} A.~A.,   {Gesicki} K.,  2010, \mn@doi [\mnras]
  {10.1111/j.1365-2966.2010.16704.x}, \href
  {https://ui.adsabs.harvard.edu/abs/2010MNRAS.406..626H} {406, 626}

\bibitem[\protect\citeauthoryear{{Heck}, {Houziaux}, {Manfroid}, {Jones}  \&
  {Andrews}}{{Heck} et~al.}{1985}]{heck1985}
{Heck} A.,  {Houziaux} L.,  {Manfroid} J.,  {Jones} D.~H.~P.,   {Andrews}
  P.~J.,  1985, \aaps, \href
  {https://ui.adsabs.harvard.edu/abs/1985A&AS...61..375H} {61, 375}

\bibitem[\protect\citeauthoryear{{Heinze} et~al.,}{{Heinze}
  et~al.}{2018}]{heinze18}
{Heinze} A.~N.,  et~al., 2018, \mn@doi [\aj] {10.3847/1538-3881/aae47f}, \href
  {https://ui.adsabs.harvard.edu/abs/2018AJ....156..241H} {156, 241}

\bibitem[\protect\citeauthoryear{{Henderson} et~al.,}{{Henderson}
  et~al.}{2016}]{henderson2016}
{Henderson} C.~B.,  et~al., 2016, \mn@doi [\pasp]
  {10.1088/1538-3873/128/970/124401}, \href
  {https://ui.adsabs.harvard.edu/abs/2016PASP..128l4401H} {128, 124401}

\bibitem[\protect\citeauthoryear{{Hermes} et~al.,}{{Hermes}
  et~al.}{2015}]{hermes2015}
{Hermes} J.~J.,  et~al., 2015, \mn@doi [\apjl] {10.1088/2041-8205/810/1/L5},
  \href {https://ui.adsabs.harvard.edu/abs/2015ApJ...810L...5H} {810, L5}

\bibitem[\protect\citeauthoryear{{Hillwig}, {Bond}, {Af{\textcommabelow s}ar}
  \& {De Marco}}{{Hillwig} et~al.}{2010}]{hillwig10}
{Hillwig} T.~C.,  {Bond} H.~E.,  {Af{\textcommabelow s}ar} M.,   {De Marco} O.,
   2010, \mn@doi [\aj] {10.1088/0004-6256/140/2/319}, \href
  {https://ui.adsabs.harvard.edu/abs/2010AJ....140..319H} {140, 319}

\bibitem[\protect\citeauthoryear{{Hillwig}, {Jones}, {De Marco}, {Bond},
  {Margheim}  \& {Frew}}{{Hillwig} et~al.}{2016}]{hillwig16}
{Hillwig} T.~C.,  {Jones} D.,  {De Marco} O.,  {Bond} H.~E.,  {Margheim} S.,
  {Frew} D.,  2016, \mn@doi [\apj] {10.3847/0004-637X/832/2/125}, \href
  {https://ui.adsabs.harvard.edu/abs/2016ApJ...832..125H} {832, 125}

\bibitem[\protect\citeauthoryear{{Hlabathe}}{{Hlabathe}}{2015}]{hlabathe15}
{Hlabathe} M.,  2015, Master's thesis, University of Cape Town

\bibitem[\protect\citeauthoryear{{Iaconi} \& {De Marco}}{{Iaconi} \& {De
  Marco}}{2019}]{iaconi19}
{Iaconi} R.,  {De Marco} O.,  2019, \mn@doi [\mnras] {10.1093/mnras/stz2756},
  \href {https://ui.adsabs.harvard.edu/abs/2019MNRAS.490.2550I} {490, 2550}

\bibitem[\protect\citeauthoryear{{Jacoby}}{{Jacoby}}{1989}]{jacoby89}
{Jacoby} G.~H.,  1989, \mn@doi [\apj] {10.1086/167274}, \href
  {https://ui.adsabs.harvard.edu/abs/1989ApJ...339...39J} {339, 39}

\bibitem[\protect\citeauthoryear{{Jacoby} \& {Van de Steene}}{{Jacoby} \& {Van
  de Steene}}{2004}]{jacoby2004}
{Jacoby} G.~H.,  {Van de Steene} G.,  2004, \mn@doi [\aap]
  {10.1051/0004-6361:20035809}, \href
  {https://ui.adsabs.harvard.edu/abs/2004A&A...419..563J} {419, 563}

\bibitem[\protect\citeauthoryear{{Jacoby} et~al.,}{{Jacoby}
  et~al.}{2010}]{jacoby2010}
{Jacoby} G.~H.,  et~al., 2010, \mn@doi [\pasa] {10.1071/AS09025}, \href
  {https://ui.adsabs.harvard.edu/abs/2010PASA...27..156J} {27, 156}

\bibitem[\protect\citeauthoryear{{Jacoby}, {De Marco}, {Davies}, {Lotarevich},
  {Bond}, {Harrington}  \& {Lanz}}{{Jacoby} et~al.}{2017}]{jacoby17}
{Jacoby} G.~H.,  {De Marco} O.,  {Davies} J.,  {Lotarevich} I.,  {Bond} H.~E.,
  {Harrington} J.~P.,   {Lanz} T.,  2017, \mn@doi [\apj]
  {10.3847/1538-4357/836/1/93}, \href
  {https://ui.adsabs.harvard.edu/abs/2017ApJ...836...93J} {836, 93}

\bibitem[\protect\citeauthoryear{{Jacoby}, {Hillwig}  \& {Jones}}{{Jacoby}
  et~al.}{2020}]{jacoby2020}
{Jacoby} G.~H.,  {Hillwig} T.~C.,   {Jones} D.,  2020, \mn@doi [\mnras]
  {10.1093/mnrasl/slaa138}, \href
  {https://ui.adsabs.harvard.edu/abs/2020MNRAS.498L.114J} {498, L114}

\bibitem[\protect\citeauthoryear{{Jones}}{{Jones}}{2020}]{jones20}
{Jones} D.,  2020, {Observational Constraints on the Common Envelope Phase}.
Springer, pp 123--153, \mn@doi{10.1007/978-3-030-38509-5_5}

\bibitem[\protect\citeauthoryear{{Jones} \& {Boffin}}{{Jones} \&
  {Boffin}}{2017a}]{jones17}
{Jones} D.,  {Boffin} H. M.~J.,  2017a, \mn@doi [Nature Astronomy]
  {10.1038/s41550-017-0117}, \href
  {https://ui.adsabs.harvard.edu/abs/2017NatAs...1E.117J} {1, 0117}

\bibitem[\protect\citeauthoryear{{Jones} \& {Boffin}}{{Jones} \&
  {Boffin}}{2017b}]{jones17b}
{Jones} D.,  {Boffin} H. M.~J.,  2017b, \mn@doi [\mnras]
  {10.1093/mnras/stw3191}, \href
  {https://ui.adsabs.harvard.edu/abs/2017MNRAS.466.2034J} {466, 2034}

\bibitem[\protect\citeauthoryear{{Jones}, {Van Winckel}, {Aller}, {Exter}  \&
  {De Marco}}{{Jones} et~al.}{2017}]{jones2017}
{Jones} D.,  {Van Winckel} H.,  {Aller} A.,  {Exter} K.,   {De Marco} O.,
  2017, \mn@doi [\aap] {10.1051/0004-6361/201730700}, \href
  {https://ui.adsabs.harvard.edu/abs/2017A&A...600L...9J} {600, L9}

\bibitem[\protect\citeauthoryear{{Jones}, {Boffin}, {Sowicka}, {Miszalski},
  {Rodr{\'\i}guez-Gil}, {Santander-Garc{\'\i}a}  \& {Corradi}}{{Jones}
  et~al.}{2019}]{jones19}
{Jones} D.,  {Boffin} H. M.~J.,  {Sowicka} P.,  {Miszalski} B.,
  {Rodr{\'\i}guez-Gil} P.,  {Santander-Garc{\'\i}a} M.,   {Corradi} R. L.~M.,
  2019, \mn@doi [\mnras] {10.1093/mnrasl/sly142}, \href
  {https://ui.adsabs.harvard.edu/abs/2019MNRAS.482L..75J} {482, L75}

\bibitem[\protect\citeauthoryear{{Keel} et~al.,}{{Keel} et~al.}{2017}]{keel17}
{Keel} W.~C.,  et~al., 2017, \mn@doi [\pasp]
  {10.1088/1538-3873/129/971/015002}, \href
  {https://ui.adsabs.harvard.edu/abs/2017PASP..129a5002K} {129, 015002}

\bibitem[\protect\citeauthoryear{{Kronberger} et~al.,}{{Kronberger}
  et~al.}{2016}]{kronberger2016}
{Kronberger} M.,  et~al., 2016, in Journal of Physics Conference Series. p.
  072012, \mn@doi{10.1088/1742-6596/728/7/072012}

\bibitem[\protect\citeauthoryear{{Lenz} \& {Breger}}{{Lenz} \&
  {Breger}}{2005}]{lenz05}
{Lenz} P.,  {Breger} M.,  2005, \mn@doi [Communications in Asteroseismology]
  {10.1553/cia146s53}, \href
  {https://ui.adsabs.harvard.edu/abs/2005CoAst.146...53L} {146, 53}

\bibitem[\protect\citeauthoryear{{Lightkurve Collaboration}
  et~al.,}{{Lightkurve Collaboration} et~al.}{2018}]{lightkurve}
{Lightkurve Collaboration} et~al., 2018, {Lightkurve: Kepler and TESS time
  series analysis in Python}, Astrophysics Source Code Library (\mn@eprint
  {ascl} {1812.013})

\bibitem[\protect\citeauthoryear{{Luger}, {Agol}, {Kruse}, {Barnes}, {Becker},
  {Foreman-Mackey}  \& {Deming}}{{Luger} et~al.}{2016}]{luger2016}
{Luger} R.,  {Agol} E.,  {Kruse} E.,  {Barnes} R.,  {Becker} A.,
  {Foreman-Mackey} D.,   {Deming} D.,  2016, \mn@doi [\aj]
  {10.3847/0004-6256/152/4/100}, \href
  {https://ui.adsabs.harvard.edu/abs/2016AJ....152..100L} {152, 100}

\bibitem[\protect\citeauthoryear{{Madappatt}, {De Marco}  \&
  {Villaver}}{{Madappatt} et~al.}{2016}]{madappatt16}
{Madappatt} N.,  {De Marco} O.,   {Villaver} E.,  2016, \mn@doi [\mnras]
  {10.1093/mnras/stw2025}, \href
  {https://ui.adsabs.harvard.edu/abs/2016MNRAS.463.1040M} {463, 1040}

\bibitem[\protect\citeauthoryear{{Manick}, {Miszalski}  \& {McBride}}{{Manick}
  et~al.}{2015}]{manick15}
{Manick} R.,  {Miszalski} B.,   {McBride} V.,  2015, \mn@doi [\mnras]
  {10.1093/mnras/stv074}, \href
  {https://ui.adsabs.harvard.edu/abs/2015MNRAS.448.1789M} {448, 1789}

\bibitem[\protect\citeauthoryear{{Martini} et~al.,}{{Martini}
  et~al.}{2014}]{martini2014}
{Martini} P.,  et~al., 2014, in {Ramsay} S.~K.,  {McLean} I.~S.,   {Takami} H.,
   eds,  Society of Photo-Optical Instrumentation Engineers (SPIE) Conference
  Series Vol. 9147, Ground-based and Airborne Instrumentation for Astronomy V.
  p. 91470Z (\mn@eprint {arXiv} {1407.4541}), \mn@doi{10.1117/12.2056834}

\bibitem[\protect\citeauthoryear{{Minniti}, {Dias}, {G{\'o}mez}, {Palma}  \&
  {Pullen}}{{Minniti} et~al.}{2019}]{minniti19}
{Minniti} D.,  {Dias} B.,  {G{\'o}mez} M.,  {Palma} T.,   {Pullen} J.~B.,
  2019, \mn@doi [\apjl] {10.3847/2041-8213/ab4424}, \href
  {https://ui.adsabs.harvard.edu/abs/2019ApJ...884L..15M} {884, L15}

\bibitem[\protect\citeauthoryear{{Miszalski}, {Acker}, {Moffat}, {Parker}  \&
  {Udalski}}{{Miszalski} et~al.}{2009a}]{miszalski09}
{Miszalski} B.,  {Acker} A.,  {Moffat} A.~F.~J.,  {Parker} Q.~A.,   {Udalski}
  A.,  2009a, \mn@doi [\aap] {10.1051/0004-6361/200811380}, \href
  {https://ui.adsabs.harvard.edu/abs/2009A&A...496..813M} {496, 813}

\bibitem[\protect\citeauthoryear{{Miszalski}, {Acker}, {Parker}  \&
  {Moffat}}{{Miszalski} et~al.}{2009b}]{miszalski09b}
{Miszalski} B.,  {Acker} A.,  {Parker} Q.~A.,   {Moffat} A.~F.~J.,  2009b,
  \mn@doi [\aap] {10.1051/0004-6361/200912176}, \href
  {https://ui.adsabs.harvard.edu/abs/2009A&A...505..249M} {505, 249}

\bibitem[\protect\citeauthoryear{{Miszalski}, {Manick}, {Miko{\l}ajewska}, {Van
  Winckel}  \& {I{\l}kiewicz}}{{Miszalski} et~al.}{2018}]{miszalski18}
{Miszalski} B.,  {Manick} R.,  {Miko{\l}ajewska} J.,  {Van Winckel} H.,
  {I{\l}kiewicz} K.,  2018, \mn@doi [\pasa] {10.1017/pasa.2018.23}, \href
  {https://ui.adsabs.harvard.edu/abs/2018PASA...35...27M} {35, e027}

\bibitem[\protect\citeauthoryear{{Moe} \& {De Marco}}{{Moe} \& {De
  Marco}}{2006}]{moe2006}
{Moe} M.,  {De Marco} O.,  2006, \mn@doi [\apj] {10.1086/506900}, \href
  {https://ui.adsabs.harvard.edu/abs/2006ApJ...650..916M} {650, 916}

\bibitem[\protect\citeauthoryear{{Otsuka} et~al.,}{{Otsuka}
  et~al.}{2017}]{otsuka17}
{Otsuka} M.,  et~al., 2017, \mn@doi [\apjs] {10.3847/1538-4365/aa8175}, \href
  {https://ui.adsabs.harvard.edu/abs/2017ApJS..231...22O} {231, 22}

\bibitem[\protect\citeauthoryear{{Paczynski}}{{Paczynski}}{1985}]{paczynski85}
{Paczynski} B.,  1985, {Evolution of Cataclysmic Binaries}.
Springer, p.~1, \mn@doi{10.1007/978-94-009-5319-2_1}

\bibitem[\protect\citeauthoryear{{Parker} et~al.,}{{Parker}
  et~al.}{2006}]{parker06}
{Parker} Q.~A.,  et~al., 2006, \mn@doi [\mnras]
  {10.1111/j.1365-2966.2006.10950.x}, \href
  {https://ui.adsabs.harvard.edu/abs/2006MNRAS.373...79P} {373, 79}

\bibitem[\protect\citeauthoryear{{Parker}, {Boji{\v{c}}i{\'c}}  \&
  {Frew}}{{Parker} et~al.}{2016}]{parker16}
{Parker} Q.~A.,  {Boji{\v{c}}i{\'c}} I.~S.,   {Frew} D.~J.,  2016, in Journal
  of Physics Conference Series. p. 032008 (\mn@eprint {arXiv} {1603.07042}),
  \mn@doi{10.1088/1742-6596/728/3/032008}

\bibitem[\protect\citeauthoryear{{Poleski}, {Penny}, {Gaudi}, {Udalski},
  {Ranc}, {Barentsen}  \& {Gould}}{{Poleski} et~al.}{2019}]{poleski2019}
{Poleski} R.,  {Penny} M.,  {Gaudi} B.~S.,  {Udalski} A.,  {Ranc} C.,
  {Barentsen} G.,   {Gould} A.,  2019, \mn@doi [\aap]
  {10.1051/0004-6361/201834544}, \href
  {https://ui.adsabs.harvard.edu/abs/2019A&A...627A..54P} {627, A54}

\bibitem[\protect\citeauthoryear{{Prinja}, {Massa}, {Urbaneja}  \&
  {Kudritzki}}{{Prinja} et~al.}{2012a}]{prinja12a}
{Prinja} R.~K.,  {Massa} D.~L.,  {Urbaneja} M.~A.,   {Kudritzki} R.~P.,  2012a,
  \mn@doi [\mnras] {10.1111/j.1365-2966.2012.20838.x}, \href
  {https://ui.adsabs.harvard.edu/abs/2012MNRAS.422.3142P} {422, 3142}

\bibitem[\protect\citeauthoryear{{Prinja}, {Massa}  \& {Cantiello}}{{Prinja}
  et~al.}{2012b}]{prinja2012}
{Prinja} R.~K.,  {Massa} D.~L.,   {Cantiello} M.,  2012b, \mn@doi [\apjl]
  {10.1088/2041-8205/759/2/L28}, \href
  {https://ui.adsabs.harvard.edu/abs/2012ApJ...759L..28P} {759, L28}

\bibitem[\protect\citeauthoryear{{Richer}, {Su{\'a}rez}, {L{\'o}pez}  \&
  {Garc{\'\i}a D{\'\i}az}}{{Richer} et~al.}{2017}]{richer2017}
{Richer} M.~G.,  {Su{\'a}rez} G.,  {L{\'o}pez} J.~A.,   {Garc{\'\i}a D{\'\i}az}
  M.~T.,  2017, \mn@doi [\aj] {10.3847/1538-3881/aa5f53}, \href
  {https://ui.adsabs.harvard.edu/abs/2017AJ....153..140R} {153, 140}

\bibitem[\protect\citeauthoryear{{Sahai}, {Morris}  \& {Villar}}{{Sahai}
  et~al.}{2011}]{sahai11}
{Sahai} R.,  {Morris} M.~R.,   {Villar} G.~G.,  2011, \mn@doi [\aj]
  {10.1088/0004-6256/141/4/134}, \href
  {https://ui.adsabs.harvard.edu/abs/2011AJ....141..134S} {141, 134}

\bibitem[\protect\citeauthoryear{{Santander-Garc{\'\i}a}, {Rodr{\'\i}guez-Gil},
  {Corradi}, {Jones}, {Miszalski}, {Boffin}, {Rubio-D{\'\i}ez}  \&
  {Kotze}}{{Santander-Garc{\'\i}a} et~al.}{2015}]{santander-garcia15}
{Santander-Garc{\'\i}a} M.,  {Rodr{\'\i}guez-Gil} P.,  {Corradi} R.~L.~M.,
  {Jones} D.,  {Miszalski} B.,  {Boffin} H.~M.~J.,  {Rubio-D{\'\i}ez} M.~M.,
  {Kotze} M.~M.,  2015, \mn@doi [\nat] {10.1038/nature14124}, \href
  {https://ui.adsabs.harvard.edu/abs/2015Natur.519...63S} {519, 63}

\bibitem[\protect\citeauthoryear{{Soker}}{{Soker}}{1997}]{soker97}
{Soker} N.,  1997, \mn@doi [\apjs] {10.1086/313040}, \href
  {https://ui.adsabs.harvard.edu/abs/1997ApJS..112..487S} {112, 487}

\bibitem[\protect\citeauthoryear{{Soszy{\'n}ski} et~al.,}{{Soszy{\'n}ski}
  et~al.}{2014}]{soszynski14}
{Soszy{\'n}ski} I.,  et~al., 2014, \actaa, \href
  {https://ui.adsabs.harvard.edu/abs/2014AcA....64..177S} {64, 177}

\bibitem[\protect\citeauthoryear{{Soszy{\'n}ski} et~al.,}{{Soszy{\'n}ski}
  et~al.}{2015}]{soszynski15}
{Soszy{\'n}ski} I.,  et~al., 2015, \actaa, \href
  {https://ui.adsabs.harvard.edu/abs/2015AcA....65...39S} {65, 39}

\bibitem[\protect\citeauthoryear{{Tappert}, {Schmidtobreick}, {Vogt}  \&
  {Ederoclite}}{{Tappert} et~al.}{2013}]{Tappert2013}
{Tappert} C.,  {Schmidtobreick} L.,  {Vogt} N.,   {Ederoclite} A.,  2013,
  \mn@doi [\mnras] {10.1093/mnras/stt1747}, \href
  {https://ui.adsabs.harvard.edu/abs/2013MNRAS.436.2412T} {436, 2412}

\bibitem[\protect\citeauthoryear{{Tonry} et~al.,}{{Tonry}
  et~al.}{2018}]{tonry18}
{Tonry} J.~L.,  et~al., 2018, \mn@doi [\apj] {10.3847/1538-4357/aae386}, \href
  {https://ui.adsabs.harvard.edu/abs/2018ApJ...867..105T} {867, 105}

\bibitem[\protect\citeauthoryear{{Toonen} \& {Nelemans}}{{Toonen} \&
  {Nelemans}}{2013}]{Toonen2013}
{Toonen} S.,  {Nelemans} G.,  2013, \mn@doi [\aap]
  {10.1051/0004-6361/201321753}, \href
  {https://ui.adsabs.harvard.edu/abs/2013A&A...557A..87T} {557, A87}

\bibitem[\protect\citeauthoryear{{Tyndall} et~al.,}{{Tyndall}
  et~al.}{2013}]{tyndall13}
{Tyndall} A.~A.,  et~al., 2013, \mn@doi [\mnras] {10.1093/mnras/stt1713}, \href
  {https://ui.adsabs.harvard.edu/abs/2013MNRAS.436.2082T} {436, 2082}

\bibitem[\protect\citeauthoryear{{Van Cleve} \& {Caldwell}}{{Van Cleve} \&
  {Caldwell}}{2016}]{vancleve2016}
{Van Cleve} J.~E.,  {Caldwell} D.~A.,  2016, {Kepler Instrument Handbook},
  Kepler Science Document KSCI-19033-002

\bibitem[\protect\citeauthoryear{{VanderPlas}}{{VanderPlas}}{2018}]{vanderplas18}
{VanderPlas} J.~T.,  2018, \mn@doi [\apjs] {10.3847/1538-4365/aab766}, \href
  {https://ui.adsabs.harvard.edu/abs/2018ApJS..236...16V} {236, 16}

\bibitem[\protect\citeauthoryear{{Vanderburg} \& {Johnson}}{{Vanderburg} \&
  {Johnson}}{2014}]{vanderburg2014}
{Vanderburg} A.,  {Johnson} J.~A.,  2014, \mn@doi [\pasp] {10.1086/678764},
  \href {https://ui.adsabs.harvard.edu/abs/2014PASP..126..948V} {126, 948}

\bibitem[\protect\citeauthoryear{{Wenger} et~al.,}{{Wenger}
  et~al.}{2000}]{wenger2000}
{Wenger} M.,  et~al., 2000, \mn@doi [\aaps] {10.1051/aas:2000332}, \href
  {https://ui.adsabs.harvard.edu/abs/2000A&AS..143....9W} {143, 9}

\bibitem[\protect\citeauthoryear{{Wesson}, {Liu}  \& {Barlow}}{{Wesson}
  et~al.}{2003}]{wesson03}
{Wesson} R.,  {Liu} X.~W.,   {Barlow} M.~J.,  2003, \mn@doi [\mnras]
  {10.1046/j.1365-8711.2003.06289.x}, \href
  {https://ui.adsabs.harvard.edu/abs/2003MNRAS.340..253W} {340, 253}

\bibitem[\protect\citeauthoryear{{Wesson}, {Jones}, {Garc{\'\i}a-Rojas},
  {Boffin}  \& {Corradi}}{{Wesson} et~al.}{2018}]{wesson18}
{Wesson} R.,  {Jones} D.,  {Garc{\'\i}a-Rojas} J.,  {Boffin} H.~M.~J.,
  {Corradi} R.~L.~M.,  2018, \mn@doi [\mnras] {10.1093/mnras/sty1871}, \href
  {https://ui.adsabs.harvard.edu/abs/2018MNRAS.480.4589W} {480, 4589}

\bibitem[\protect\citeauthoryear{{Yungelson}, {Tutukov}  \&
  {Livio}}{{Yungelson} et~al.}{1993}]{Yungelson1993}
{Yungelson} L.~R.,  {Tutukov} A.~V.,   {Livio} M.,  1993, \mn@doi [\apj]
  {10.1086/173436}, \href
  {https://ui.adsabs.harvard.edu/abs/1993ApJ...418..794Y} {418, 794}

\makeatother
\end{thebibliography}


\onecolumn

\appendix

\section{Table of Non-Variable Targets}

\begin{deluxetable}{clcccc}
\tablecaption{Non-Variable {\it K2} PNe
\label{tab:nonvar-PN-list}} 
\tablewidth{0.75\linewidth}

    \tablehead{\colhead{Campaign} & \colhead{PN Name} & \colhead{{\it K2}} & \colhead{RA (2000)} & \colhead{Dec (2000)} & \colhead{HASH} \\
\colhead{} & \colhead{}  & \colhead{KIC} & \colhead{} & \colhead{} & \colhead{Status}}
\startdata
0 &  J 900  & 202060054 & 96$\fdg$4885 & +17$\fdg$7910 &  True \\
0 &  K 3-71  & 202065169 & 93$\fdg$4791 & +26$\fdg$8825 &  True \\
0 &  HoCr 1  & 202065170 & 95$\fdg$4208 & +23$\fdg$5869 &  True \\
\\
2 &  MPA J1640-2757  & 202899888 & 250$\fdg$0441 & -27$\fdg$9594 &  Likely \\
2 &  KnPa 2  & 205465931 & 244$\fdg$6546 & -17$\fdg$4043 &  True \\
2 &  Pa 43  & 210282533 & 247$\fdg$1025 & -20$\fdg$3856 &  True \\
\\
7 &  M 3-33  & 214857824 & 282$\fdg$0505 & -25$\fdg$4812 &  True \\
7 &  M 3-32  & 214910524 & 281$\fdg$1797 & -25$\fdg$3594 &  True \\
7 &  Kn J1857$\fdg$7-2349  & 215636537 & 284$\fdg$4433 & -23.8276 &  Possible\\
7 &  V 348 Sgr  & 216129500 & 280$\fdg$0830 & -22$\fdg$9081 &  Likely \\
7 &  Kn J1838$\fdg$0-2202  & 216606732 & 279$\fdg$5069 & -22.0469 &  Possible\\
7 &  Abell 51  & 218761322 & 285$\fdg$2558 & -18$\fdg$2042 &  True \\
7 &  SB 21  & 229228185 & 282$\fdg$0457 & -18$\fdg$4939 &  True \\
7 &  SB 20  & 229228214 & 282$\fdg$3513 & -19$\fdg$8711 &  True \\
7 &  PHR J1852-2749  & 229228269 & 283$\fdg$2125 & -27$\fdg$8192 &  True \\
\\
11 &  MPA J1748-2511    & 223903889 & 267$\fdg$1735 & -25$\fdg$1926 &   True  \\
11 &  Th 3-27   & 224338141 & 263$\fdg$9934 & -24$\fdg$4248 &   True  \\
11 &  NGC 6369   & 224716713 & 262$\fdg$3352 & -23$\fdg$7597 &   True  \\
11 &  IRAS 17292-1704   & 227817039 & 263$\fdg$0244 & -17$\fdg$1143 &   True  \\
11 &  PM 2-22   & 231201639 & 257$\fdg$9122 & -24$\fdg$1258 &   Poss-Pre  \\
11 &  M 4-3   & 232133129 & 257$\fdg$6741 & -27$\fdg$1456 &   True  \\
11 &  PM 1-130   & 233281006 & 257$\fdg$9206 & -30$\fdg$4804 &   Likely  \\
11 &  MaC 1-4   & 233571636 & 261$\fdg$6587 & -16$\fdg$8083 &   True  \\
11 &  M 2-12   & 234669191 & 261$\fdg$0061 & -25$\fdg$9898 &   True  \\
11 &  BMP J1716-2313   & 235312774 & 259$\fdg$0869 & -23$\fdg$2323 &   Likely  \\
11 &  M 3-41   & 235785996 & 261$\fdg$4992 & -29$\fdg$3639 &   True  \\
11 &  M 3-38   & 235887408 & 260$\fdg$2686 & -29$\fdg$0499 &   True  \\
11 &  H 2-10   & 236056035 & 261$\fdg$8869 & -28$\fdg$5186 &   True  \\
11 &  M 3-10   & 236072455 & 261$\fdg$8341 & -28$\fdg$4642 &   Likely  \\
11 &  PHR J1724-2716   & 236426765 & 261$\fdg$2268 & -27$\fdg$2741 &   Likely  \\
11 &  IRAS 17195-2710   & 236439999 & 260$\fdg$6819 & -27$\fdg$2269 &   Poss-Pre  \\
11 &  PM 1-145   & 236703809 & 261$\fdg$2368 & -26$\fdg$2493 &   Poss-Pre  \\
11 &  K 5-28   & 237747285 & 259$\fdg$0594 & -30$\fdg$9268 &   True  \\
11 &  Th 3-13   & 237830758 & 261$\fdg$3307 & -30$\fdg$6783 &   True  \\
11 &  H 1-9   & 237942517 & 260$\fdg$3830 & -30$\fdg$3469 &   True  \\
11 &  Terz N 140   & 237944026 & 258$\fdg$7606 & -30$\fdg$3424 &   True  \\
11 &  G 4$\fdg$4+6$\fdg$4   & 238779606 & 262.9695 & -21.8252 &   Likely  \\
11 &  M 3-11   & 238941350 & 263$\fdg$8393 & -20$\fdg$9565 &   True  \\
11 &  PHR J1728-1844   & 239347893 & 262$\fdg$0585 & -18$\fdg$7420 &   True  \\
11 &  PHR J1743-2538   & 239546272 & 265$\fdg$9143 & -25$\fdg$6378 &   True  \\
11 &  Bl B   & 240349772 & 264$\fdg$2497 & -29$\fdg$6690 &   True  \\
11 &  H 1-18   & 240385695 & 262$\fdg$4282 & -29$\fdg$5473 &   True  \\
11 &  Th 3-23   & 240491097 & 262$\fdg$5890 & -29$\fdg$1702 &   True  \\
11 &  M 4-6   & 240525009 & 263$\fdg$8083 & -29$\fdg$0529 &   True  \\
11 &  IRAS 17253-2831   & 240654613 & 262$\fdg$1373 & -28$\fdg$5572 &   Poss-Pre  \\
11 &  Th 3-19   & 240678451 & 262$\fdg$1742 & -28$\fdg$4555 &   True  \\
11 &  RPZM 25   & 240750595 & 263$\fdg$8996 & -28$\fdg$1237 &   Possible \\
11 &  JaSt 11   & 242165458 & 264$\fdg$7516 & -30$\fdg$1931 &   True  \\
11 &  Th 3-1   & 251248498 & 256$\fdg$4354 & -25$\fdg$4172 &   True  \\
11 &  M 3-40   & 251248499 & 260$\fdg$6179 & -27$\fdg$1451 &   True  \\
11 &  M 3-7   & 251248500 & 261$\fdg$1433 & -29$\fdg$4054 &   True  \\
11 &  PPA J1727-2653   & 251248501 & 261$\fdg$9933 & -26$\fdg$8958 &   Likely  \\
11 &  H 2-15   & 251248503 & 263$\fdg$6117 & -22$\fdg$8888 &   True  \\
11 &  IRAS 17332-2215   & 251248504 & 264$\fdg$0713 & -22$\fdg$2889 &   Poss-Pre  \\
11 &  M 3-12   & 251248505 & 264$\fdg$0942 & -21$\fdg$5201 &   True  \\
11 &  H 2-20   & 251248507 & 266$\fdg$4158 & -25$\fdg$6667 &   True  \\
11 &  K 6-20   & 251248509 & 257$\fdg$1871 & -27$\fdg$5172 &   True  \\
11 &  K 6-21   & 251248510 & 257$\fdg$6808 & -30$\fdg$5411 &   True  \\
11 &  PHR J1710-2946   & 251248511 & 257$\fdg$6992 & -29$\fdg$7772 &   True  \\
11 &  Terz N 5   & 251248512 & 257$\fdg$8117 & -27$\fdg$1919 &   True  \\
11 &  PHR J1712-3114   & 251248514 & 258$\fdg$0421 & -31$\fdg$2444 &   True  \\
11 &  Terz N 8   & 251248515 & 258$\fdg$1400 & -26$\fdg$4226 &   True  \\
11 &  K 6-22   & 251248516 & 258$\fdg$1629 & -30$\fdg$3610 &   True  \\
11 &  Terz N 139   & 251248517 & 258$\fdg$2225 & -30$\fdg$6690 &   True  \\
11 &  PHR J1712-3049   & 251248518 & 258$\fdg$2246 & -30$\fdg$8312 &   True  \\
11 &  PHR J1715-2905   & 251248521 & 258$\fdg$9421 & -29$\fdg$0845 &   True  \\
11 &   PHR J1716-3002   & 251248522 & 259$\fdg$0171 & -30$\fdg$0378 &   True  \\
11 &   RPZM 46   & 251248523 & 259$\fdg$1625 & -27$\fdg$8641 &   Poss-Pre  \\
11 &   PHR J1716-2916   & 251248524 & 259$\fdg$1729 & -29$\fdg$2723 &   True  \\
11 &   PHR J1717-2759   & 251248525 & 259$\fdg$2563 & -27$\fdg$9920 &   True  \\
11 &   Th 3-3   & 251248527 & 259$\fdg$3354 & -28$\fdg$9911 &   True  \\
11 &   CBF J1717-3038   & 251248528 & 259$\fdg$3608 & -30$\fdg$6400 &   Likely  \\
11 &   IRAS 17149-3053   & 251248529 & 259$\fdg$5454 & -30$\fdg$9444 &   Poss-Pre  \\
11 &   PHR J1718-3019   & 251248530 & 259$\fdg$6550 & -30$\fdg$3236 &   True  \\
11 &   PHR J1718-3055   & 251248532 & 259$\fdg$6946 & -30$\fdg$9247 &   True  \\
11 &   Th 3-5   & 251248533 & 259$\fdg$7642 & -30$\fdg$8982 &   True  \\
11 &   Terz N 41   & 251248535 & 260$\fdg$0917 & -24$\fdg$8645 &   True  \\
11 &   PHR J1720-2719   & 251248536 & 260$\fdg$1283 & -27$\fdg$3277 &   True  \\
11 &   PHR J1721-3027   & 251248538 & 260$\fdg$3821 & -30$\fdg$4556 &   True  \\
11 &   Terz N 18   & 251248539 & 260$\fdg$4083 & -28$\fdg$9207 &   True  \\
11 &   BMP J1721-2554   & 251248540 & 260$\fdg$4921 & -25$\fdg$9067 &   True  \\
11 &   PHR J1723-2856   & 251248542 & 260$\fdg$7841 & -28$\fdg$9398 &   Likely  \\
11 &   PPA J1723-3038   & 251248543 & 260$\fdg$8479 & -30$\fdg$6445 &   True  \\
11 &   H 2-7   & 251248544 & 260$\fdg$8537 & -28$\fdg$9850 &   True  \\
11 &   PHR J1724-2543   & 251248545 & 261$\fdg$0179 & -25$\fdg$7205 &   True  \\
11 &   H 2-8   & 251248546 & 261$\fdg$1896 & -21$\fdg$5598 &   True  \\
11 &   Bica 1   & 251248547 & 261$\fdg$2242 & -24$\fdg$3225 &   True  \\
11 &   PPA J1724-3043   & 251248548 & 261$\fdg$2429 & -30$\fdg$7178 &   True  \\
11 &   PHR J1725-2534   & 251248549 & 261$\fdg$3046 & -25$\fdg$5711 &   True  \\
11 &   MPA J1725-3033   & 251248551 & 261$\fdg$3892 & -30$\fdg$5658 &   Likely  \\
11 &   M 3-9   & 251248553 & 261$\fdg$4308 & -26$\fdg$1987 &   True  \\
11 &   PPA J1725-2915   & 251248554 & 261$\fdg$4771 & -29$\fdg$2523 &   Likely  \\
11 &   PBOZ 18   & 251248555 & 261$\fdg$5983 & -30$\fdg$7608 &   True  \\
11 &   IRAS 17233-2602   & 251248556 & 261$\fdg$6196 & -26$\fdg$0827 &   Poss-Pre  \\
11 &   SaWe 2   & 251248558 & 261$\fdg$7517 & -27$\fdg$6764 &   True  \\
11 &   PHR J1727-2747   & 251248560 & 261$\fdg$9979 & -27$\fdg$7986 &   True  \\
11 &   Sab 49   & 251248561 & 262$\fdg$0321 & -30$\fdg$6384 &   True  \\
11 &   PHR J1728-2936   & 251248562 & 262$\fdg$0371 & -29$\fdg$6150 &   Likely  \\
11 &   PHR J1728-3032   & 251248563 & 262$\fdg$0592 & -30$\fdg$5372 &   True  \\
11 &   PHR J1728-2743   & 251248564 & 262$\fdg$1833 & -27$\fdg$7275 &   Likely  \\
11 &   PHR J1729-2354   & 251248565 & 262$\fdg$2942 & -23$\fdg$9084 &   True  \\
11 &   MPA J1729-3016   & 251248566 & 262$\fdg$3463 & -30$\fdg$2801 &   True  \\
11 &   MGE 357$\fdg$9893+02$\fdg$8341   & 251248567 & 262.4362 & -29.1158 &   Possible \\
11 &   K 6-25   & 251248568 & 262$\fdg$4721 & -29$\fdg$9529 &   True  \\
11 &   Al 2-F   & 251248569 & 262$\fdg$6266 & -28$\fdg$5986 &   True  \\
11 &   Th 3-24   & 251248571 & 262$\fdg$7137 & -30$\fdg$2868 &   True  \\
11 &   PHR J1731-2927   & 251248572 & 262$\fdg$8408 & -29$\fdg$4644 &   Likely  \\
11 &   PPA J1731-2955   & 251248573 & 262$\fdg$8946 & -29$\fdg$9272 &   True  \\
11 &   PHR J1732-2408   & 251248574 & 263$\fdg$2425 & -24$\fdg$1459 &   True  \\
11 &   K 6-26   & 251248575 & 263$\fdg$2792 & -23$\fdg$4667 &   True  \\
11 &   PHR J1733-1836   & 251248576 & 263$\fdg$3163 & -18$\fdg$6045 &   True  \\
11 &   PPA J1733-2415   & 251248577 & 263$\fdg$3421 & -24$\fdg$2512 &   True  \\
11 &   PHR J1733-2407   & 251248578 & 263$\fdg$4075 & -24$\fdg$1295 &   Likely  \\
11 &   PPA J1733-2908   & 251248579 & 263$\fdg$4188 & -29$\fdg$1428 &   Likely  \\
11 &   PPA J1734-2954   & 251248580 & 263$\fdg$5071 & -29$\fdg$9098 &   True  \\
11 &   PHR J1734-2944   & 251248583 & 263$\fdg$6208 & -29$\fdg$7414 &   Likely  \\
11 &   JaSt 2-1   & 251248584 & 263$\fdg$6217 & -29$\fdg$0345 &   Possible \\
11 &   JaSt 1   & 251248585 & 263$\fdg$6817 & -29$\fdg$7847 &   True \\
11 &   PPA J1734-3004   & 251248588 & 263$\fdg$6942 & -30$\fdg$0725 &   True  \\
11 &   JaSt 2   & 251248589 & 263$\fdg$7538 & -29$\fdg$3710 &   True  \\
11 &   PPA J1735-2809   & 251248590 & 263$\fdg$8000 & -28$\fdg$1586 &   True  \\
11 &   Th 3-32   & 251248591 & 263$\fdg$8146 & -28$\fdg$1171 &   Likely  \\
11 &   JaSt 3   & 251248592 & 263$\fdg$8446 & -29$\fdg$3715 &   True  \\
11 &   K 5-6   & 251248593 & 263$\fdg$8800 & -23$\fdg$1966 &   True  \\
11 &   MGE 358$\fdg$2600+01$\fdg$4563   & 251248595 & 263.9316 & -29.6394 &   Possible \\
11 &   JaSt 5   & 251248596 & 263$\fdg$9679 & -28$\fdg$9743 &   True  \\
11 &   K 6-27   & 251248597 & 263$\fdg$9742 & -22$\fdg$3338 &   True  \\
11 &   Al 2-K   & 251248598 & 264$\fdg$0592 & -28$\fdg$0129 &   True  \\
11 &   MGE 358$\fdg$8567+01$\fdg$6271   & 251248599 & 264.1350 & -29.0450 &   Possible \\
11 &   PPA J1736-2804   & 251248601 & 264$\fdg$2367 & -28$\fdg$0783 &   Likely  \\
11 &   K 5-7   & 251248603 & 264$\fdg$3342 & -24$\fdg$0582 &   True  \\
11 &   PN G359$\fdg$6+01$\fdg$9   & 251248604 & 264.3363 & -28.1686 &   Possible \\
11 &   PPA J1737-2341   & 251248605 & 264$\fdg$3683 & -23$\fdg$6919 &   Possible \\
11 &   MGE 358$\fdg$8655+01$\fdg$2862   & 251248606 & 264.4696 & -29.2203 &   Possible \\
11 &   PPA J1738-2800   & 251248607 & 264$\fdg$5492 & -28$\fdg$0020 &   True  \\
11 &   JaSt 8   & 251248609 & 264$\fdg$6154 & -28$\fdg$8671 &   True  \\
11 &   JaSt 9   & 251248611 & 264$\fdg$6896 & -29$\fdg$1493 &   True  \\
11 &   JaSt 20   & 251248615 & 264$\fdg$9608 & -29$\fdg$0801 &   Possible \\
11 &   JaSt 23   & 251248617 & 265$\fdg$0958 & -27$\fdg$8200 &   True  \\
11 &   JaSt 26   & 251248618 & 265$\fdg$1396 & -29$\fdg$7708 &   True  \\
11 &   PBOZ 25   & 251248619 & 265$\fdg$1454 & -28$\fdg$6931 &   True  \\
11 &   PBOZ 26   & 251248620 & 265$\fdg$1608 & -28$\fdg$0171 &   True  \\
11 &   JaSt 27   & 251248621 & 265$\fdg$1758 & -28$\fdg$2088 &   True  \\
11 &   JaSt 2-3   & 251248622 & 265$\fdg$2892 & -28$\fdg$4650 &   True  \\
11 &   JaSt 36   & 251248624 & 265$\fdg$6046 & -27$\fdg$9267 &   True  \\
11 &   GRM 1   & 251248625 & 265$\fdg$8621 & -29$\fdg$7920 &   Possible \\
11 &   JaSt 47   & 251248626 & 265$\fdg$8892 & -28$\fdg$3121 &   Possible \\
11 &   JaFu 1   & 251248627 & 265$\fdg$9883 & -26$\fdg$1982 &   True  \\
11 &   PHR J1744-2545   & 251248628 & 266$\fdg$0013 & -25$\fdg$7600 &   True  \\
11 &   PPA J1744-2605   & 251248629 & 266$\fdg$0071 & -26$\fdg$0959 &   Likely  \\
11 &   JaSt 50   & 251248630 & 266$\fdg$1396 & -26$\fdg$4150 &   True  \\
11 &   JaSt 51   & 251248632 & 266$\fdg$1671 & -29$\fdg$3734 &   Possible \\
11 &   PPA J1745-2542   & 251248633 & 266$\fdg$3287 & -25$\fdg$7014 &   True  \\
11 &   JaSt 55   & 251248635 & 266$\fdg$4054 & -27$\fdg$0216 &   True  \\
11 &   PPA J1745-2514   & 251248636 & 266$\fdg$4167 & -25$\fdg$2349 &   Possible \\
11 &   MPA J1745-2626   & 251248638 & 266$\fdg$4787 & -26$\fdg$4453 &   Likely  \\
11 &   MGE 003$\fdg$2610+01$\fdg$8522   & 251248639 & 266.5338 & -25.1881 &   Possible \\
11 &   MGE 003$\fdg$0836+01$\fdg$6435   & 251248640 & 266.6292 & -25.4481 &   Possible \\
11 &   MGE 003$\fdg$5004+01$\fdg$8442   & 251248641 & 266.6783 & -24.9878 &   Possible \\
11 &   MGE 002$\fdg$8530+01$\fdg$4514   & 251248642 & 266.6787 & -25.7447 &   Possible \\
11 &   PHR J1746-2611   & 251248643 & 266$\fdg$7167 & -26$\fdg$1973 &   True  \\
11 &   MPA J1747-2649   & 251248646 & 266$\fdg$8679 & -26$\fdg$8302 &   True  \\
11 &   Terz N 2111   & 251248648 & 267$\fdg$1183 & -24$\fdg$6900 &   True  \\
11 &   PPA J1748-2427   & 251248649 & 267$\fdg$1346 & -24$\fdg$4528 &   Possible \\
11 &   Terz N 2337   & 251248650 & 267$\fdg$1879 & -26$\fdg$7235 &   True  \\
11 &   K 6-29   & 251248652 & 267$\fdg$4154 & -24$\fdg$5371 &   True  \\
11 &   PN PM 1-182   & 251248653 & 267$\fdg$8454 & -25$\fdg$0314 &   Possible \\
11 &   PHR J1752-2527   & 251248654 & 268$\fdg$0358 & -25$\fdg$4633 &   Possible \\

\hline
\enddata
\end{deluxetable}


\bsp	
\label{lastpage}
\end{document}